# Did Mars possess a dense atmosphere during the first ~400 million years?

M. Scherf[1], H. Lammer[1]

[1]Austrian Academy of Sciences, Space Research Institute, Graz, Austria (manuel.scherf@oeaw.ac.at, helmut.lammer@oeaw.ac.at);

**Abstract** It is not yet entirely clear whether Mars began as a warm and wet planet that evolved towards the present-day cold and dry body or if it always was cold and dry with just some sporadic episodes of liquid water on its surface. An important clue into this question can be gained by studying the earliest evolution of the Martian atmosphere and whether it was dense and stable to maintain a warm and wet climate or tenuous and susceptible to strong atmospheric escape. In this review we therefore discuss relevant aspects for the evolution and stability of a potential early Martian atmosphere. This contains the EUV flux evolution of the young Sun, the formation timescale and volatile inventory of the planet including volcanic degassing, impact delivery and removal, the loss of the catastrophically outgassed steam atmosphere, atmosphere-surface interactions, as well as thermal and non-thermal escape processes affecting a potential secondary atmosphere at early Mars. While early non-thermal escape at Mars before 4 billion years ago is poorly understood, in particular in view of its ancient intrinsic magnetic field, research on thermal escape processes and the stability of a $CO_2$-dominated atmosphere around Mars against high EUV fluxes indicate that volatile delivery and volcanic degassing cannot counterbalance the strong thermal escape. Therefore, a catastrophically outgassed steam atmosphere of several bars of $CO_2$ and $H_2O$, or CO and $H_2$ for reduced conditions, through solidification of the Martian magma ocean could have been lost within just a few million years. Thereafter, Mars likely could not build up a dense secondary atmosphere during its first ~400 million years but might only have possessed an atmosphere sporadically during events of strong volcanic degassing, potentially also including $SO_2$. This indicates that before ~4.1 billion years ago Mars indeed might have been cold and dry. A denser $CO_2$- or CO-dominated atmosphere, however, might have built up afterwards but must have been lost later-on due to non-thermal escape processes and sequestration into the ground.

## 1 Introduction

Today Mars only possesses a very tenuous atmosphere of only about 6 mbar consisting mainly of $CO_2$ (see e.g., Mahaffy et al., 2013; Webster et al., 2013; Fränz et al., 2017). Early Mars, however, in particular during the Noachian eon – roughly the period from ~4.0 to ~3.7 billion years ago (e.g., Werner and Tanaka, 2011) – and probably even later, is believed to have had periods with at least sporadic episodes of liquid water on its surface (e.g., J. Bibring et al., 2006; Ehlmann et al., 2016; Mangold et al., 2016, and references therein; Kite, 2019). Several recent studies are in addition suggesting that Mars could have had a denser atmosphere during the Noachian eon with an approximate pressure of around 500 mbar to 1 bar ( Hu et al. 2015; Amerstorfer et al. 2017; Cravens et al. 2017; Jakosky et al. 2018; Kurokawa et al. 2018) but not more than about 2 bar (Kite et al., 2014; Warren et al., 2019). In the Hesperian eon, at 3.5 billion years ago (Ga) atmospheric pressure did already not exceed a few 10s of mbar (Bristow et al., 2016).

For the pre-Noachian eon, however, the time before 4.0 to 4.1 Ga, almost no geological records as well as empirical constraints on Mars' atmosphere exist. But there are some studies that modelled and discussed the escape of its early atmosphere (Tian et al., 2009; Lammer et al., 2013; N. V Erkaev et al., 2014a; Yoshida and Kuramoto, 2020). Odert et al. (2018) studied, in addition, the escape of steam atmospheres from Mars-sized planetary embryos at different orbital locations around the Sun.

There are several key factors that have to be taken into account to get a better understanding of the very early evolution of the Martian atmosphere and, thus, to evaluate whether Mars possessed an atmosphere during the first ~400 million years or not. The main questions to be resolved are the following:

- Did Mars form early, i.e. within the disk lifetime (e.g., Dauphas and Pourmand, 2011)? If yes, did it accrete a hydrogen atmosphere which must have been lost later on (e.g., N. V Erkaev et al., 2014a)?
- It is very likely that Mars outgassed a steam atmosphere from its solidifying magma ocean (Elkins-Tanton, 2008; Cannon et al., 2017). How much $H_2O$ and $CO_2$, predominantly in case of oxidized, or CO and $H_2$, in case of reduced conditions, was outgassed from the solidifying Martian magma ocean and could subsequently be lost to space before the steam atmosphere condensed (e.g., N. V Erkaev et al., 2014a; Odert et al., 2018)? To prevent it from condensation a shallow magma ocean below the steam atmosphere might be needed.
- What are the initial volatile content and outgassing history of Mars? Could outgassing counterbalance atmospheric loss (Tian et al., 2009; Lammer et al., 2013; N. V Erkaev et al., 2014a)?
- What is the role of early impactors; did they remove part of the atmosphere or deliver volatiles? Did they increase or decrease the atmospheric pressure (De Niem et al., 2012; Lammer et al., 2013; Pham and Karatekin, 2016)? Could impacts further help to maintain a shallow magma ocean over several million years (Maindl et al., 2015)?
- How strong was thermal escape on early Mars (Tian et al., 2009; Yoshida and Kuramoto, 2020)? Was an early $CO_2$ atmosphere stable against the high EUV flux of the young Sun?
- What role did non-thermal escape processes played on Mars during the first ~400-600 million years (Terada et al., 2009; Lammer et al., 2013; Amerstorfer et al., 2017; Zhao et al., 2017; B. M. Jakosky et al., 2018; Dong et al., 2018a; Sakata et al., 2020)? In case Mars had an intrinsic magnetic field early on (Acuna et al., 1999; Lillis et al., 2013), was it decreasing or increasing atmospheric escape (e.g., Blackman and Tarduno, 2018; Gunell et al., 2018; Egan et al., 2019)?

- Could a $CO_2$-dominated atmosphere be stable on early Mars? Under which conditions can it collapse to form permanent $CO_2$ ices (e.g., Forget et al., 2013; Wordsworth et al., 2013; Soto et al., 2015), or will be converted to CO (Zahnle et al., 2008)?
- The Martian surface evolution is coupled with the evolution of its atmosphere. How do atmosphere-surface interactions affect a potential early atmosphere and what can we learn from the remaining Martian water inventory, its carbonates and the oxidation of its soil (e.g., Lammer et al., 2003b; Niles et al., 2013; Hu et al., 2015)?
- And finally: How did the EUV flux and mass loss evolution of the young Sun evolve (e.g. Wood et al. 2005; Guedel 2007; Ribas et al. 2005; Tu et al. 2015, Johnstone et al. 2015b; Airapetian and Usmanov 2016)? This is particularly important to understand and quantify the early escape processes in the solar system and most importantly of early Mars, which was due to its small size highly susceptible to atmospheric escape (Tian et al., 2009; Lammer et al., 2013; N. V Erkaev et al., 2014a; Odert et al., 2018).

In Section 2 we will briefly review our current understanding of the EUV flux and solar wind evolution of the young Sun which will be followed by Section 3 on the Martian formation timescales and the therewith connected potential accumulation and loss of a primordial hydrogen atmosphere. In Section 4 we will discuss the initial Martian volatile content together with its catastrophically outgassed steam atmosphere and how it might have escaped to space. Section 5 will review the influence of impactors and volcanic degassing on the early atmospheric pressure, whereas Section 6 and 7 will discuss the influence of thermal and non-thermal escape processes, respectively, and the potential influence of the intrinsic magnetic field of Mars. Section 8 will discuss other factors that might affect the stability and evolution of an early atmosphere such as atmospheric collapse and interactions with the surface. Section 9 will discuss most likely scenarios for the evolution of the early Martian atmosphere and potential future steps to better constrain its early atmospheric history. Section 10 will conclude the review.

## 2 The EUV flux and solar wind evolution of the young Sun

It is already known for a long time that the solar bolometric luminosity is increasing over time (e.g., Sackmann and Boothroyd, 2002; Guedel, 2007). The x-ray and EUV flux evolution on the other hand steadily decreased since the Sun arrived at the main sequence (e.g., Ribas et al., 2005; Guedel, 2007; Claire et al., 2012). In addition, observations of solar analogues show that the X-ray and the therewith connected EUV flux is initially saturated at very high levels before starting to continuously decay to lower values (e.g., Tu et al., 2015). This has significant importance for the evolution of early atmospheres since high EUV fluxes have a pronounced effect on the atmospheric structure of terrestrial planets and thus on atmospheric escape (see e.g., Johnstone et al. 2019, 2020; Kulikov et al. 2006, 2007; Tian et al. 2008a, 2008b, 2009). A reconstruction of the EUV flux of the young Sun is therefore crucial for a better understanding of the evolution and escape of an early Martian atmosphere.

Recent observations of solar-like stars of different ages found that the magnetic and EUV flux evolution of these stars are strongly dependent on their initial rotation rate (Johnstone et al., 2015a; Tu et al., 2015), as can be seen in Fig. 1a. Initially fast rotating solar analogues have a much longer saturation phase (i.e. about 6, 25, and 230 million years for a slow, moderate, and fast rotators) with their EUV flux declining much slower over time than initially slow or moderate rotating solar-like stars. After approximately one billion years, however, all of these different tracks converged towards one (Tu et al., 2015). This makes it difficult to precisely reconstruct the EUV flux history of the Sun, since its past evolution cannot be simply derived from its present-day behaviour.

It has further to be emphasized that, before converging, the difference in EUV

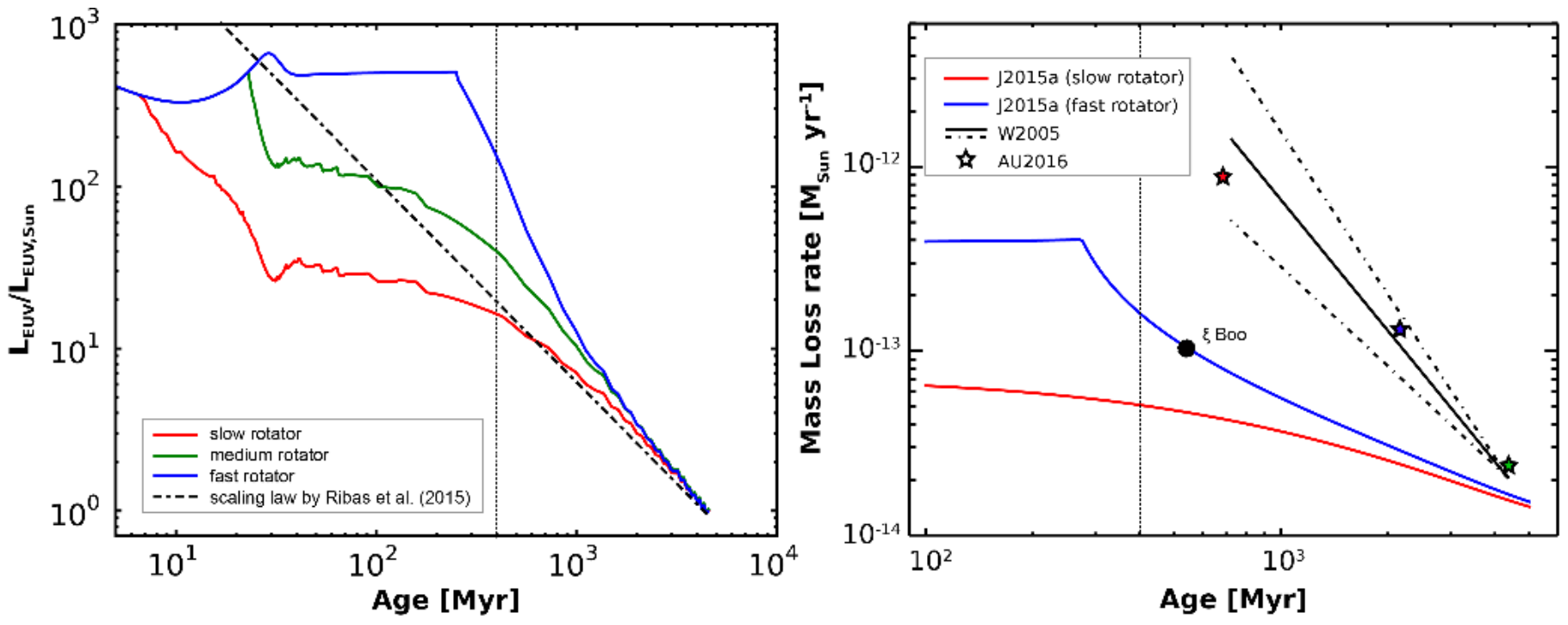

**Fig. 1. Left panel: EUV flux evolution of the young Sun according to Tu et al. 2015 (scaling law by Ribas et al. (2015) for comparison). Depending on whether the Sun was a slow, moderate or fast rotator, the EUV flux could have been different by an order of magnitude for the first >100 Myr. Right: Evolution of the solar mass loss according to Johnstone et al. (2015b), Wood et al. (2005), and Airapetian and Usmanov (2006). The stars illustrate simulations by Airapetian and Usmanov (2006), while the circle depicts observations of the star ξ Boo. Left figure adopted from Tu et al. (2015) and Amerstorfer et al. (2017), right figure adopted from Johnstone et al. (2015) and Airapetian and Usmanov (2016).**

fluxes is likely to be higher than an order of magnitude between a slow and a fast rotator. At 400 million years after formation of the solar system, for instance, the EUV flux of a fast rotating Sun would have been up to 150 times larger than at present-day, but only about 15 times in the case of a slow rotator. Which of these different tracks the early Sun followed on its evolution is not yet very well established. There are, however, significant hints that the Sun was likely a slow rotator, such as reproduction attempts of Ar and Ne isotopic ratios in the atmospheres of Venus and Earth (Odert et al., 2018; Lammer et al., 2020), the lunar K/U ratio (Saxena et al., 2019) and the stability of the terrestrial atmosphere (Johnstone et al., 2020).

For non-thermal escape processes, the solar mass loss evolution of the Sun has to be taken into account as well, since the solar wind can significantly affect atmospheric erosion (Dong et al., 2018a). Observational studies of stellar astrospheres indicate that the solar wind might have been higher in the past, but in particular for the very early stages, the evolution of the solar and of stellar winds are not very well established (e.g., Wood et al., 2002, 2005). The black solid, and dash-dotted lines in Fig. 1b illustrate the mass loss evolution of the Sun as inferred by Wood et al. (2005), through the observation of stellar astrospheres of different ages. The solar mass loss evolution is only extrapolated back to about 700 Myr, since observations of the young star ξ Boo indicated a suddenly weaker mass loss rate as one goes back in time for solar-like stars (Wood et al. 2005).

Johnstone et al. (2015a, 2015b) developed a 1D thermal pressure driven hydrodynamic stellar wind evolution model for solar-like stars that is calibrated with present-day solar wind data. Their model assumes that the mass loss rate $\dot{M}_\star$ of a star depends on its rotation rate $\Omega_\star$ through $\dot{M}_\star \propto \Omega_\star^a$, and found $a$ to be ~ 1.33. Fig. 1b shows the related solar mass loss evolution by Johnstone et al. (2015b), with red and blue lines illustrating the slow and fast rotator tracks, respectively. One can clearly see that the evolution as inferred by Johnstone et al. (2015a) indicates significantly lower mass loss rates in the past than the one by Wood et al. (2005). Their results suggest a much weaker dependence of the mass loss rate on time than the $t^{-2.33}$ dependence found by Woods et al. (2005). Johnstone et al. (2015b) argue that a reason for this could be that Wood et al. (2005) derived their power law through a combination of a wider stellar mass range that also included more active K stars. Interestingly, ξ Boo falls within the range of a fast-rotating star in the model of Johnstone et al. (2015b).

Another study (Airapetian and Usmanov, 2016) that simulated the mass los of solar-like stars with different age, however, found similar values than Wood et al. (2005). Airapetian and Usmanov (2016) applied a three-dimensional magnetohydrodynamic Alfvén wave driven solar wind model, called ALF3D, and modelled stars with ages of 0.7, 2, and 4.65 Gyr. Their results are well within the extrapolation by Wood et al. (2005), and they found that for 0.7 Gyr the solar wind was about 50 times as dense as at present-day. This is about 25 times denser than for a slow rotator in the model of Johnstone et al. (2015b), and yet about 20 and 15 times as dense as for a moderate and fast rotator, respectively.

Finally, one also has to note that extreme solar events were likely significantly more frequent in the early solar system, including CMEs, solar flares and potentially even superflares (e.g., Notsu et al., 2013; Shibata et al., 2013; Airapetian et al., 2016; Odert et al., 2017; Kay et al., 2019; Saxena et al., 2019). The high EUV flux of flares and superflares might have driven extreme thermal escape, higher than for nominal early solar wind conditions, while frequent and strong CMEs can likewise substantially influence non-thermal loss processes.

Two of the important drivers of non-thermal escape, as we will discuss in Chapter 7, are solar wind density and velocity. Regarding the present-day solar wind evolution scenarios discussed above, it will, therefore, be crucial to better constrain the solar mass loss history of the Sun –including the frequency distribution of interplanetary CMEs – if one wants to understand the influence of non-thermal escape processes on an early Martian atmosphere. For thermal escape, on the other hand, the above discussed evolution of the Sun's EUV flux and flaring rate is a particularly crucial factor. Its influence on the early Martian atmosphere will be discussed in more detail in Chapter 6.

## 3 Martian formation timescale and its potential primordial hydrogen atmosphere

The formation of terrestrial planets is generally believed to take place within ≤100 million years (e.g. (Chambers and Wetherill, 1998; Kokubo and Ida, 1998; Chambers, 2001; Raymond et al., 2009; Morbidelli et al., 2012). A problem within these classical models, however, constituted the formation of Mars, which is known as the "small Mars-problem" (Chambers, 2001; Raymond et al., 2009; Walsh et al., 2011). Several solutions for this challenge have been proposed (see e.g. Brasser (2013) for a review) such as the "Nice model" (O'Brien et al. 2006), the "Grant-Tack scenario" (Walsh et al., 2011) or the formation of the terrestrial planets within a narrow annulus which prevented Mars from growing any further (Hansen, 2009).

All these models have in common that Mars formed very early and that it constitutes a planetary embryo which did not grow to a full terrestrial planet. A fast formation timescale is also consistent with $^{182}Hf$-$^{182}W$- and $^{60}Fe$-$^{60}Ni$-chronologies, which strongly support the idea that Mars formed very early, growing to about half of its mass within about 1.8 million years (Dauphas and Pourmand, 2011; Tang and Dauphas, 2014). It can therefore be considered to be likely that Mars grew to its almost full size within the lifetime of the solar nebula which was recently estimated to be about 3 to 4 million years (Bollard et al., 2017; Wang et al., 2017).

If the planet almost finished its growth before the evaporation of the solar nebula, it could have accreted a thin hydrogen-dominated atmosphere (N. V Erkaev et al., 2014a; Stökl et al., 2015, 2016). Recently, Saito and Kuramoto (2018) even suggested that Mars could have accumulated a hybrid-type proto-atmosphere of up to several kbar due to accretion of solar nebula gases. Such a potential primordial envelope around Mars, however, would not be stable and is expected to be lost completely immediately after the dissipation of the solar nebula. When the gas disk evaporates its pressure onto the accreted atmosphere diminishes and the system consequently moves out of equilibrium. This results in a short phase of extreme thermal hydrodynamic loss which is termed as the so-called "blow-off" phase (Owen and Wu, 2015; Fossati et al., 2017). Even if a tiny hydrogen envelope could remain afterwards, it would be lost within less than 1 million years or even faster due to strong EUV-driven hydrodynamic escape (N. V. Erkaev et al., 2014; Massol et al., 2016). Strong escape rates of a hypothetical proto-atmosphere were recently also found by Yoshida and Kuramoto (2020) who investigated the loss of a primordial Martian atmosphere that consisted of a hydrogen-dominated solar and a degassed reduced component, together consisting mainly of $H_2$, $CH_4$ and CO. While in the case of Erkaev et al. (2014), all the $CO_2$ molecules were assumed to be dissociated by the strong solar EUV flux, a significant fraction of the $CH_4$ and CO molecules survived in the lower region of the atmosphere, as studied by Yoshida and Kuramoto (2020), thereby leading to radiative cooling. This effect increased the time of the escape of the whole hydrogen atmosphere by one order of magnitude to about 10 million years if they assumed the same atmospheric mass as Erkaev et al. (2014).

For significantly higher masses of up to 300 bar $H_2$, the whole primordial atmosphere was lost within about 25 Myr. In addition to the hydrogen, the Martian atmosphere also lost carbon with an equivalent value of up to 20 bar $CO_2$ (Yoshida and Kuramoto, 2020) which is in agreement with Erkaev et al. (2014).

To summarize, Mars likely was too small to accrete a hydrogen-dominated primordial atmosphere from the solar nebula that survived the blow-off. If, however, such an atmosphere could have been accreted, it would have been lost rapidly. A primordial hydrogen atmosphere around Mars was therefore – if existing – not stable and can thus be considered negligible for the subsequent atmospheric evolution of the planet.

# 4 Initial Martian volatile inventory and its catastrophically outgassed steam atmosphere

## *4.1 Outgassing of the steam atmosphere*

To evaluate whether Mars could have sustained an atmosphere after the disk evaporated one has to estimate the initial volatile content of the planet. This will affect the amount of a catastrophically outgassed steam atmosphere during magma ocean solidification and the subsequent volcanic degassing later in its history.

Estimates for the initial volatile amount of Mars are varying. Lunine et al. (2003) found that the initial water inventory of the planet should have been about 0.06-0.27 times an Earth Ocean which would in gaseous form be equivalent to a surface pressure of about 10-100 bar. Another study by Hansen (2009) found that Mars could have been much drier. If Mars' formation, however, was affected by the migration of the giant planets, as in the Grand Tack scenario (Brien et al., 2014), then water should have been delivered to Mars from farther outside of the solar system making the planet more volatile rich than previously thought (Brasser 2013). Based on the Grant-Tack scenario, Rubie et al. (2015) found water contents on the Mars-analogues within the simulations varying between ~ 0.04 wt% and ~ 0.48 wt%. From another perspective, McCubbin et al. (2012), for instance, found through petrological studies of magmatic apatites in geochemically depleted and enriched Martian meteorites, the shergottites, that their parent magma contained 0.073 - 0.287 wt% of $H_2O$ prior to degassing and 0.0073 – 0.029 wt% afterwards. In a more recent study, they estimated the present-day water content of the Martian mantle and crust through the study of 12 different meteorites (Mccubbin et al., 2016). They found ~0.14 wt%, or an equivalent global layer (EGL) of ~ 229 EGL, of $H_2O$, still residing in the crust, while the Martian mantle seems to be significantly dryer with up to only 0.0073 wt% $H_2O$ in the geochemical sources identified by McCubben et al. (2016). They further point out, however, that the present information and samples available are severely limited and that the estimates of the present-day Martian volatile content, therefore, remain, vague. Similarly, any estimates on its initial volatile content can only represent a broad, but uncertain spectrum.

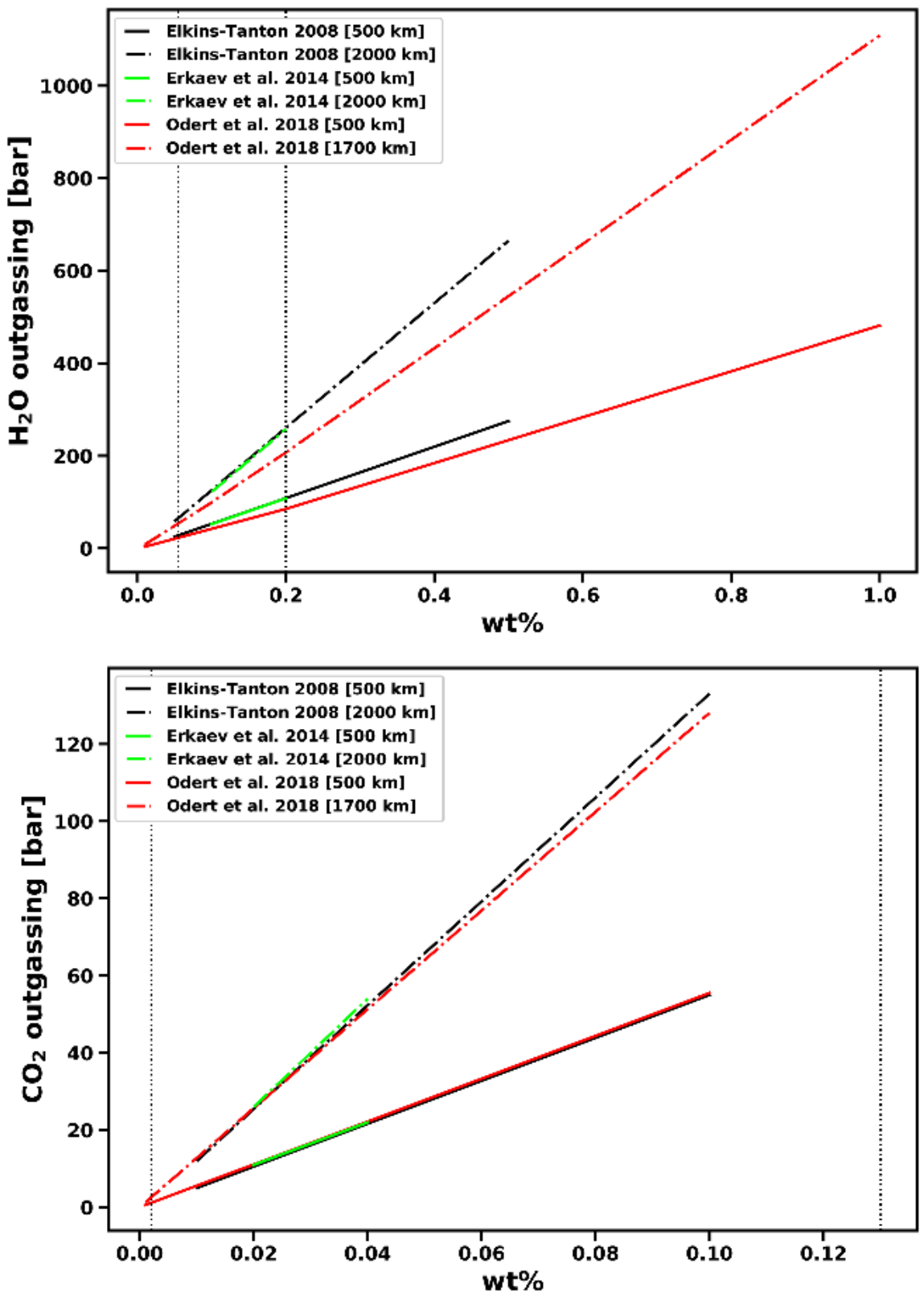

**Fig. 2. Different estimates of the Martian volatile content for $H_2O$ (a) and $CO_2$ (b), and the therewith connected outgassed steam atmospheres. The vertical dotted lines illustrate minimum and maximum estimates for the H2O (Jambon and Zimmermann, 1990) and CO2 (Dasgupta and Hirschmann, 2010) contents of the primitive terrestrial mantle.**

There is evidence from U-Pb and $^{146}$Sm-

$^{142}Nd$ chronology that Mars possessed a magma ocean that subsequently solidified (Debaille et al., 2007; Bouvier et al., 2018). A substantial amount of the above mentioned inventory could have been catastrophically outgassed during the solidification of its magma ocean (Elkins-Tanton, 2008). Depending on its depth (~500-2000 km) and on the initial volatile inventory (~0.05-0.1 wt% for $H_2O$ and ~0.01-0.02 wt% for $CO_2$), Mars could have thus likely outgassed between 30-120 bar $H_2O$ and 7-25 bar $CO_2$, or even up to 665 bar $H_2O$ and 133 bar $CO_2$ if one considers a significantly larger, but likely unrealistic, volatile content of 0.5 wt% $H_2O$, and 0.1 wt% $CO_2$, respectively (Elkins-Tanton, 2008; Lebrun et al., 2013; N. V Erkaev et al., 2014a). Some of the $H_2O$ and $CO_2$, however, would have remained in the mantle being outgassed subsequently via volcanic degassing (Elkins-Tanton, 2008).

**Table 1**. Examples of estimates of the Martian volatile content for $H_2O$ and $CO_2$, and the therewith connected outgassed steam atmosphere. Note that realistic estimates for $H_2O$ and $CO_2$ are in the range of ~0.05-0.1 wt% and ~0.01-0.02 wt%, respectively; potentially less realistic values are written in italic.

| | magma ocean depth [km] | $H_2O$ [wt%] | $CO_2$ [wt%] | $pH_2O$ [bar] | $pCO_2$ [bar] |
|---|---|---|---|---|---|
| Elkins-Tanton (2008) | 500 | 0.05 | 0.01 | 25 | 5 |
| | | *0.5* | *0.1* | *275* | *55* |
| | 2000 | 0.05 | 0.01 | 58 | 12 |
| | | *0.5* | *0.1* | *665* | *133* |
| Erkaev et al. (2014) | 500 | 0.1 | 0.02 | 52 | 11 |
| | | 0.2 | 0.04 | 108 | 22 |
| | 2000 | 0.1 | 0.02 | 122 | 26 |
| | | 0.2 | 0.04 | 257 | 54 |
| Odert et al. (2018)* | 500 | 0.01 | 0.001 | 3.3 | 0.6 |
| | | 0.1 | 0.01 | 42 | 5.6 |
| | | 0.2 | 0.02 | 85 | 11.1 |
| | | *1* | *0.1* | *481.6* | *55.5* |
| | 1700 | 0.01 | 0.001 | 7.9 | 1.3 |
| | | 0.1 | 0.01 | 99.2 | 12.8 |

| | | | | |
|---|---|---|---|---|
| | 0.2 | 0.02 | 207 | 25.7 |
| | *1* | *0.1* | *1108.7* | *128* |

*Estimates for a Mars-sized embryo

Recently Odert et al. (2018) further estimated a brought range of different volatile inventories that could have theoretically been outgassed from the magma ocean of a Mars-sized embryo (see Table 1 and Fig. 2 for different estimates of the initial Martian volatile inventory in particular, and for Mars-sized embryos in general). The parameter space in this study ranged from 0.01 – 1 wt% for $H_2O$ and 0.001-0.1 wt% for $CO_2$ and magma ocean depths from 500 to 1700 km with the latter being equivalent to a globally molten mantle. For comparison, the present-day Earth's upper mantle water concentration is expected to be between 0.005 – 0.02 wt% (Saal et al., 2002); the initial terrestrial primitive mantle between 0.055 – 0.19 wt% (Jambon and Zimmermann, 1990). Carbonaceous chondrites, on the other hand, can have up to 25 wt% $H_2O$ (Garenne et al., 2014). In case of $CO_2$, Earth's mantle has an expected abundance of 0.002 – 0.13 wt% (Dasgupta and Hirschmann, 2010). Based on these assumptions, Odert et al. (2018) retrieved lower and upper limits for the catastrophically outgassed steam atmosphere of 3.3 bar $H_2O$ and 0.6 bar $CO_2$ for a magma ocean depth of 500 km and the lowest volatile inventory, and 1109 bar $H_2O$ and 128 bar $CO_2$ for 1 700 km and the highest volatile inventory, respectively. Possible estimates for the Martian magma ocean, however, modeled in agreement with siderophile-element abundances and silicate partitioning data range from 500 (Agee and Draper 2004; Elkins-Tanton 2012; Righter et al. 1998) to 1 000 km (Elkins-Tanton 2012; Righter and Chabot 2011). If one further assumes a reasonable volatile amount of ~ 0.1 wt% for $H_2O$ and ~0.02 wt% for $CO_2$, a realistic outgassed steam atmosphere based on Odert et al. (2018) might be in the range of 50 to 150 bar $H_2O$ and 10 to 20 bar $CO_2$, or even lower, which is in very good agreement with the estimates of Elkins-Tanton (2008). The different outgassing rates of these studies, including minimum and maximum estimates for the Earth's primitive mantle are also illustrated in Fig. 2.

Table 2. Different estimates of $CO_2$, CO, and $CH_4$ degassing through the formation of a 50 km thick protocrust at Mars for different oxygen fugacities $fO_2$ in log units of the iron-wüstite buffer (IW).

| | $CO_2$ [bar] | CO [bar] | $CH_4$ [bar] | $fO_2$ |
|---|---|---|---|---|
| Hirschmann and Withers (2008) | 0.07 | | | IW-1 |
| | 13 | | | IW+1 |
| Wetzel et al. (2013) | 11.7 | | | >IW-0.55 |
| | | 1 | 1.3 | <IW-0.55 |
| Ramirez et al. (2014) | 5.2 | | | <IW-0.55 |

Another study by Hirschmann and Withers (2008) that investigated the degassing of $CO_2$ from a solidifying Martian magma ocean and the formation of a 50 km thick crust found significantly lower numbers (see also Table 2). Depending on the oxygen fugacity $fO_2$ of the mantle, the degassed amount of $CO_2$ would range from 70 mbar at reducing conditions of one $\log_{10}$ unit below the iron-wüstite (IW-1) buffer to 13 bar for oxidizing conditions of IW+1. Wetzel et al. (2013) also studied the formation of a 50 km thick crust and found outgassing rates of 11.7 bar $CO_2$ for $fO_2$ > IW-0.55 and a mixture of 1 bar CO and 1.3 bar $CH_4$ for $fO_2$ < IW-0.55. In the second case, however, CO and $CH_4$ could have been oxidized by the byproducts of water vapor photolysis, yielding about 5.2 bar of $CO_2$ (Ramses M. Ramirez et al., 2014).

As can be seen, the outgassing rate is strongly depending on the oxygen fugacity of the Martian magma ocean but also on temperature, pressure and $H_2/H_2O$ fugacity (Hirschmann and Withers, 2008; Grott et al., 2011; Wetzel et al., 2013). In addition, magmas under reduced conditions preferentially outgas CO and $CH_4$, whereas oxidized conditions preferentially degas $CO_2$; similarly $H_2$ is favored to be degassed at reduced conditions and $H_2O$ at oxidized.

It is not entirely clear whether the early Martian mantle was more or less reduced than at present-day, even though the bulk of evidence favours a reduced early mantle (see e.g. Ramirez et al. 2014, for a brief discussion, and, for later stages in the Martian history, also Section 8.2 within this review) with reducing conditions at around IW-1 best reproducing the Martian mantle characteristics (Rai and Van

Westrenen, 2013). On the basis of experimentally retrieved spectroscopic data, Grewal et al. (2020) recently predicted that under such conditions a solidifying magma ocean would degas a CO and $NH_3$-rich atmosphere. Another recent theoretical study by Deng et al. (2020), using first-principles simulations combined with thermodynamic modeling, additionally suggests that the Martian magma ocean likely degassed under more reducing conditions than the deeper magma ocean of the Earth, thereby resulting predominantly in outgassed $H_2$ and $H_2O$.

Besides this, a study by Zhang et al. (2017) that is in agreement with Deng et al. (2020) indicates that oxygen fugacities at the surface of shallow magma oceans are more reduced than at their depths which would lead to the outgassing of highly reduced $H_2$ and CO atmospheres. Low mantle oxygen fugacity in addition also traps most of the carbon as graphite within the mantle (Grott et al., 2011; Stanley et al., 2011) which in turn leads to very low outgassing rates and hence also to low atmospheric pressure (Hirschmann and Withers, 2008; Grott et al., 2011; Stanley et al., 2011). Based on these considerations, it can be expected that the degassed $CO_2$ content is at maximum at the lower values of the estimates of Elkins-Tanton (2008), Erkaev et al. (2014) and Odert et al. (2018). The main proportion of outgassed hydrogen and carbon molecules might have even been $H_2$ and CO instead of $H_2O$ and $CO_2$ with a pressure that would in total be lower than for oxidized conditions. This would even result in faster escape, because of a reduced amount of the thermospheric cooler $CO_2$ being present in the upper atmosphere.

## *4.2 Escape of the steam atmosphere*

A steam atmosphere that catastrophically outgassed from a solidifying magma ocean at Mars' orbit has a water condensation time of only about 0.1 million years after the magma ocean solidifyed (Lebrun et al., 2013) which would be below the typical escape time of such an atmosphere (N. V. Erkaev et al., 2014; Odert et al., 2018). This, however, does not take into account that a shallow magma ocean might have remained up to about 100 Myr (Debaille et al., 2007) below the steam atmosphere due to frequent impacts onto the young planet (Maindl et al., 2015). Whether Mars had such a long protracted shallow magma ocean (Debaille et al., 2007) is though debated. A recent study (Bouvier et al., 2018) indicates that Mars did already possess a primordial crust 20 million years after the formation of the solar system. This crust, however, might have been melted after around 100 million years due to frequent impacts (Bouvier et al., 2018) and solidifyed again at about 4.43 billion years ago (Humayun et al., 2013; Bouvier et al., 2018) That there might have been a shallow magma ocean on Mars either during the whole first ~100 million years or at least sporadically due to impact heating might also be supported by a late veneer and the therewith connected impacting material that, after the main accretion of the planet finished, delivered approximately 0.5% of the total Martian mass until within about 4.5 billion years ago and still about 0.1% afterwards (Brasser et al., 2016). Furthermore, another recent study by Brasser and Mojzsis (2017) suggests that at around 4.43 billion years ago a giant impactor of a size of about 1,200 km hit Mars, being probably responsible for the Martian dichotomy. It might therefore be possible that an early steam atmosphere around Mars was not able to condense, or at least not able to stay in condensed form for a longer time, and therefore was susceptible to strong atmospheric escape.

Due to the high EUV flux, $H_2$ and $H_2O$ molecules will mostly be dissociated in the thermosphere so that the upper part of such an atmosphere would mainly be dominated by hydrogen atoms (Kasting and Pollack, 1983; Chassefière, 1996; Yelle, 2004; Koskinen et al., 2010; Lammer et al., 2013; N. V. Erkaev et al., 2014; Guo, 2019). In addition, $CO_2$ will be dissociated as well at high EUV fluxes (Tian et al., 2009) with its constituents subsequently populating the lower thermosphere. These atoms can then be efficiently dragged by the hydrodynamically outflowing hydrogen particles (Zahnle and Kasting, 1986; Hunten et al., 1987; Zahnle et al., 1990; Chassefière, 1996; Chassefière and E., 1996; N. V. Erkaev et al., 2014; Odert et al., 2018).

Erkaev et al. (2014) found that, if $CO_2$ gets dissociated due to the high EUV flux of the young Sun, a ~50 bar $H_2O$ and ~10 bar $CO_2$ atmosphere can completely escape within less than 2.5 million years for low heating efficiences and surface temperatures of 1 500 K. For higher heating efficiencies and temperatures of up to 3000 K such an atmosphere could even escape within less than 0.4 million years. Even an atmosphere with ~260 bar $H_2O$ and ~55 bar $CO_2$ would have been lost within 2.5 to 12 million years which might have been even before the formation of the first primordial crust (Bouvier et al., 2018), if one assumes that the Martian magma ocean solidifyed fast after the solar nebula dissipated, the steam atmosphere outgassed quickly and only a shallow magma ocean remained until the first crust finally formed. According to Erkaev et al. (2014) Mars would therefore not be able to sustain a dense atmosphere very early on. It would either be lost or would have to condense very fast.

It has, however, to be noted that these simulations were performed with an EUV flux evolution based on the study of Ribas et al. (2005). In this pioneering work the saturation phase of the solar EUV flux lasts around 100 million years being as high as about a hundred times the present-day EUV flux (see Fig. 1). This saturation phase is very much longer than a steam atmosphere could survive around Mars. But as already discussed earlier, the more recent study by Tu et al. (2015) found that the EUV flux evolution of the young Sun is strongly dependend on its initial rotational period. For a slowly rotating solar-like star the saturation phase would therefore only be about 5.7 million years with a steep decline in the EUV flux afterwards, whereas it would be up to 23 and 226 million years for a moderate and fast rotating solar-like star, respectively. The complete erosion of a steam atmosphere is therefore, besides its initial mass and its condensation time, strongly dependent on the initial rotation rate of the young Sun.

Odert et al. (2018) studied how the different rotational evolutionary tracks of Tu et al. (2015) affect the escape of a steam atmosphere around Mars-like embryos and the therewith connected dragging of heavier species like C, $CO_2$ and O with a similar model as was applied by Erkaev et al. (2014). They started with their simulation at 10 million years after formation of the solar system. Since Mars likely formed within less than 5 million years and the magma ocean solidification time is expected to be in the order of 0.1 million years (Elkins-Tanton, 2008 e.g. Lebrun et al., 2013), the escape of the steam atmosphere might even have started earlier than assumed in their model.

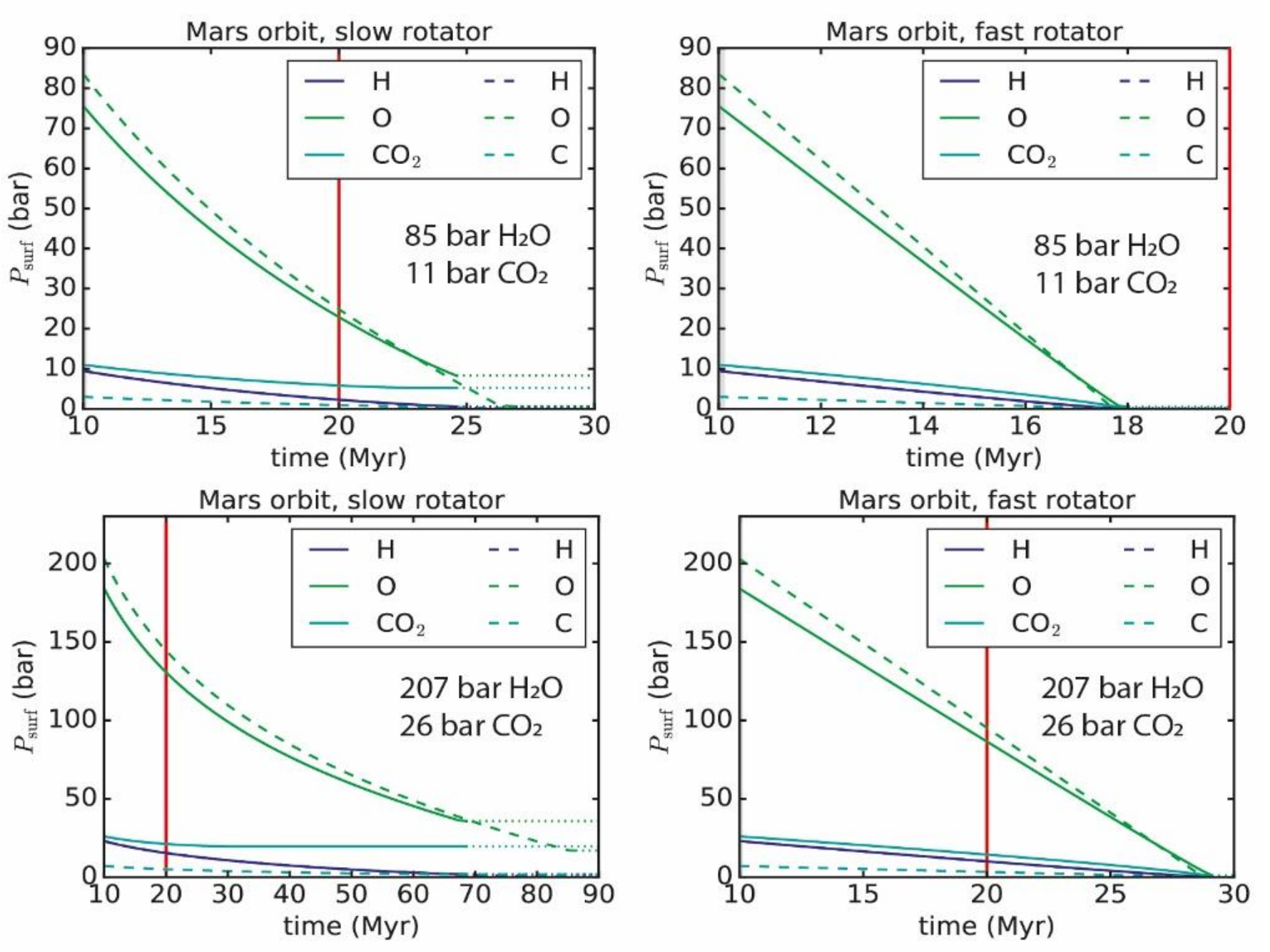

**Fig. 3. Escape of a catastrophically outgassed steam atmosphere according to Odert et al. (2018). For a slow rotating Sun (left panels) part of the $CO_2$ and O will remain in the atmosphere. For a moderate and fast rotating Sun (right panels; here, a moderate rotator yields the same results than a fast rotator due to the long saturation phase) everything will be lost as long as the steam atmosphere does not condense. The red vertical lines illustrate the time of the formation of the first Martian protocrust according to Bouvier et al. (2018). The dashed lines illustrate runs where it was assumed that the CO2 was completely dissociated. Figure adopted from Odert et al. (2018)[1].**

In case that the steam atmosphere will not condense, Odert et al. (2018) showed that for an 85 bar $H_2O$ and 11 bar $CO_2$ atmosphere the hydrogen will escape completely within about 25 million years in case the Sun was a slow rotating solar-like star (see Fig. 3[1], upper left panel). $CO_2$, however, will not be dragged away entirely if it will not dissociate, with about 5 bar remaining at the embryo. In addition, up to about 8 bars of oxygen will remain at the protoplanet due to the dissociation of $H_2O$. For a more massive steam atmosphere, i.e. 207 bar $H_2O$ and 26 bar $CO_2$, the hydrogen will be completely escaped after about 70 million years; 36 bar of O and 19 bar of $CO_2$ will remain (Fig. 3, lower left panel). For the moderate and fast rotator cases (Fig. 3, right panels) all of the outgassed steam atmosphere will be lost within less than 18 million years for the 96 bar atmosphere and before about 30 million years for the much denser 233 bar atmosphere. Due to the long saturation phases there is no significant difference between moderate and fast rotator. A more recent study about the loss of moderately volatile elements from planetary embryos (Benedikt et al., 2020) retrieved similar results for Mars-sized bodies at 1.5 AU. They found that for a moderate to fast rotating Sun a steam atmosphere with at least 150 bar $H_2O$ and 30 bar $CO_2$ around such a body will be lost efficiently within the time of the first crust formation on Mars. Only if the Sun were a slow rotator and the steam atmosphere was not too hot, i.e. below 2000 K, a small fraction of the 150 bar $H_2O$ and 30 bar $CO_2$ could have survived the hydrodynamic escape.

[1] Reprinted from Icarus, 307, P. Odert, H. Lammer, N.V. Erkaev, A. Nikolaou, H.I.M. Lichtenegger, C.P. Johnstone, K.G. Kislyakova, M. Leitzinger, N. Tosi, Escape and fractionation of volatiles and noble gases from Mars-sized planetary embryos and growing protoplanets, 327-346, Copyright (2018), with permission from Elsevier.

Even though the recent study by Yoshida and Kuramoto (2020) did not explicitly study the escape of a catastrophically outgassed steam atmosphere but a mixture of an accreted and degassed proto-atmosphere, their simulations nevertheless illustrate the potential for such an atmosphere to be lost. Since the $H_2O$ of a steam atmosphere will be dissociated into hydrogen and oxygen, it will result in a comparable setup than the studied proto-atmosphere of Yoshida and Kuramoto (2020), particularly if one assumes reducing conditions. An escaping reservoir of carbon species equivalent to up to 20 bar of $CO_2$ is significantly higher for such reducing conditions than suggested by any of the studies that were outlined in the previous subsection.

From these findings it can be concluded that a steam atmosphere would not have been able to survive the strong escape due to the high EUV flux of the young Sun. If the Sun was a slowly rotating solar-like star, and depending on the amount of the outgassed volatiles and the redox conditions of the solidifying magma ocean, at maximum a few bars of $CO_2$ and O, most probably below 10 and 5 bar, respectively, might not have been dragged away by hydrogen and would therefore either be susceptible to further thermal (Section 6) or non-thermal escape processes (Section 7), condensation, or, in case of $CO_2$ and O, sequestration into the crust or oxidation of the soil (see Section 8.2). The $CO_2$ reservoir in carbonates in the Martian crust, however, is estimated to be likely less than 1 bar (Wray et al., 2016), as we will discuss further in Section 8.2. Since not more than approximately 1 bar of $CO_2$ could have additionally been lost into space since the Noachian eon, i.e. ~4 Gyr ago, any other $CO_2$ that might have remained in the atmosphere after the loss of the initially outgassed steam atmosphere must have been further lost to space within the pre-Noachian, i.e. within the first ~400-500 million years by other relevant escape processes. Thermal (Section 6) and non-thermal escape processes (Section 7) likely were strong enough to remove these remnants from the atmosphere later-on.

# 5 The influence of impacts and volcanic degassing

## *5.1 Impacts*

While giant impacts into primordial atmospheres are generally believed to strip away part of the envelope (Genda and Abe, 2003; L.B.S. Pham et al., 2011; Schlichting et al., 2015; Lammer et al., 2020), it is less clear what happens if smaller impactors collide with a planet. Whether such impacting material leads to atmospheric erosion or delivers volatiles has to be studied via complex hydrocode simulations (Melosh and Vickery, 1989; Ahrens and J., 1993; Shuvalov and Artemieva, 2001; Svetsov, 2007; Pham et al., 2009; Shuvalov and V., 2009; L. B.S. Pham et al., 2011; Lammer et al., 2013; Pham and Karatekin, 2016) which, depending on the various assumptions, often lead to different results.

Since these studies need significant computational resources and cannot be used to directly simulate atmospheric evolution, the influence of the different parameters that determine atmospheric erosion and volatile delivery have been parameterized. A common parameterization to study the influence of impactors on atmospheric evolution was developed by Melosh and Vickery (1989) and is the so-called "tangent plane model", which was subsequently adopted by several different authors who studied the evolution of the Martian atmosphere (e.g., Pham et al., 2009; L. B.S. Pham et al., 2011; Lammer et al., 2013; Pham and Karatekin, 2016). Within the "tangent plane model", the total mass $m_{tan}$ above the plane tangent to the surface at the impacting point of an impactor with the critical mass $m_{crit}$ is considered to be lost to space (see, e.g, Lammer et al., 2013). Here, $m_{crit}$ is the minimum impactor mass that can eject $m_{tan}$, with $m_{crit}$ being proportional to $m_{tan}$ through the so-called impactor efficiency $n$ by $n = m_{crit}/m_{tan}$. Fig. 4 illustrates for comparison different Martian atmospheric evolution scenarios from different studies that are based on the tangent plane model but use different impactor efficiencies.

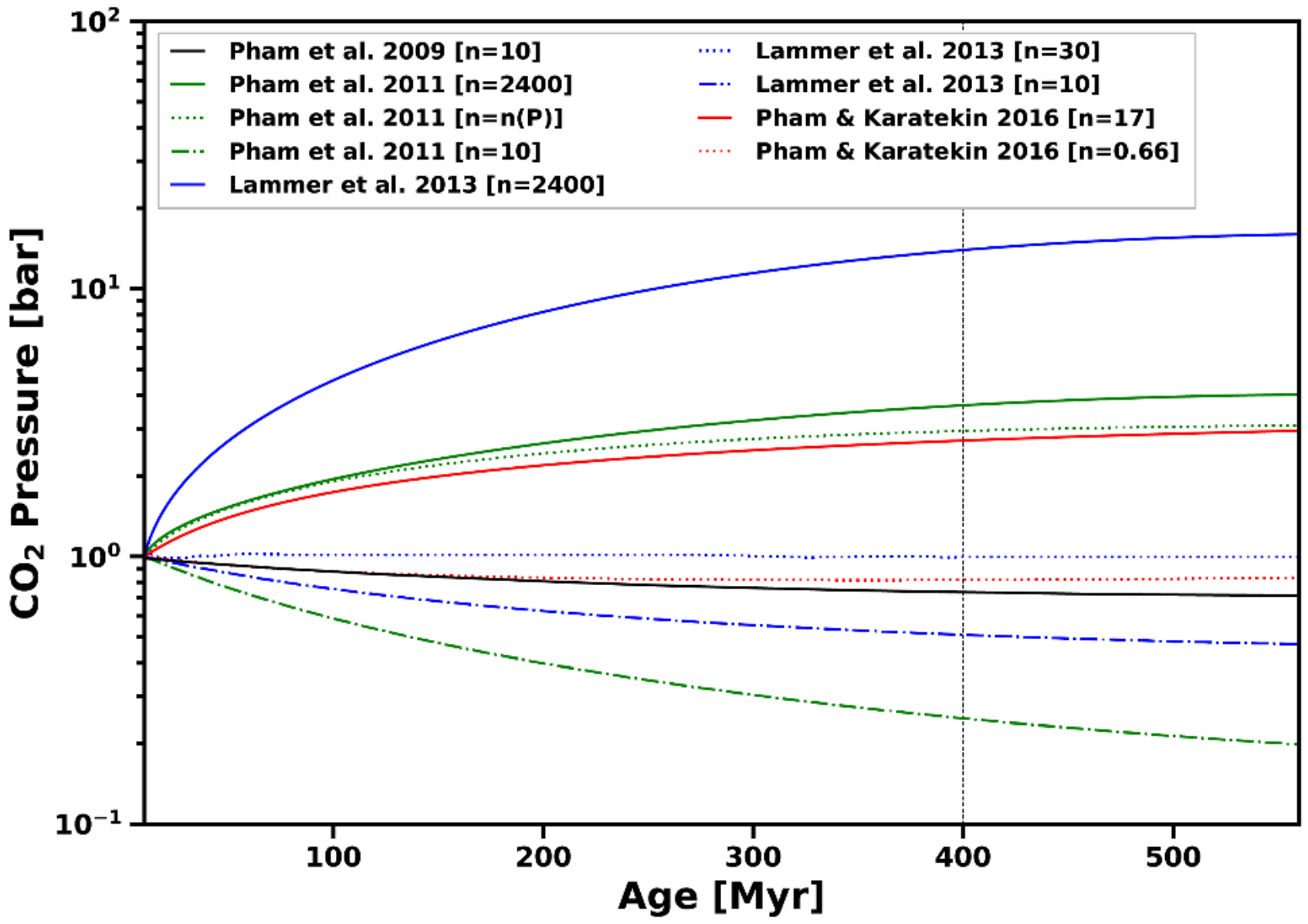

**Fig. 4. Different simulations on the evolution of the Martian atmosphere based on several adaptations of the so-called "tangent plane model" for different impactor efficiencies *n* and for an initial surface pressure of 1 bar. Note that the assumed models of different studies are not entirely identical (thereby for instance resulting in different evolutions for n = 10). Even though, for comparison, all simulations are set to a starting time of 0 Myr in this plot, Pham and Karatekin (2016) studied the Noachian eon, beginning at 500 Myr after the formation of Mars; all other studies start at 0 Myr. The vertical dotted line illustrates 400 Myr since this review is mainly interested in the evolution prior to this age.**

By applying this model to Mars, Melosh and Vickery (1989) and Pham et al. (2011) found a slight erosion of the Martian atmosphere which could have accounted to a loss of about 1 bar during a potential Late Heavy Bombardment (LHB). A study by Svetsov (2007) on the other hand found impactors to preferentially be a source of volatile delivery instead of erosion which is in agreement with simulations performed by De Niem et al. (2012). They found that in extreme cases of up to 30% volatile content within the impacting material, the Martian atmosphere could have been grown by as much as 4 bar. A more recent study by Pham and Karatekin (2016) investigated the effects of impacts during the Noachian eon and found in dependence on the impact efficiency scenarios ranging from an increase in pressure by 2 bar ($n = 17$) to a decrease by 0.2 mbar ($n = 0.66$). They, however, did not model impacts during the pre-Noachian eon, i.e. the first 500 million years of Martian history, the time frame relevant for our discussion; for comparison, however, their simulations are nevertheless illustrated in **Fehler! Verweisquelle konnte nicht gefunden werden.** to cover a brought parameter space of potential impactor related evolutionary paths. Similarly, Pham et al. (2009) and Lammer et al. (2013) found scenarios that either erode the atmosphere by up to more than 0.9 bar ($n = 10$; maximum erosion by Pham et al. 2009) or increase the atmospheric pressure by up to even 10 bar for n = 2400 (maximum delivery by Lammer et al. 2013). In Fig. 4**Fehler! Verweisquelle konnte nicht gefunden werden.**, the simulation by Pham et al. (2009) denoted with $n = n(P)$ is a scenario in which $n$ varies with pressure.

Another recent study (Brasser et al., 2020), finally, has to be mentioned which investigated the impact bombardment chronology of the terrestrial planets over the first billion years. Brasser et al. (2020) track the atmospheric erosion in their Monte Carlo simulation based on the tangent plane model by Melosh and Vickery (1989) and following Schlichting et al. (2015), who studied atmospheric mass loss through planetesimal impacts during the formation of the terrestrial planets. Similar to Schlichting et al. (2015), they found that impact erosion could have been significant, with Mars being able to lose about 50% of its atmosphere. Even though Brasser et al. (2020) point out that their estimate should be interpreted with great caution due to the simple assumptions made in their model, it illustrates that besides volatile delivery through impacts, giant impactors early-on had the potential to significantly erode any primordial and secondary atmosphere. The giant impactor which likely formed the Martian dichotomy, for instance certainly had a significant effect on any atmosphere. It, generally, can be expected that impactors with diameters > 500 km that formed the Martian basins would have evaporated any potential ocean and, by additionally ejecting rock vapor, building a transient short-period steam-rock vapor-atmosphere with surface temperatures of several 100 K that survived for up to a few millennia (Sleep and Zahnle, 1998; Segura et al., 2002).

All in all, these different results indicate that volatile delivery through impactors might have been more important than loss later-on, while giant impacts early-on could have significantly eroded any potential atmosphere. But these results are, however, strongly dependent on the initial conditions of the respective scenarios.

**Table 3.** Different simulated estimates on the total amount of outgassed $CO_2$, $SO_2$, and $H_2O$ during the pre-Noachian, i.e. during the first 400 million years and after the catastrophically outgassing of the steam atmosphere.

| | | $CO_2$ [mbar] | $SO_2$ [mbar] | $H_2O$ [m EGL] |
|---|---|---|---|---|
| Amerstorfer et al. (2017) | $T_m$ = 1700 K; $T_b$ = 2300 K (earliest outgassing) | 380 | | |
| | $T_m$ = 1700 K; $T_b$ = 2100 K | 235 | | |
| | $T_m$ = 1600 K; $T_b$ = 1900 K[a] | 0 | | |
| | $T_m$ = 1500 K; $T_b$ = 1900K (latest outgassing)[b] | 0 | | |
| Carr (1999) | | 550 | | |
| Grott et al. (2011) | global | 744 | | 44.3 |
| | only plumes | 293 | | 5.6 |
| Hirschmann and Withers (2008) | $fO_2$ = IW+1; $T_m$ = 1540° C | 450 | | |
| | $fO_2$ = IW+1; $T_m$ = 1320° C | 55 | | |
| | $fO_2$ = IW; $T_m$ = 1540° C | 140 | | |
| | $fO_2$ = IW; $T_m$ = 1320° C | 25 | | |
| Kurokawa et al. (2014) | based on polar layered deposits | | | 41 – 99 |
| | based on Carr and Head (2003) | | | 53 – 208 |
| | based on oxygen escape[c] | | | 31 – 133 |

| | | | | | |
|---|---|---|---|---|---|
| Manning et al. (2006) | | 1730 | | | |
| Stanley et al (2011) | $fO_2$ = IW+1; $T_m$ = 1320° C | 92 | | | |
| | $fO_2$ = IW; $T_m$ = 1320° C | 15 | | | |
| Craddock and Greeley (2009) | | 100 | 80 | | 0.3 |
| Chassefiére et al. (2013a) | Scaled to global model (Grott et al. 2011) with $f_{SO2}/f_{CO2}$ = 1 | | 945 | | |
| | Scaled to plume model (Grott et al. 2011); $f_{SO2}/f_{CO2}$ = 0.5 | | 225 | | |

[a]Outgassing starts at 480 Myr; [b]at 740 Myr; [c]estimated through oxygen escape simulated by Terada et al. (2009).

## *5.2 Volcanic outgassing*

Besides impacts that might have delivered volatiles into an early atmosphere, there is ample evidence that significant volcanism took place on early Mars (see, e.g., Grott et al., 2013); episodic activity was reported even until the late Amazonian (Plescia, 1990; Neukum et al., 2004; Vaucher et al., 2009). Here, it is generally assumed that the volcanic rate was intense early on and that it steadily declined over time with episodic eruptions of higher intensity later in the Martian history (Grott et al., 2013). Evidence for intense volcanism during the early Noachian at around 4 billion years ago can, e.g., be found in the heavily cratered southern highlands, which are rich in volcanic remnants (Xiao et al., 2012), or in crystallization products of liquids produced by partial melting during the earliest Noachian, which can be interpreted as petrological expressions of intense early Martian volcanism (Baratoux et al., 2013). Even though geological evidence from before 4.0 billion years ago is scarce, these findings together with numerical thermal evolution models (Grott et al., 2011; Morschhauser et al., 2011), which are predicting high crust production rates during the first ~500 Myr of Martian history, indeed suggest relatively strong early volcanism during the pre-Noachian and Noachian eons.

Several studies (Craddock and Greeley, 1995, 2009; Carr, 1999; Manning et al., 2006; Musselwhite et al., 2006; Hirschmann and Withers, 2008; Grott et al., 2011;

Stanley et al., 2011; Lammer et al., 2013) investigated the outgassing rates of $CO_2$ over the history of the planet (see also Fig. 5a and Table 3) or at specific events such as during the first crust formation (Wetzel et al., 2013) or the Tharsis event (Phillips et al., 2001; Craddock and Greeley, 2009). While Pepin (1994) yet arbitrarily assumed that volcanic outgassing during early Mars accumulated between 1 and 3 bar $CO_2$, first models that included early Mars retrieved amounts of ~0.5 bar (Carr, 1999) to almost 2 bar (Manning et al., 2006) for the pre-Noachian. Based on numerical models of the thermal evolution of Mars, Grott et al. (2011) estimated that depending on the oxygen fugacity about 293 to 744 mbar of $CO_2$ could have been accumulated in the atmosphere via volcanic outgassing until 4.1 Ga. Two other studies (Hirschmann and Withers, 2008; Stanley et al., 2011) investigated the outgassing rates from 4.5 Ga until today with similar models. Hirschmann and Withers (2008) found that post-4.5 Ga magmatism could have contributed between 40 mbar (for IW-1) and 1.4 bar (for IW+1) of $CO_2$ to the total atmospheric pressure whereas Stanley et al. (2011) retrieved a similar range of 30 mbar to 1.2 bar for the same time period. In both studies most of the outgassed $CO_2$ accumulates within the first 500 Myr. Another study that is based on the same model (Amerstorfer et al., 2017) finds cumulative rates of up to about 1 bar $CO_2$ over the whole history of Mars and up to 380 mbar during the first 400 Myr. The outgassing time, however, depends on the assumed initial mantle temperature $T_m$ and the initial bottom temperature $T_b$ at the core-mantle boundary. For temperatures of $T_m < 1600$ K and $T_b < 1900$ K most of the outgassing takes place around and after 500 Myr whereas for higher temperatures most of the $CO_2$ is outgassed earlier. Another study by Craddock and Greely (2009) estimated the volatile content of Martian lavas based on data collected from terrestrial basaltic eruptions, and found typical contents of ~0.7 wt% $CO_2$, ~0.5 wt% $H_2O$, and ~0.14 wt% $SO_2$. For $CO_2$, this inventory resulted in the outgassing of about 400 mbar over the Martian history with < 100 mbar degassing over the first ~400 - 500 Myr.

In addition to $CO_2$, volcanic degassing should have also led to the accumulation of $H_2O$ (Craddock and Greeley, 2009; Grott et al., 2011; Morschhauser et al., 2011; Kurokawa et al., 2018) which would either form liquid water on the surface or dissociate to H and O. This process could therefore accumulate a substantial amount of oxygen in the atmosphere which then should either be sequestered into the crust or escape to space (Lammer et al., 2003a, 2003b). Grott et al. (2011) estimated that for a realistic initial bulk water content of 100 ppm in the Martian mantle an equivalent global layer (EGL) of about 5.6 m to 44.3 m of water could have been outgassed during the pre-Noachian until 4.1 Ga (see Fig. 5b). If all $H_2O$ would dissociate and accumulate in the atmosphere this would be equivalent to approximately 0.1 – 1 bar of O. The present-day Martian D/H indeed suggests significant loss of water from the surface, either through atmospheric escape of hydrogen and oxidation of the soil (e.g., Lammer et al., 2003b), and/or through serpentinization of the water into the ground (e.g., Chassefière and Leblanc, 2011a). Kurokawa et al. (2014) simulated the evolution of the Martian water reservoirs and retrieved loss rates during the pre-Noachian of about 41 – 99 m EGL which would be equivalent to the loss and/or sequestration of ~0.9 – 2.2 bar of O. The whole water loss over the history of Mars was estimated in the same study to be 51 – 152 m EGL (Kurokawa et al., 2014) which is in agreement with another study by Villanueva et al. (2015) who found that early Mars had at least 137 m EGL. Older studies retrieved values in comparable ranges such as 17 – 61 m EGL (Lammer et al., 2003a), and 65 – 120 EGL (Krasnopolsky and Feldman, 2001), while some estimated higher values such as 100 – 800 m EGL (Donahue, 2004) and ~400 m EGL (Chassefière and Leblanc, 2011a). It has to be noted, however, that estimates based on the present-day D/H ratio might only represent lower values for the initial water inventory since part of D and H most likely escaped unfractionated via hydrodynamic escape early on.

Craddock and Greeley (2009) also estimated the amount of degassed H2O based on their inferred water content of ~0.5 wt%. They found that only about 290 mbar, equivalent to ~ 1.1 m EGL, might have been outgassed over the Martian history with only about 70 mbar over the first ~ 400 Myr. Within their study they also investigated the volcanic degassing of other volatile species, among them most im-portantly $SO_2$ (Fig. 6a and Table 3). Craddock and Greeley (2009) inferred the total amount of outgassed $SO_2$ to be over 300 mbar while being ~ 80 mbar within the first ~ 400 Myr. Chassefière et al. (2013a) scaled the outgassing of $SO_2$ with the volcanic degassing model of Grott et al. (2011) and found significantly higher values than Craddock and Greeley (2009). For the plume model of Grott et al. (2011), Chassefière et al. (2013a) assumed a ratio of $SO_2$ to $CO_2$ of $f_{SO2}/f_{CO2}$ ~ 0.5, while for the global model of Grott et al. (2011) they took $f_{SO2}/f_{CO2}$ ~ 1. Based on these values, they derived a total $SO_2$ outgassing amount of ~ 600 mbar, and ~ 1400 mbar respectively. For the pre-Noachian eon, these outgassing rates result in ~ 225 mbar, and ~ 950 mbar, respectively (see also Fig. 6a and Table 3).

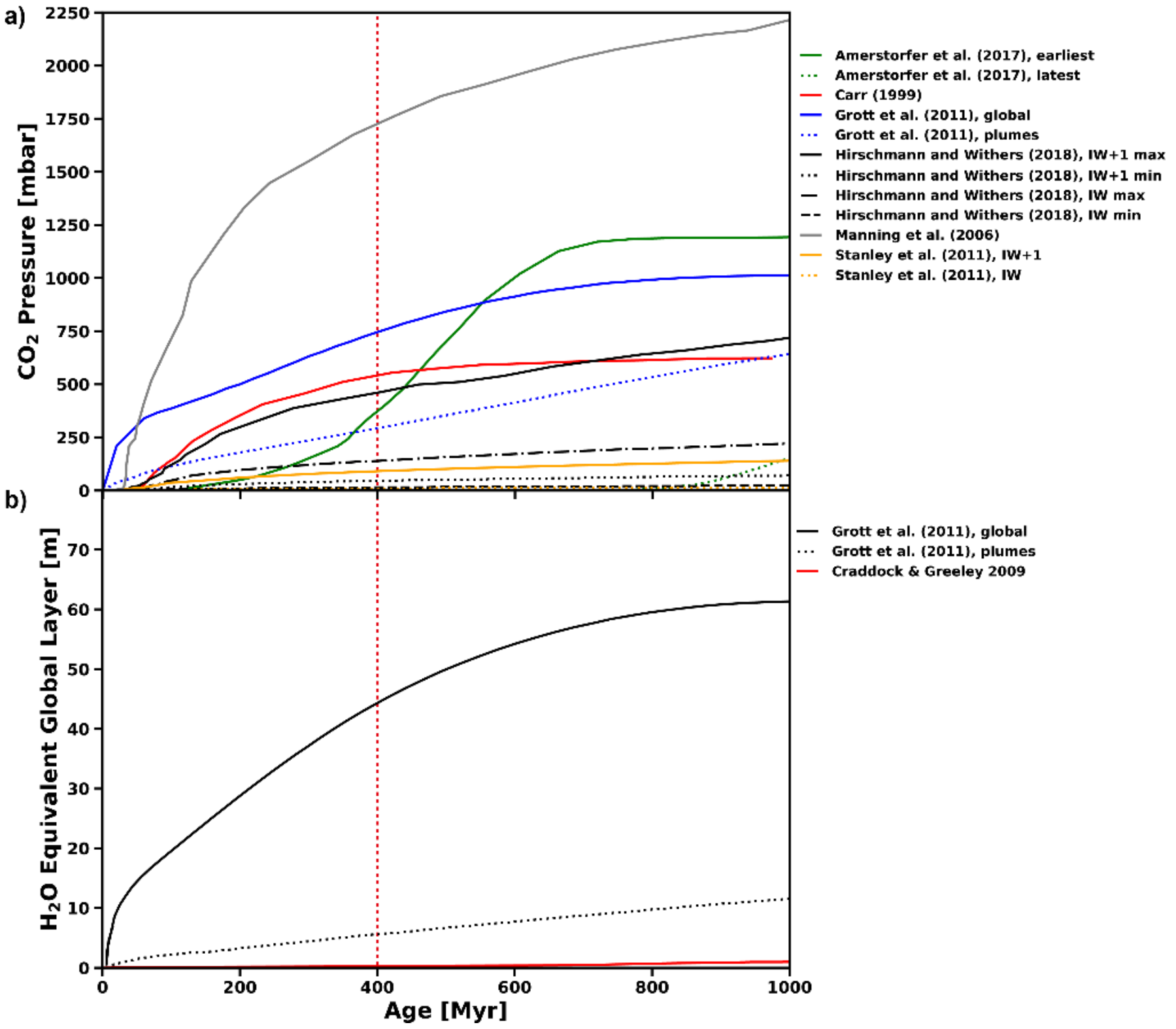

**Fig. 5. Different simulations for the outgassing of $CO_2$ (a) and $H_2O$ (b) over time. For Amerstorfer et al. (2017) the earliest and latest outgassing models were taken; the Grott et al. (2011) simulations illustrate global and localized outgassing over plumes. The “max”-values of Hirschmann and Withers (2018) are for higher temperatures (1540°C) of mantle melting (from Musselwhite et al. 2006), “min for 1320°C.**

For the degassing of other relevant volatile species, Craddock and Greeley (2009) found significantly lower values than for $CO_2$, $H_2O$, and even $SO_2$ (Fig. 6b). The most notable molecule in terms of outgassed volume is $H_2S$ with ~ 20 mbar over the Martian history and ~ 5 mbar in the first ~ 400 Myr. Other gases they investigated were $H_2$, $N_2$, HCl, HF, and CO. All of these molecules, however, only degassed in the range of below 1 mbar to about 3 mbar (Craddock and Greeley 2009), as can be seen in Fig. 6b.

To summarize, volcanic outgassing of $CO_2$, $H_2O$, but also $SO_2$, likely led to the accumulation of a certain amount of C, O, and potentially to a minor part S, within the pre-Noachian atmosphere. Other volatiles such as $H_2S$ or $N_2$ likely only played a negligible role in terms of total potential atmospheric inventory. In addition, impacts might have also delivered some volatiles to the planet. In total, however, these processes could not have cumulated much more than about 1 bar of $CO_2$, 2 bar of O, and potentially up to a few 100 mbar of $SO_2$ into the escaping atmosphere; if impactors mainly delivered volatiles, this value might have been somewhat higher.

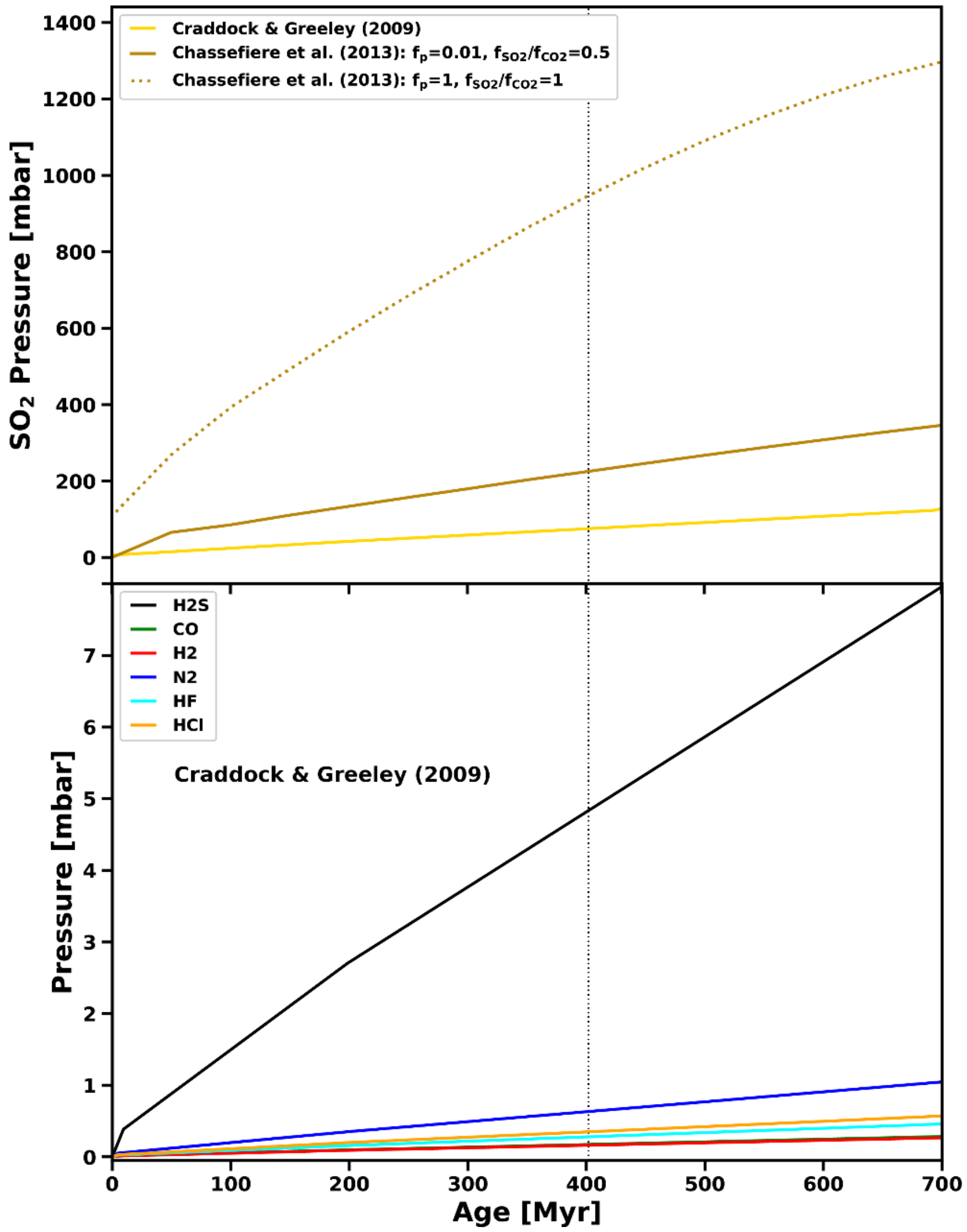

**Fig. 6. Upper panel: Volcanic degassing of $SO_2$ from Craddock and Greeley (2009) and Chassefière et al. (2013). The curves by Chassefière et al. (2009) are scaled by the outgassing cases of Grott et al. (2009), see** Fehler! Verweisquelle konnte nicht gefunden werden.**a. Here, *f*p = 1 denoted the global and $f_p$ = 0.01 the plume model by Grott et al. (2009). Lower panel: Volcanic degassing of several other molecules as simulated by Craddock and Greeley (2009). The vertical line depicts 400 Myr.**

# 6 Thermal escape processes after the escape of the steam atmosphere

Since thermal escape at early Mars is one of the most important processes that determine the stability of an atmosphere during the pre-Noachian eon, it is warranted to first outline the circumstances under which thermal escape and the therewith connected hydrodynamic escape prevails, to discuss the reliability of predicting strong hydrodynamic loss rates and whether these may indeed exist in reality.

Thermal escape is basically determined by the gravity and atmospheric temperature at the exobase level of a planetary body which is characterized by the Jeans escape parameter $\lambda$ that is defined as the ratio between the gravitational energy of the planetary body to the thermal energy of the respective particle, and can be written as

$$\lambda = \frac{GMm}{kTr}.$$

Here, *G* is the gravitational constant, *M* the mass of the planet, *m* the mass of the escaping particle, *k*, the Boltzmann constant, *T* the exobase temperature, and *r* is the radial distance at the exobase. As long as $\lambda$ exceeds a critical value of $\lambda{\sim}2.5$ a steady hydrodynamic solution can be found with a smooth transition from a subsonic to a supersonic outflow, whereas for smaller values no stationary hydrodynamic solution can exist (Volkov et al., 2011; Volkov and Johnson, 2013; N. V Erkaev et al., 2015). An atmosphere is no longer gravitationally bound if $\lambda$ falls below the critical value; as a result, a fast non-stationary atmospheric expansion might occur that leads to extreme thermal loss rates. This is often called “hydrodynamic escape”. Such a regime can be realized when the bulk velocity of the gas reaches a supersonic value at the exobase, and it can be driven by the absorption of the incident EUV flux at the so-called EUV absorption radius below the exobase level (e.g., Watson et al., 1981), but also through the thermal energy of an underlying magma ocean (e.g., Benedikt et al., 2020), or through both (e.g., Kubyshkina et al., 2020). For values of $\lambda$ between ~2.5 and ~6, Volkov et al. (2011) has shown that hydrodynamic escape transitions into normal thermal escape, whereas the latter constitutes the parameter space for $\lambda > 6$.

Doubts on the viability of the theory on and the existence of hydrodynamic escape came up, after Cassini investigated Titan and New Horizons measured the escape rates of $N_2$ at Pluto. For both satellites, Strobel (2008; 2009) proposed a so-called "slow hydrodynamic escape model" which yielded strong $N_2$ escape rates in the order of $\sim 2 \times 10^{27} \mathrm{s}^{-1}$ which significantly overestimated thermal escape at Titan and, as measured by New Horizons, at Pluto (Gladstone et al., 2016). The model by Strobel (2008; 2009) requires an extended quasi-collisional region on top of the exobase at which efficient energy transfer can occur. However, Erkaev et al. (2020) studied thermal escape at present-day Titan with a 1D upper atmosphere EUV radiation absorption and hydrodynamic escape model and found that the flux of molecules at the exobase level with velocities above the escape velocity is just too small to provide the extracted escape flux by Strobel (2008; 2009). The authors of Strobel (2008; 2009) continued modelling above the exobase with the assumption that the energy is supplied through upward conduction. However, it was shown that conduction fails above the exobase and that the collision probability of nitrogen above Titan's exobase is decreasing rapidly with height (Tucker and Johnson, 2009; Schaufelberger et al., 2012). An extended quasi-collisional region above the exobase, as assumed by Strobel (2008; 2009), is, therefore, not supported and hydrodynamic escape does, consequently, not take place neither at present-day Titan nor Pluto.

Thermal escape of $N_2$ at Titan was recently simulated to be only about $\sim 8 \times 10^{17} \mathrm{s}^{-1}$ (Erkaev et al., 2020), while New Horizons measured a loss rate of $N_2$ at Pluto of about $\sim 1 \times 10^{23} \mathrm{s}^{-1}$ (Gladstone et al., 2016). That Pluto is not in a hydrodynamic escape regime at present-day can also be seen if one calculates $\lambda_{\mathrm{N2}}$ at its exobase level $r_{\mathrm{exo}}$. Wit $r_{\mathrm{exo}}$ ~ 2850 km from the centre of Pluto and an exobase temperature of $T_{\mathrm{exo}}$ ~70 K (Gladstone et al., 2016), one yields $\lambda_{\mathrm{N2}} \sim 15$, which is clearly above the critical value of $\lambda \sim 2.5$. For present-day Titan, the same calculation even yields $\lambda_{\mathrm{N2}} \sim 50$, if one assumes an exobase altitude of $r_{\mathrm{exo}} \sim 4050$ km, and an exobase temperature of $T_{\mathrm{exo}} \sim 155$ K, as taken from Erkaev et al. (2020).

It must be further noted, however, that hydrodynamic escape cannot only occur for hydrogen but also for heavier species. Erkaev et al. (2020), for instance, found strong hydrodynamic escape rates for a potential early nitrogen-dominated atmosphere at Titan with escape rates as high as $\sim 5 \times 10^{29} \mathrm{s}^{-1}$ for 400, and $\sim 1.5 \times 10^{28} \mathrm{s}^{-1}$ for 100 times the present-day EUV flux at Titan's orbit. Moreover, hydrodynamic escape might for certain circumstances also escape for carbon (e.g., Tian et al. 2009) which is significantly lighter than nitrogen (6 and 14 amu, respectively), as we will see below in more detail.

That hydrodynamic escape is, finally, indeed existing for inflated atmospheres under high EUV fluxes can also be seen, if one regards the so-called "evaporation valley" in the distribution of exoplanets (e.g., Owen and Wu, 2015, 2017; Fulton and Petigura, 2018; Owen et al. 2020, this issue; Gupta and Schlichting, 2019). The radius distribution of close-in low-mass exoplanets shows a "valley" that is separated by two peaks in radii at 1.3 $R_{Earth}$ and 2.6 $R_{Earth}$ (Owen and Wu, 2017). This gap can best be explained through the hydrodynamic escape of primordial atmospheres from these planets. Exoplanets above a certain mass are massive enough (i.e., their ratio of gravitational to thermal energy exceed the value for a critical $\lambda$) close-in to their respective star to sustain their accreted hydrogen envelope, while smaller planets (with $\lambda$ below the critical value) lose it due to the strong incident EUV flux. The first peak at 1.3 $R_{Earth}$ depicts the most massive planets that were not able to retain their atmospheres, while the second peak at 2.6 $R_{Earth}$ illustrates the least massive planets that yet retained their hydrogen envelopes. Their larger radius is thus a consequence of the expanded atmosphere that is yet existing around these exoplanets. The evaporation valley furthermore gets smaller for decreasing EUV fluxes, i.e. for exoplanets that circumvent their respective host stars at orbits with lower EUV flux values (e.g., Gupta and Schlichting, 2019).

The same mechanism that forms the evaporation valley, consequently, should have also prevailed in the early Solar System, and hydrodynamic escape should have been sculpting the early evolution of the terrestrial planets. This can also be seen in the noble gas isotope fractionation at Venus and Earth which can both be reconstructed if these planets would have hydrodynamically lost a small primordial hydrogen envelope that they first accreted from the solar nebula (Lammer et al., 2020). It seems, therefore, reasonable that similar conditions were also active on early Mars, as we will discuss in the rest of this chapter.

As already described in Section 2 and as can be seen in Fig. 1, the EUV flux from the young Sun was significantly higher in the past than at present-day (Ribas et al., 2005; Tu et al., 2015), which has a profound effect on the structure of the upper atmosphere (Tian et al., 2009; Johnstone et al., 2018) as can be seen in Fig. 7. While for the present-day the Martian exobase level which separates the collision-dominated from the collisionless part of the atmosphere is located at a height of around 250 km, it significantly rises with increasing EUV flux. According to Tian et al (2009) the level of the exobase would non-linearly rise to about 400 km for 3 times, 900 km for 10 times, and 10 000 km for 20 times the present-day EUV flux. Such a strong non-linear expansion of a $CO_2$-dominated atmosphere under strong irradiation from the Sun is in agreement with several other studies covering different planets such as Kulikov et al. (2006, 2007) for Mars and Venus, and Smithro and Sojka (2005), Tian et al. (2008a, 2008b), and Johnstone et al. (2018) for the Earth.

Tian et al. (2009) used a 1-D multi-component, hydrodynamic, planetary-thermosphere model (Tian et al., 2008a, 2008b), in which an electron transport-energy deposition model was coupled to a thermosphere-model, to self-consistently reproduce the structure of a $CO_2$-atmosphere under different incident radiation levels. For the EUV flux of the young Sun, they used the extrapolation by Ribas et al. (2005), i.e. $F = 29.7 \cdot t^{-1.23}$ergs cm$^{-2}$s$^{-1}$, with $t$ being the solar age in billion years.

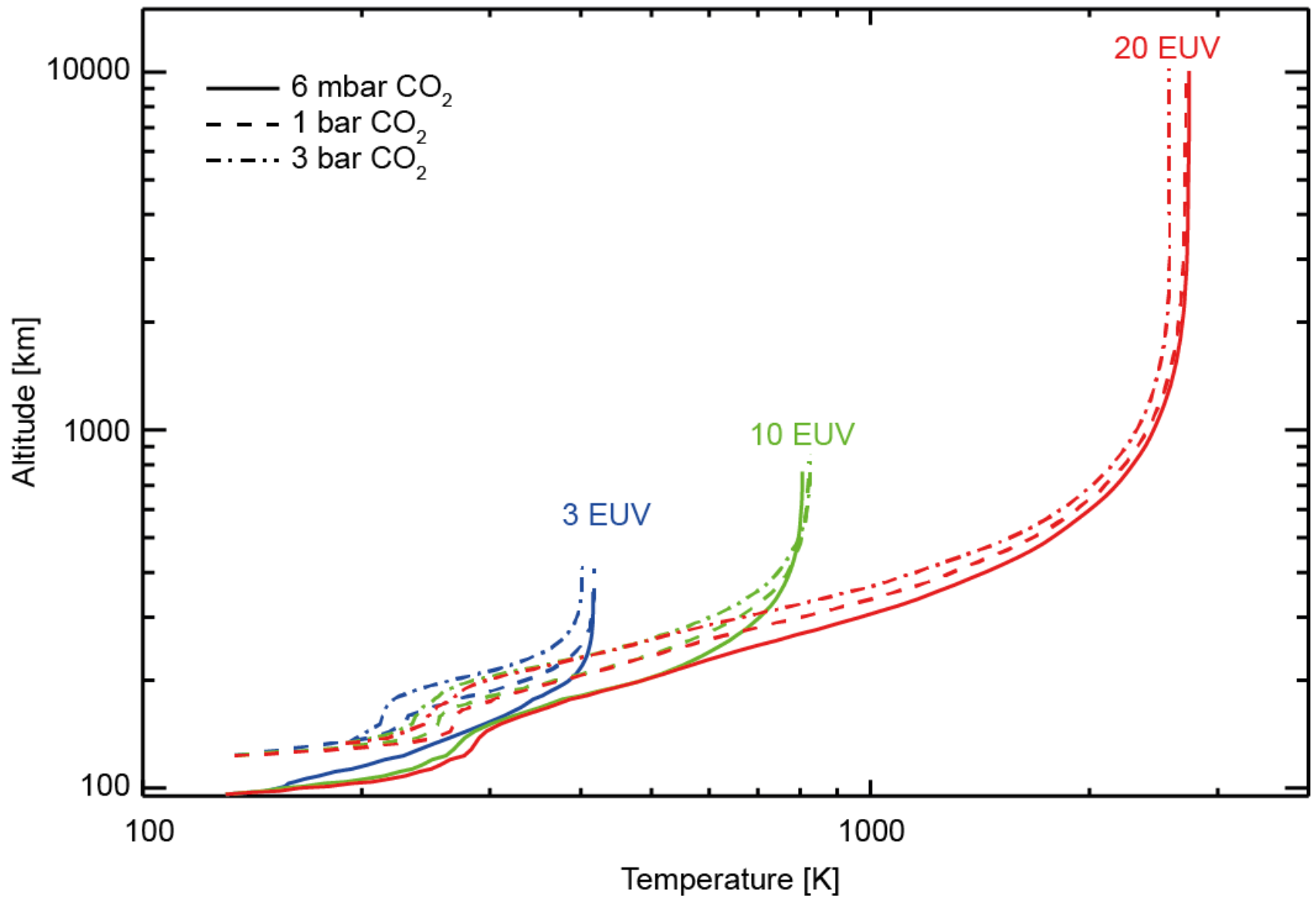

**Fig. 7. Structure of the Martian CO2 atmosphere for different EUV fluxes and pressures. Figure adopted from Tian et al. 2009.**

Here, the different solar EUV fluxes correspond to 3.8 Ga for 10 times the present-day level (10 EUV$_{\odot}$) and 4.1 Ga for 20 times the present-day level (20 EUV$_{\odot}$). While for 10 EUV$_{\odot}$, in which the exobase temperature rose to 800 K at an altitude of ~ 900 km and no efficient thermal escape took place, the upper atmosphere was significantly expanded and warmer for 20 EUV, with an exobase altitude of ~ 10 000 km and an exobase temperature > 2 500 K. This non-linear response of the upper atmosphere is due to $CO_2$ – the main IR cooler in the atmosphere – being dissociated under such high EUV fluxes (Smithtro and Sojka, 2005; Tian et al., 2009). As cooling decreases, the exobase level considerably increases and the Jeans escape parameter $\lambda$ for atomic carbon and oxygen at the exobase significantly drops to $\lambda_c = 1.8$ and $\lambda_O = 2.4$ – both below the critical value of $\lambda \sim 2.5$. For 20 EUV, i.e. 4.1 Ga according to Ribas et al. (2005), the thermal escape flux of carbon from the Martian atmosphere in the simulation of Tian et al. (2009) was simulated to be ~ 1.4 x $10^{29}$ $s^{-1}$ and could have risen up to ~ 1.4 x $10^{30}$ $s^{-1}$ for 4.5 Ga, if one considers the EUV flux scaling law of Ribas et al. (2005). Such strong escape fluxes would lead to the loss of 1 bar of $CO_2$ from the Martian atmosphere within 10 Myr at 4.1 Ga, and within 1 Myr at 4.5 Ga, respectively. It is therefore likely that a dense $CO_2$ atmosphere during the pre-Noachian could not have been maintained under such harsh conditions, unless volcanic outgassing would have been able to balance these extreme loss rates which seems to be unlikely.

As discussed earlier, such extreme escape is also in agreement with hydrodynamic escape studies by Erkaev et al. (2015) and Volkov et al. (2011), who both independently found that particles with a critical Jeans escape parameter $\lambda_{cr} < 2.5 - 3$ are not bound gravitationally to the planet any more, but will escape hydrodynamically into space.

Tian et al. (2009) estimated whether volcanic outgassing at Mars could counterbalance the strong escape rates during the pre-Noachian eon. They assumed that if Mars formed from the same material than Earth, then its initial total $CO_2$ reservoir could have been around 20 bar (see also Carr, 1986), which is also in good agreement with estimates from other studies on the Martian volatile content (such as Elkins-Tanton (2008), see also Section 4 and Table 1). Such reservoir would have been lost within 20 Myr thermally if all $CO_2$ would have been outgassed into the atmosphere ~4.5 Gyr ago. Even if Mars would have had twice as much volatiles as the Earth, the whole reservoir would have been lost within 40 Myr, respectively, under such an assumption (Tian et al., 2009). They further estimated the atmospheric evolution, in case that the whole $CO_2$ reservoir was not outgassed immediately but with an exponentially decaying rate from 4.56 Ga until the Tharsis outgassing event during the late Noachian eon (see, e.g., Phillips et al., 2001). Neither for reservoirs of 20 bar, nor 40 bar of $CO_2$, a dense atmosphere could have built up before ~4.0 Ga, since in any case the escape rates would have been higher than the respective outgassing rates. Even for strong episodic outgassing events instead of continuous degassing, a dense atmosphere could only have been maintained for very short periods < 1 – 10 Myr (Tian et al., 2009).

Since $H_2$ is also released into the atmosphere via volcanic degassing, Tian et al. (2009) simulated the effect of different $H_2$ mixing ratios onto the escape of carbon. They found that, although higher mixing ratios slightly reduced the escape, the respective loss would nevertheless be significantly higher than any of the assumed $CO_2$ outgassing fluxes. Besides the escape of atomic C they also investigated the escape of atomic O. Since oxygen is heavier than C, the Martian atmosphere could have accumulated $O_2$ over time at a rate proportional to the excess of C escape over O escape (Tian et al., 2009). Whether any net $O_2$ will accumulate in the Martian atmosphere, however, was not investigated further in the study of Tian et al. (2009).

If one considers the EUV flux evolution of the young Sun according to the model of Tu et al. (2015), the respective times for 10 $EUV_\odot$ and 20 $EUV_\odot$ from Tian et al. (2009) slightly change. For a slow rotating young Sun 20 $EUV_\odot$ shifts backwards to 4.3, and 10 $EUV_\odot$ remains at 3.8 Gyr ago, while for a moderate rotator these shift forward to 3.9, and 3.5 Gyr ago, respectively. In the unlikely scenario of a fast rotating young Sun, 20 and 10 $EUV_\odot$ would even be shifted towards 3.7 and 3.4 Gyr ago, respectively. Based on the results and scenarios of Tian et al. (2009), a $CO_2$-dominated atmosphere could therefore either start to build up earlier in the Martian history for a slow, or later for a moderate or fast rotating Sun, respectively.

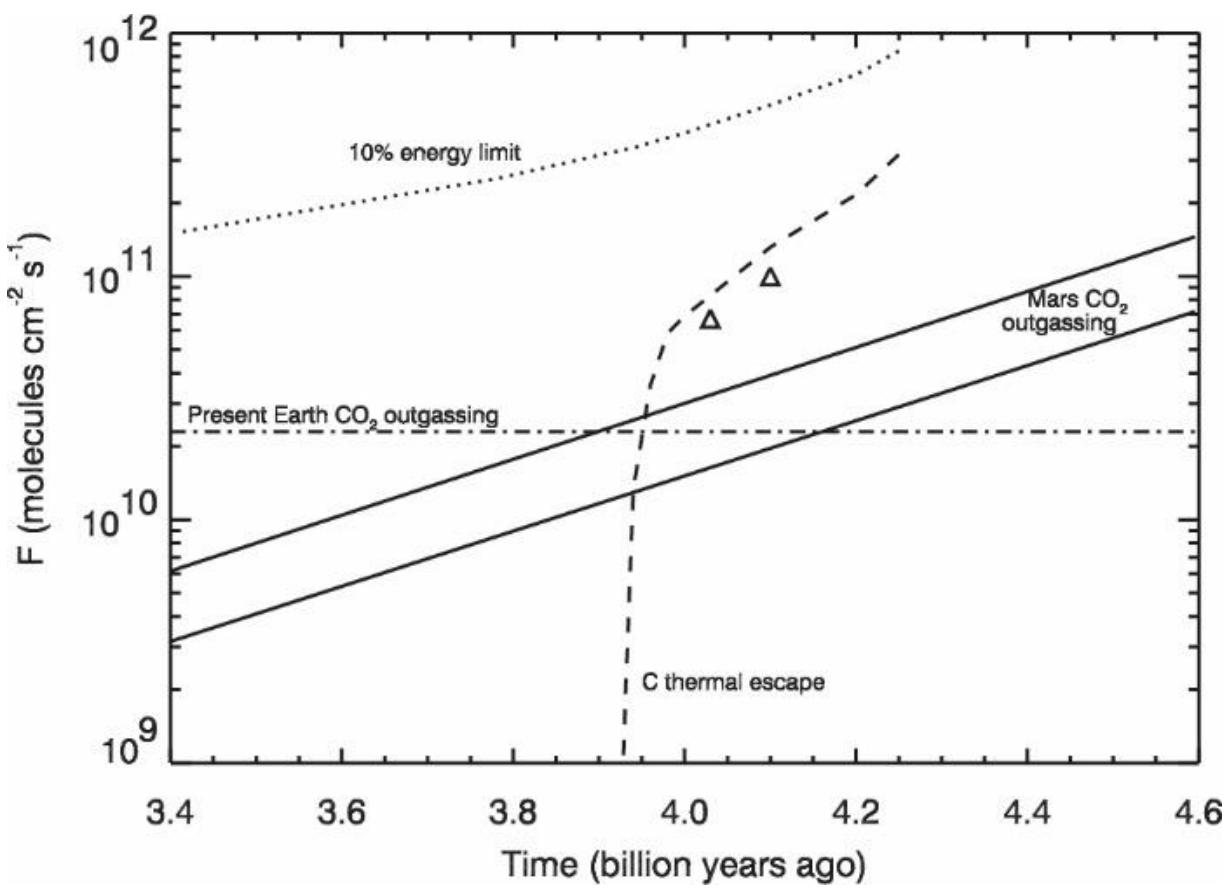

**Fig. 8. $CO_2$ outgassing and escape at early Mars according to Tian et al. (2009). The strong thermal escape of C on early Mars cannot be counterbalanced through outgassing of $CO_2$ before ~4 billion years, neither if one assumes that Mars was as volatile rich (lower solid line) or even as twice as rich (upper solid line) as Earth. Here, the thermal escape of C as simzlated by Tian et al. (2009) corresponds to an EUV flux evolution based on Ribas et al. (2005). Figure from Tian et al. (2009).**

## 7 Non-thermal escape processes and the role of the magnetic field

Non-thermal or suprathermal escape can basically be divided into several different types (Shizgal and Arkos, 1996; Chassefière and Leblanc, 2004; Chassefière et al., 2007; Kulikov et al., 2007; Lundin et al., 2007; e.g., Catling and Kasting, 2017), i.e.

1) *Photochemical escape.* Neutrals that were produced through dissociative recombination – the photoionization of neutrals by the solar EUV irradiation with subsequent recombination of the ions into energetic neutrals – can eventually gain enough energy to reach escape velocity (Lammer and Bauer, 1991; Luhmann et al., 1992; Lammer et al., 1996, 2006; Fox and Hać, 1997; Chassefière and Leblanc, 2004; Zhao and Tian, 2015; Amerstorfer et al., 2017; Zhao et al., 2017).
2) *Ion pick-up escape*. Neutrals that are ionized in the exosphere by photoionization or charge exchange are accelerated by the electric field of the magnetized solar wind, reaching either escape velocity or re-impacting onto the atmosphere (Spreiter and Stahara, 1980; Lundin et al., 1989; Luhmann and Kozyra, 1991; Kallio and Janhunen, 2002; Kallio et al., 2006; Chassefière et al., 2013b; Curry et al., 2013, 2015; Yamauchi et al., 2015).
3) *Sputtering*. Ions that were produced in the upper atmosphere but are subsequently not picked-up by the solar wind can re-impact onto the neutral atmosphere, thereby ejecting neutrals from the planet (Luhmann et al., 1992; e.g. Jakosky et al., 1994, 2017a; Luhmann, 1997; Leblanc and Johnson, 2001, 2002; Wang et al., 2014; Leblanc et al., 2018).

4) *Polar and ionospheric outflow*. Ions that follow the magnetic field lines of a planet can eventually escape to space if the field lines open up to the solar magnetic field. This is particularly important for a planet with an intrinsic magnetic field such as the Earth. But also on non- or weakly-magnetized planets such as Mars or Venus, ions that are produced in the ionosphere through shocked solar wind plasma can be transported to the ionopause and dragged away by the solar wind (e.g., Pérez-de-Tejada, 1987; Chassefière and Leblanc, 2004; Ma et al., 2004; Fox, 2009; Lundin et al., 2011b, 2011a; Liemohn et al., 2013; Collinson et al., 2015; Fig. 6 from Futaana et al., 2017; Kislyakova et al., 2020). Furthermore, ion outflows that play an important role at the Earth are likely connected with energized auroral electron precipitation as well as heating and acceleration of ions in the upper ionosphere through wave-particle interactions (see, e.g., Moore and Horwitz, 2007 for a comprehensive review on ion outflow processes). Similar phenomena have been observed at Mars (e.g., Lillis and Brain, 2013; Xu et al., 2016; Girazian et al., 2017), but whether these processes also trigger ion outflow at an unmagnetized planet remains poorly understood.
5) *Bulk removal processes*. Finally, instabilities at the solar wind-ionosphere boundary can lead to a sudden detachment of significant portions of the ionized atmosphere from planets without an intrinsic magnetic field. This process includes the so-called Kelvin-Helmholtz instability that occurs at the boundary layer between two fluids flowing relative to each other, thereby forming vortices at this boundary layer (e.g., Chandrasekhar, 1961). This phenomenon can lead to significant short-term escape at unmagnetized planets such as Mars (e.g., Penz et al., 2004; Ruhunusiri et al., 2016) and Venus (e.g., Möstl et al., 2011). Bulk removal processes further include sudden atmospheric responses to time-dependent solar wind conditions such as interplanetary CMEs or CIRs (e.g., Perez-de-Tejada, 1992; Brain et al., 2010; Edberg et al., 2011). Although often neglected, these escape mechanisms can lead to significant atmospheric losses, particularly if considering the harsher space weather conditions early in the history of the solar system, or more specifically, in the evolution of early Mars.

While polar and ionospheric outflow, as well as ion-pickup escape is associated with ions escaping the atmosphere, photochemical escape describes the escape of neutral particles. Neutrals are also escaping via sputtering; pick-up ions, however, are needed to trigger their escape.

Studying ion-pickup and sputtering during the first 400 Myr might therefore be more challenging than for the later periods. For both escape processes, as well as for bulk removal, and ionospheric and polar outflow, it is important whether a planet has an intrinsic magnetic field or not, which might shield the planet from ion-pickup and sputtering on the one hand, and provide another source of escape through polar outflow on the other hand. Whether an intrinsic magnetic field is really a protection against or a funnel for non-thermal escape is not yet entirely clear. While the common paradigm believed the magnetosphere to be a shield, some recent studies suggest that at least for certain circumstances, the intrinsic magnetic field could serve as an energy funnel enforcing atmospheric escape mainly through polar outflow (e.g. Blackman and Tarduno, 2018; Gunell et al., 2018; Egan et al., 2019).

A magnetized compared to an unmagnetized planet significantly enhances the surface area of interaction with the solar wind and dissipates its energy towards the polar caps. For Earth, with an entire magnetosphere cross-section of 15 to 20 $R_{Eearth}$, this effect leads to an energy flux to the auroral zone that is about 10 to 300 times larger than onto the ionospheres of the unmagetized planets Venus and Mars. (e.g., Moore and Horwitz, 2007; Tarduno et al., 2014). This effect can drive significant loss through the polar caps as illustrated by the comparable present-day escape rates of these three planets. This argues in favour of the “magnetic field as a funnel”-theory. But one must be cautious with such a comparison since i) the characteristics of these three terrestrial planets are rather different and ii) environmental characteristics in the early solar system were crucially diverging from present-day. For instance, atmospheres likely were significantly more extended than at present-day. Without a strong intrinsic magnetic field, this would lead to a larger surface area interacting with the solar wind, thereby reducing the difference in incident energy flux between an unmagnetized and magnetized planet described above. In addition, any magnetosphere would at the same time be more compressed than at present-day due to the stronger incident solar wind early in the history of the solar system, thereby leading to significantly broader polar caps which in turn, again, increases atmospheric escape. Comparative research on such scenarios would clearly enhance our knowledge on this subject.

The role of an intrinsic magnetic field for non-thermal atmospheric escape is indeed a relevant question for the early evolution of the Martian atmosphere. Mars Global Surveyor found remnant magnetization in the crust of Mars (see Fig. 9) which is generally believed to be the remnant of an ancient intrinsic magnetic field (Acuna et al., 1999; Connerney et al., 1999, 2005). It ceased around 4.1 to 4.0 Ga (Lillis et al., 2013; Robbins et al., 2013; Vervelidou et al., 2017) or later at about 3.7 Ga (Langlais and Purucker, 2007; Hood et al., 2010; Milbury et al., 2012; Mittelholz et al., 2020), and could have been as strong as up to the value of the present-day terrestrial field (Stevenson, 2001; Quesnel et al., 2009), although the dynamo may have been weaker in its earlier stage that towards its end (Vervelidou et al., 2017).

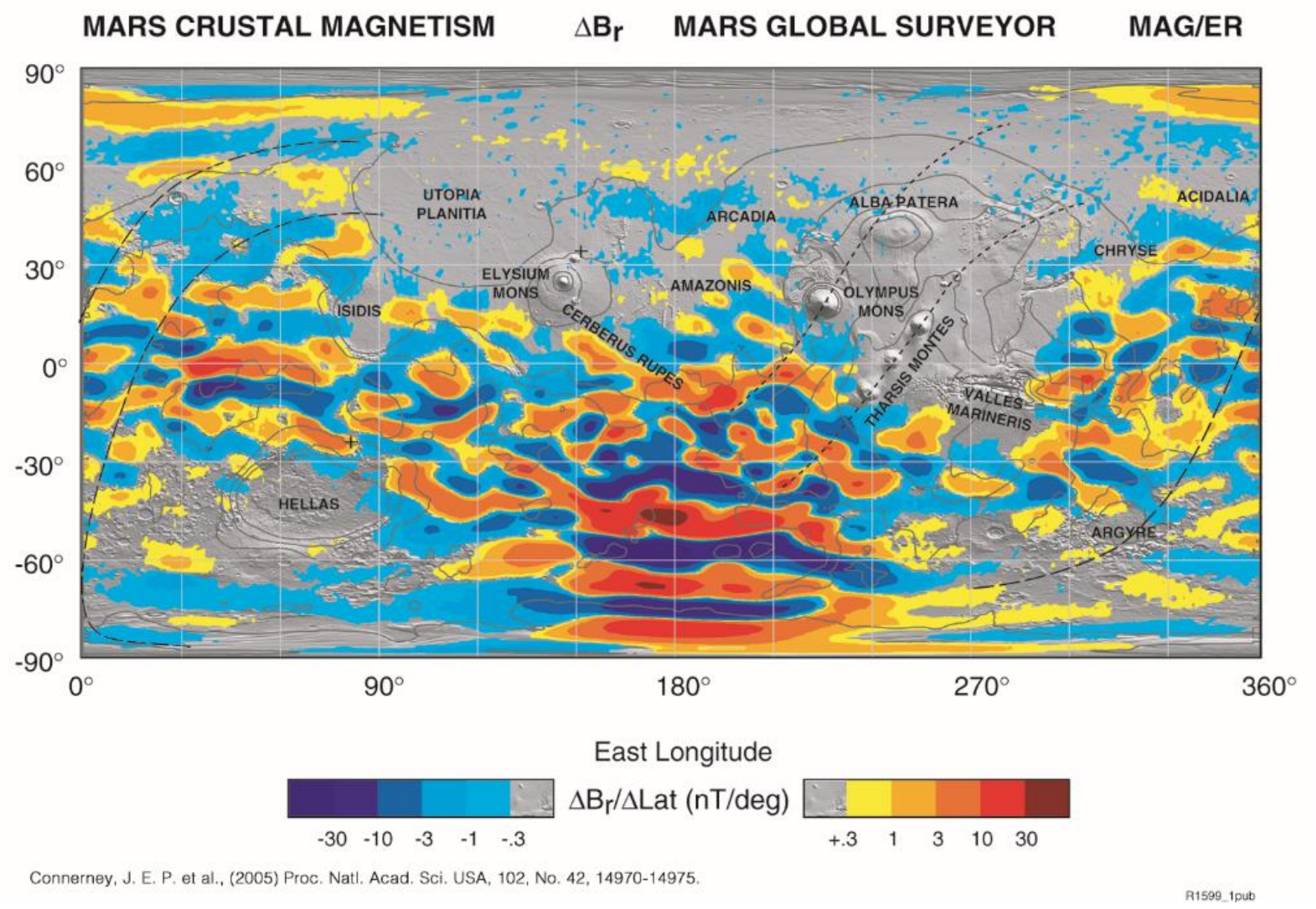

**Fig. 9. The famous map of the crustal magnetic field of Mars as observed with the satellite Mars Global Surveyor at an altitude of 400 km. This strong remnant magnetization is believed to be an indication for an ancient intrinsic magnetic field at Mars which ceased about 4 billion years ago. The estimated age of the cessation is mainly based on crater dating; younger basins and craters such as Hellas do not show a magnetization whereas older do. Re-evaluating the crater ages might therefore also change our picture about the ancient Martian magnetosphere. Figure from Connerney et al. (2005), Copyright (2005) National Academy of Sciences.**

There are several studies which investigated non-thermal escape, i.e. photochemical and ion-pickup escape, sputtering, and ionospheric outflow, over the Martian history. Most of these studies, however, do not cover the early period in which the intrinsic magnetic field of Mars likely was present since the early magnetosphere would have changed the behaviour of non-thermal escape. Scaling laws for non-thermal escape from the time after the magnetic dynamo ceased are therefore most likely not applicable to the time when it was yet active. However, it makes sense to briefly discuss these studies about non-thermal escape at Mars to get a sense on how escape changed over the past.

## *7.1 Non-thermal escape processes over the last ~4 Gyr*

Due to current and past space missions such as Phobos, Mars Express, and MAVEN orbiting Mars, there are various different measurements and simulations on the present-day escape rates of C, O, $O_2$, CO, $CO_2$ (e.g., Carlsson et al., 2006; Barabash et al., 2007; Nilsson et al., 2011; Ramstad et al., 2013, 2015; B. M. M. Jakosky et al., 2018; Leblanc et al., 2018), but also of N, or Ar (e.g., Leblanc et al., 2018) – Tables 5 and 6 list some of them for oxygen and carbon species together with simulated loss estimates over the history of Mars. For present-day Mars, photochemical escape of O and C species shows broadly comparable loss rates than ion escape processes, as depicted in Tables 5 and 6. Generally, loss rates for O species (Table 5) are at present-day in the order of ~ $10^{24}$ – $10^{25}$ $s^{-1}$ for nominal solar wind conditions, while they are lower for C species, including $CO_2$, with rates around ~ $10^{23}$ – $10^{24}$ $s^{-1}$ for most of these studies (Table 6). The reason for the lighter element escaping from Mars with lower rates is due to the heavy $CO_2$ binding most of C in the Martian atmosphere, while O is predominantly bound in the lighter molecule $O_2$ (e.g., Amerstorfer et al., 2017; Lichtenegger et al., 2020). For higher EUV fluxes, however, it can be expected that C escape rates will significantly rise since $CO_2$ becomes more and more dissociated by the incident radiation (e.g., Tian et al., 2009), as already discussed in Section 6. In addition, it can be expected that ion escape might become the dominant non-thermal escape rate at Mars further back in time which is due to higher ionization rates induced by the increasing incident EUV flux (e.g., Lichtenegger et al., 2020). All in all, it can be expected that the escape rates will rise back into the Martian past and that the pattern of these losses will significantly change; a process that was already simulated by various different authors for the last ~ 4 billion years.

**Table 4.** Simulated and measured escape rates for oxygen for different EUV fluxes for photochemical loss of O (Hot O in the table), $O^+$, and sputtered neutral O (Sp. O in the table). The table also includes measuredcombined escape of heavy ions ($O^+$,$O_2^+$,$CO_2^+$) from present-day Mars. The corresponding times (Ga = Giga years ago) to the EUV fluxes are for a slow rotator according to Tu et al. (2015).

| | | 1 EUV<br>0 Ga | 3 EUV<br>1.9 Ga | 6 EUV<br>2.6 Ga | 10 EUV<br>3.8 Ga | 20 EUV<br>4.2 Ga |
|---|---|---|---|---|---|---|
| Hot O | Luhmann et al. (1992)[a] | $8 \times 10^{25}$ | $5 \times 10^{26}$ | $1 \times 10^{27}$ | | |
| | Zhang et al. (1993)[a] | $7 \times 10^{25}$ | $4 \times 10^{26}$ | $1 \times 10^{27}$ | | |
| | Luhmann et al. (1997) | $6 \times 10^{24}$ | $3 \times 10^{25}$ | $8 \times 10^{25}$ | | |
| | Chassefiére et al. (2007) | $3 \times 10^{24}$ | $3 \times 10^{25}$ | $8 \times 10^{25}$ | | |
| | Valeille et al. (2010), HSA[b] | $1.3 \times 10^{26}$ | $2.6 \times 10^{26}$ | $6.9 \times 10^{26}$ | | |
| | Valeille et al. (2010), LSA[c] | $3.8 \times 10^{25}$ | $2 \times 10^{26}$ | $4.9 \times 10^{26}$ | | |
| | Zhao and Tian (2015) | $1.7 \times 10^{25}$ | $8.4 \times 10^{25}$ | | $1.8 \times 10^{25}$ | $2 \times 10^{25}$ |
| | Chassefière & Leblanc (2011a,b) | $9.2 \times 10^{25}$ | $2 \times 10^{26}$ | $6 \times 10^{26}$ | | |
| | Cravens et al. (2017)[d] | $4 \times 10^{25}$ | $1.2 \times 10^{26}$ | $2.5 \times 10^{26}$ | $4 \times 10^{26}$ | |
| | Amerstorfer et al. (2017)[e] | $5.9 \times 10^{25}$ | $2.1 \times 10^{26}$ | | $3.8 \times 10^{26}$ | $6.1 \times 10^{26}$ |
| | Zhao et al. (2017), FH14[f] | $5.2 \times 10^{25}$ | $1.2 \times 10^{26}$ | | $8.2 \times 10^{26}$ | $3 \times 10^{25}$ |
| | Zhao et al. (2017), G14[g] | $2.5 \times 10^{25}$ | $8.5 \times 10^{25}$ | | $5.6 \times 10^{25}$ | $3 \times 10^{24}$ |
| | Lillis et al. (2017)[d] | $4.3 \times 10^{25}$ | $6.5 \times 10^{26}$ | $5 \times 10^{27}$ | $1.5 \times 10^{28}$ | |
| | Dong et al. (2018) | $2.7 \times 10^{25}$ | $8.5 \times 10^{25}$ | $9.9 \times 10^{25}$ | $1 \times 10^{26}$ | |
| | Leblanc et al. (2018)[h] | $2.7 \times 10^{23}$ | | | | |
| $O^+$ | Luhmann et al. (1992) | $3 \times 10^{23}$ | $3 \times 10^{26}$ | $3 \times 10^{27}$ | | |
| | Zhang et al. (1993) | $4 \times 10^{24}$ | $3 \times 10^{26}$ | $2 \times 10^{27}$ | | |
| | Lammer et al. (2003a,b) | $3 \times 10^{24}$ | $4 \times 10^{26}$ | $8.3 \times 10^{26}$ | | |

| | | | | | |
|---|---|---|---|---|---|
| | Carlsson et al. (2006)[i] | $5\times10^{25}$ | | | |
| | Fang et al. (2013)[j] | $4.1\times10^{24}$-$4.8\times10^{25}$ | | | |
| | Barabash et al. (2017)[i] | $1.6\times10^{23}$ | | | |
| | Chassefiére et al. (2007)[k] | $3.9\times10^{24}$ | $2\times10^{26}$ | $8.3\times10^{26}$ | |
| | Bösswetter et al. (2010)[l] | $4\times10^{24}$ | | | $1.5\times10^{26}$ |
| | Jakosky et al. (2018)[h] | $5\times10^{24}$ | | | |
| | Dong et al. (2018) | $4.4\times10^{24}$ | $3.1\times10^{25}$ | $2.5\times10^{26}$ | $1.1\times10^{27}$ |
| | Lichtenegger et al. (2020) | $1.0\times10^{25}$ | $6.6\times10^{26}$ | | $1.2\times10^{27}$ |
| $O_2^+$ | Carlsson et al. (2006)[i] | $4.5\times10^{25}$ | | | |
| | Barabash et al. (2007)[i] | $1.5\times10^{23}$ | | | |
| Sp. O | Luhmann et al. (1992) | $3\times10^{23}$ | $3\times10^{26}$ | $3\times10^{27}$ | |
| | Luhmann et al. (1992)[m] | $7\times10^{23}$ | $1.6\times10^{26}$ | $4.2\times10^{27}$ | |
| | Kass and Yung (1995,1996) | $3\times10^{24}$ | $7\times10^{26}$ | $7\times10^{27}$ | |
| | Leblanc and Johnson (2002) | $5\times10^{23}$ | $1.4\times10^{26}$ | $1.8\times10^{27}$ | |
| | Chassefiere et al. (2007) | $5\times10^{23}$ | $1\times10^{26}$ | $1.6\times10^{27}$ | |
| | Chassefiere & Leblanc (2011a,b) | $5\times10^{23}$ | $5.5\times10^{24}$ | $2.3\times10^{25}$ | |
| | Fang et al. (2013)[j] | $1.1\times10^{24}$-$5.4\times10^{25}$ | | | |
| | Wang et al. (2015)[j] | $2\times10^{24}$-$1\times10^{26}$ | | | |
| | Leblanc et al. (2015)[h] | $1\times10^{24}$ | | | |
| | Jakosky et al. (2018)[h] | $3\times10^{24}$ | | | |
| | Leblanc et al. (2018)[h] | $4\times10^{23}$ | | | |
| | Lichtenegger et al. (2020) | $1.4\times10^{25}$ | $6.6\times10^{25}$ | | $4.5\times10^{25}$ |
| Heavy ions | Nilsson et al. (2011)[i] | $2\times10^{24}$ | | | |
| | Ramstad et al. (2013)[n] | $(2-3)\times10^{25}$ [l] | | | |
| | Ramstad et al. (2015)[i,o] | $1.9\times10^{24}-5.6\times10^{24}$ | | | |

[a]Escape rates corrected by Luhmann et al. 1997; [b]high solar activity; [c]low solar activity; [d]Only dissoc. recombination; [e]More source reactions, see text and Fig. 10; [f]Collision cross section from Fox and Hać (2014); [g]Collision cross section from Gröller et al. (2014); [h]Based on Maven measurements; [i]measured with MEX/APSERA-3; [j]From quiet to extreme solar wind conditions; [k]Ionospheric and pick-up ion escape; [l]for 30 EUV: $1.9\times10^{26}$, and for 100 EUV: : $6.6\times10^{26}$ for normal to $-1.5\times10^{29}$ for extreme solar wind conditions; [m]Luhmann et al. (1992) as corrected by Jakosky et al. (1994); [n]Measured with Phobos 2/ASPERA; [o]Seven years of MEX observations, minimum to maximu rates;

**Table 5**. Simulated and measured escape rates for carbon and $CO_2$ for different EUV fluxes for photochemical loss of C (Hot C in the table), $C^+$ and $CO_2^+$, and sputtered neutral C and $CO_2$ (Sp. O/$CO_2$ in the table). The Table also shows the combined sputter results for O, $CO_2$, CO, and C by Wang et al. (2015). The corresponding times (Ga = Giga years ago) to the EUV fluxes are for a slow rotator according to Tu et al. (2015).

| | | 1 EUV | 3 EUV | 6 EUV | 10 EUV | 20 EUV |
|---|---|---|---|---|---|---|
| | | 0 Ga | 1.9 Ga | 2.6 Ga | 3.8 Ga | 4.2 Ga |
| Hot C | Chassefiére et al. (2007) | $8.2 \times 10^{23}$ | $8.8 \times 10^{24}$ | $1.8 \times 10^{25}$ | | |
| | Kulikov et al. (2007) | $8 \times 10^{23}$ | $4.2 \times 10^{24}$ | $1.3 \times 10^{25}$ | | |
| | Amerstorfer et al. (2017)[a] | $9.5 \times 10^{24}$ | $5.6 \times 10^{25}$ | | $2.3 \times 10^{26}$ | $8.6 \times 10^{26}$ |
| | Zhao et al. (2017), FH14[b] | $5.2 \times 10^{23}$ | $2.4 \times 10^{24}$ | | $2.7 \times 10^{25}$ | $1.9 \times 10^{25}$ |
| | Zhao et al. (2017), G14[c] | $7 \times 10^{22}$ | $2 \times 10^{24}$ | | $2.9 \times 10^{25}$ | $2.7 \times 10^{25}$ |
| Hot CO | Leblanc et al. (2018)[d] | $7.1 \times 10^{22}$ | | | | |
| $C^+$ | Chassefiere et al. (2007)[d] | $6.8 \times 10^{23}$ | $1.3 \times 10^{25}$ | $3.2 \times 10^{25}$ | | |
| | Chassefiere & Leblanc (2011a,b)[e] | $3.5 \times 10^{23}$ | $7.8 \times 10^{24}$ | $1 \times 10^{24}$ | | |
| | Lichtenegger et al. (2020) | $1.9 \times 10^{23}$ | $2.0 \times 10^{26}$ | | $4.7 \times 10^{27}$ | |
| $CO_2^+$ | Carlsson et al. (2006)[f] | $4 \times 10^{24}$ | | | | |
| | Barabash et al. (2007)[f] | $8 \times 10^{22}$ | | | | |
| | Bösswetter et al. (2010)[g] | $3 \times 10^{24}$ | | | $3 \times 10^{26}$ | |
| | Dong et al. (2018) | $3.6 \times 10^{23}$ | $1.4 \times 10^{24}$ | $2.7 \times 10^{24}$ | $4.1 \times 10^{24}$ | |
| Sp. C | Chassefiere et al. (2007) | $1.8 \times 10^{23}$ | $2.5 \times 10^{25}$ | $3.5 \times 10^{25}$ | | |
| | Chassefiere & Leblanc (2011a,b) | $8.2 \times 10^{23}$ | $1 \times 10^{24}$ | $4.4 \times 10^{24}$ | | |
| | Wang et al. (2015)[h] | $2 \times 10^{23}$- $8 \times 10^{24}$ | | | | |
| | Leblanc et al. (2018) | $1.3 \times 10^{23}$ | | | | |
| | Lichtenegger et al. (2020) | $5.5 \times 10^{23}$ | $8.9 \times 10^{24}$ | | $8.5 \times 10^{25}$ | |
| Sp. CO | Leblanc et al. (2018)[d] | $4.7 \times 10^{23}$ | | | | |
| | Wang et al. (2015)[h] | $4 \times 10^{20}$- $2 \times 10^{23}$ | | | | |
| | Lichtenegger et al. (2020) | $5.6 \times 10^{23}$ | $1.8 \times 10^{24}$ | | $7.4 \times 10^{23}$ | |
| Sp. $CO_2$ | Luhmann et al. (1992) | $3 \times 10^{23}$ | $6 \times 10^{25}$ | $3 \times 10^{26}$ | | |
| | Wang et al. (2015)[h] | $9 \times 10^{21}$- $4 \times 10^{23}$ | | | | |
| | Leblanc et al. (2018)[d] | $9.6 \times 10^{22}$ | | | | |
| | Lichtenegger et al. (2020) | $2.4 \times 10^{23}$ | $1.6 \times 10^{23}$ | | $3.7 \times 10^{21}$ | |
| Sp. all | Wang et al. (2015)[h] | $6 \times 10^{23}$- $3 \times 10^{25}$ | $1 \times 10^{26}$- $6 \times 10^{27}$ | $1 \times 10^{27}$- $9 \times 10^{28}$ | | |

[a]More source reactions compared to other studies, see text and Fig. 10Fig. 10

[b]Collision cross section from Fox and Hać (2014)

[c]Collision cross section from Gröller et al. (2014)

[d]Based on MAVEN measurements

[e]Sum of ionospheric and pick-up ion escape

[f]Measured by MEX/ASPERA-3

[g]For 30 EUV: $4.2 \times 10^{26}$, and for 100 EUV: $1.7 \times 10^{27}$ for normal to $1.5 \times 10^{29}$ for extreme solar wind conditions
[h]From quiet to extreme solar wind conditions

Luhmann et al. (1992) investigated the evolutionary impact of sputtering by $O^+$ pick-up ions on the Martian atmosphere from present-day to 3.5 Ga. They found that the sputtering escape rates were significantly rising with increasing EUV fluxes, and to a lesser extend also with increasing solar wind parameters such as velocity and magnetic field. While they retrieved an escape rate for $CO_2$ of $3 \cdot 10^{23}$ $s^{-1}$ at present-day, they found it to be 3 magnitudes higher ($3 \cdot 10^{26}$ $s^{-1}$) for 6 $EUV_\odot$. The integrated loss of $CO_2$ was calculated to be up to 0.14 bar, which was later corrected to be ~15-30% lower (Luhmann, 1997). A similar study by Zhang et al. (1993) investigated photochemical and ion pick-up escape of oxygen until 3 Ga. They found the same behaviour as Luhmann et al. (1992), i.e. a significant increase in escape rates for higher EUV fluxes. Here, the loss of $O^+$ increased from < $10^{25}$ $s^{-1}$ for the present-day to > $10^{27}$ $s^{-1}$ for 6 $EUV_\odot$. The numbers of Zhang et al. (1993), however, were later also corrected by Lumann et al. (1997) to be lower, i.e. ~$6 \cdot 10^{24}$ $s^{-1}$ for present-daay, ~$3 \cdot 10^{25}$ $s^{-1}$ for 3 times, and ~$8 \cdot 10^{25}$ $s^{-1}$ for 6 timaes the present-day EUV flux.

The evolution of all relevant non-thermal escape processes from the present-day to ~3.5 Gyr ago was studied by Chassefière and Leblanc (2004), and Chassefière et al. (2007). Similar to Luhmann et al. (1992) and Zhang et al. (1993), these studies also found a significant increase in ion pick-up escape for rising EUV fluxes. For 6 times the present-day EUV flux, which is around 3.5 Gyr ago in these studies, they retrieved an ion pick-up escape rate for O of ~$10^{27}s^{-1}$ which is close to the rates retrieved by Zhang et al (1993). Also for sputtering, both studies retrieved an increase over time from < $10^{24}$ $s^{-1}$ at present-day to > $10^{27}$ $s^{-1}$ ~ 3.5 Gyr ago. In these studies loss rates of O are generally higher than of C; ionospheric outflow seems further to be less important. A similar behaviour was found by Gillmann et al. (2011) who – based on the work of Chassefière et al. (2007) – also estimated non-thermal escape of C and O during the last ~4 Gyr. It has to be noted, however, that Chassefière and Leblanc (2011a, 2011b) and Chassefière et al. (2013b) re-evaluated their earlier models and found significantly lower rates for sputtering at ~ 3.5 Gyr ago of only > $10^{25}$ $s^{-1}$ compared to >$10^{27}$$s^{-1}$ in their earlier studies (Chassefière and Leblanc, 2004; Chassefière et al., 2007). These rates would accumulate to a loss of only ~7 mbar of $CO_2$ and ~5 m EGL since ~4.1 Gyr ago. The present-day escape rates in the model of Chassefière and Leblanc (2011a, 2011b) and Chassefière et al. (2013b), however, are somewhat below the measured present-day values from MAVEN; correlating their model with MAVEN measurements would lead to higher cumulative losses over time (B. M. M. Jakosky et al., 2018).

Another study (Boesswetter et al., 2010) investigated non-thermal water loss through the history of Mars with 3D multi-ion hybrid simulations and found loss rates for $O^+$ and $O_2^+$ of $4 \cdot 10^{24}$ $s^{-1}$ and $5 \cdot 10^{23}$ $s^{-1}$ for the present-day, and $1.5 \cdot 10^{26}$ $s^{-1}$ and $5 \cdot 10^{25}$ $s^{-1}$ for 10 times the present-day EUV flux, respectively. In addition, they also investigated the loss of $CO_2^+$, for which they retrieved with their model escape rates of $3 \cdot 10^{24}$ $s^{-1}$ for today, and $4.2 \cdot 10^{26}$ $s^{-1}$ for 10 times the present-day EUV flux. They also studied non-thermal escape for 30 and 100 times the present-day EUV flux, which corresponds in their simulations to 4.33 and 449 billion years ago. These will be discussed further in Section 7.2.

Comparatively low values for ion escape over time were found by Barabash et al. (2007) and Ramstad et al. (2018) who extrapolated present-day solar wind induced ion loss rates back in time until ~3.5 Gyr and ~3.9 Gyr ago, respectively. They found a significantly lower integrated loss of only 0.2-4 mbar $CO_2$ through ion pick-up and charge exchange (Barabash et al., 2007) and up to 9 mbar $CO_2$ including cold ion outflow (Ramstad et al. 2018). It has to be noted, however, that sputtering and photochemical escape of neutrals were not included in these studies. Furthermore these extrapolations did not take into account that the exobase of the Martian atmosphere will be significantly expanded under high EUV fluxes (e.g., Lammer et al., 2008; Tian et al., 2009). Ramstad et al. (2015), however, took into account the EUV variability, but found that the dependence of the escape on the EUV flux is relatively weak, if all other parameters (such as the extension of the exobase) stay fixed.

Another recent study by Dong et al. (2018a) made use of sophisticated 3D numerical simulations to model the ion and photochemical losses from Mars over time until about 4 Gyr ago. They retrieved significantly higher atmospheric ion escape rates earlier in the past. While at present-day photochemical escape is dominating, ion escape clearly takes over for higher EUV fluxes (Dong et al., 2018a). At ~4 Gyr ago, which coincides in their study with an EUV flux 10 times the present-day value, the ion escape rate exceeds $10^{27}$ $s^{-1}$, while photochemical escape is only ~$10^{26}$ $s^{-1}$. In total, Dong et al. (2018) found a loss of 0.1 bar $CO_2$ during the last 4 Gyr. It has to be noted, that the losses of $CO_2^+$, which Dong et al. (2018) found for 10 times the present-day EUV flux (i.e. $4.1 \cdot 10^{24}$ $s^{-1}$) is significantly lower than the value retrieved by Boesswetter et al. (2010), i.e. $4.2 \cdot 10^{26}$ $s^{-1}$ (see also Table 5), while the escape of $O^+$ is an order of magnitude higher ($1.1 \cdot 10^{27}$ $s^{-1}$ in the study of Dong et al. 2018, vs $1.4 \cdot 10^{26}$ $s^{-1}$ in the study of Boesswetter et al. 2010). Since the solar wind densities $n_{sw}$ and velocities $v_{sw}$ that were used in both studies for a EUV flux of 10 times the present-day value are identical ($n_{sw} = 46 - 47$ $cm^{-3}$, and $v_{sw} = 858$ km/s), the main reason for this discrepancy is that Boesswetter et al. (2010) used predefined atmospheric profiles by Kulikov et al. (2007) in which $CO_2$ was fixed and was not dissociated by the high EUV flux, while Dong et al. (2018) simulated them self-consistently which results in much higher dissociation rates of $CO_2$. There was simply less $CO_2^+$ but more $O^+$ available than in the simulation of Boesswetter et al. (2010).

Jakosky et al. (2018) used scaling laws to extend the present-day Martian escape rates back into the past until 4.2 Gyr ago, the time at which they assume the cessation of the magnetic dynamo. Their scaling laws are based on the studies of Luhmann et al. (1992), Chassefière et al. (2013b) and Lillis et al. (2017). While the latter only includes photochemical escape, the other two scale photochemical, ion pick-up, and sputtering escape of O. To match the present-day loss rates as measured by MAVEN they multiply each loss rate in the model of Chassefière et al. (2013b) by a constant factor. They retrieve a total loss through time of 0.79 bar $CO_2$ based on Chassefière et al. (2013) and 7.7 bar $CO_2$ based on Luhmann et al. (1992). For $H_2O$ they find a loss of 23 EGL and 253 EGL, respectively. Jakosky et al. (2018)

point out that for the loss prior to 3.5 Gyr ago the uncertainties in the extrapolations of the present-day loss rates are getting very large due to the increasing uncertainties in past solar properties and also due to increasing uncertainties due to non-linearities in extrapolating atmospheric composition and characteristics back in time.

Here, it finally has to be emphasized that changes in the solar activity, either through the solar cycle or through extreme space weather events, can provide an important window into the turbulent distant past of Mars (e.g., Lundin et al., 2008; Nilsson et al., 2010; Bruce M. Jakosky et al., 2015; Ramstad et al., 2017). Besides EUV flux, other external drivers such as the solar wind, CMEs, solar energetic particles, and corotating interactive regions, strongly influence non-thermal escape and studying variations in all these drivers can give an important inside in how any changes affect atmospheric escape (Bruce M. Jakosky et al., 2015). Jakosky et al. (2015) and Ma et al. (2017), for instance, found through analysis of MAVEN data that during incident CMEs the escape at Mars can increase by an order of magnitude. Wang et al. (2014) further found that sputtering at Mars can increase from $2 \times 10^{24} s^{-1}$at nominal solar wind conditions fifty-fold if the interplanetary magnetic field (IMF) strength and solar wind pressure become stronger.

One of the major sputter sources at Mars are incident $O^+$ pick up ions that were ionized by the solar wind but re-entered the Martian atmosphere (e.g., Rahmati et al., 2015). Masunaga et al. (2017) found that the rate of $O^+$ that do not re-enter the Martian atmosphere is strongly dependent on IMF, solar wind dynamic pressure and the gyroradius of the $O^+$ ions, but rather not on the EUV flux. Under extreme solar wind conditions, Masunaga et al. (2017), therefore, deduce, sputtering might even be reduced due to an increase of $O^+$ reflection into the solar wind. Also, Ramstad et al. (2015) found a statistically relevant decrease in ion escape rates of up to a factor of 3 with increasing averaged solar wind density, while the escape rates rise for low solar wind density and increased solar wind velocity; both effects are particularly strong for low EUV fluxes whereas high EUV fluxes tend to show strong variations in the escape, probably due to extreme outflow events or short periods of increased ionization. An explanation for higher escape rates during decreased solar wind densities might be that such conditions result in weaker induced magnetic fields, thereby enhancing the interaction region between the ionosphere and the solar wind (Ramstad et al., 2015). It, however, illustrates that the interaction between the solar wind and the atmosphere does not simply extrapolate towards "higher solar wind densities meaning higher escape rates".

One also has to note that the Sun at maximum activity can only serve as a proxy for ancient average quiet conditions back to about 1.5 – 2.0 Ga (e.g., Lammer et al., 2012). Earlier, particularly during the first few 100 Myr, the plasma and radiation environment of the early Sun was clearly more extreme than can be observed at present-day (e.g, Guedel, 2007; Johnstone et al., 2015b, 2015a; Tu et al., 2015; Airapetian and Usmanov, 2016; see also Section 2 and Fig. 1). Extrapolating back in time farther than ~ 2 Ga is, therefore, highly uncertain and one must constrain the Sun's early activity through solar analogues and semi-empirical evolution models, but not through observations of direct effects on planetary atmospheres, at least within our solar system.

To summarize, several different models, many of them based on present-day observations, extrapolate non-thermal escape rates from today's Martian atmosphere back in time until the cessation of its magnetic dynamo. The vast majority of these models predict increased non-thermal escape rates in the past that get particularly stronger towards 3.5 – 4.0 Ga, thereby leading to significant cumulative losses of $CO_2$ ranging from several mbar to even a few bar. However, extrapolating back in time farther than about 3.5 Ga comes along with significant uncertainties, and one must be extremely cautious. Even though the Sun can serve as a proxy, many interconnected processes, such as the EUV flux, IMF, solar energetic particles, solar wind density and velocity must be taken into account that do not necessarily have a linear effect on non-thermal escape processes. Before 4 Ga, the situation gets even more complicated due to an even harsher solar environment and the existence of an intrinsic magnetic field at Mars. This will be discussed in more detail in the next chapter.

## *7.2 Non-thermal escape before 4 Gyr ago*

Even though none of the afore described studies estimated ion pick-up and sputtering rates for the time when the Martian intrinsic magnetic field was yet existing, most of them indicate that non-thermal escape after the dynamo ceased was significantly higher than at present-day. If one would simply extrapolate these loss rates back to earlier times by neglecting the potential Martian magnetosphere before ~4 Gyr ago, the escape of $CO_2$ and $H_2O$ due to non-thermal escape would most likely be highly significant. This, however, is clearly not possible since the role of the magnetic field has to be taken into account. But this specific role is not entirely clear. And even though some recent studies tried to investigate the effect of the magnetic field on atmospheric escape in more detail (Blackman and Tarduno, 2018; Gunell et al., 2018; Egan et al., 2019), the effect of the intrinsic magnetic field on the atmospheric escape of an extended atmosphere under high EUV fluxes, which was the case at early Mars, was never taken into account in great detail.

There are, however, a few attempts to study non-thermal escape at Mars before ~4 Gyr ago. In particular Terada et al. (2009) studied the non-thermal escape at early Mars under extreme EUV and solar wind conditions. They assumed an EUV flux of 100 times the present-day value for 4.5 Gyr ago with a solar wind density being about 300 times higher than today. Furthermore, they supposed a late onset of the Martian intrinsic magnetic field. They found $O^+$ ion pick-up loss rates of up to $1.5 \times 10^{28}$ $s^{-1}$ which would lead to a water loss of only 8 m EGL during the first $\leq$ 150 Myr, while cold ion outflow into the Martian tail could at the same time account for a loss of 10 – 70 m EGL. Besides of neglecting a potential intrinsic magnetic field of Mars at this time this study also neglected the strong expansion of the atmosphere due to the high solar EUV flux. Including this effect would clearly lead to different escape rates, potentially to much higher values.

Also, Boesswetter et al. (2010) studied non-thermal escape for 30 and 100 times the present-day EUV flux, corresponding to 4.33 and 4.49 billion years ago. However, they did not include the Martian intrinsic magnetic field in their work, which should have been present at these ancient times. They found $O^+$ and $O_2^+$ escape rates for 30 and 100 times the present-day EUV flux of $\sim 1.9 \times 10^{26}$ and $\sim 7.0 \times 10^{25}$, and $\sim 6.6 \times 10^{26}$ and $\sim 3.0 \times 10^{26}$, respectively, for their nominal solar wind cases. In addition, they simulated an extreme solar wind for 100 times the present-day EUV flux, which resulted in escape rates of $\sim 1.5 \times 10^{29}$ for $O^+$ and $\sim 3.0 \times 10^{28}$ for $O_2^+$. For this extreme case, they chose a solar wind density of $n_{sw}$ = 10 000 $cm^{-3}$, while for their nominal case the density was chosen to be $n_{sw}$ = 679.95 $cm^{-3}$. The solar wind velocity was the same in both cases, i.e. $v_{sw}$ = 1717 km/s. Both cases, however, might overestimate the actual solar wind values of the early Sun significantly, as at least the study by Johnstone et al. (2015a, 2015b), a solar wind evolution model that is based on observations of stellar clusters with solar-like stars, indicate. Moreover, and as already briefly discussed in Section 7.1, the simulations by Boesswetter et al. (2010) and Terada et al. (2009) did not take into account the dissociation of $CO_2$ under such high EUV fluxes, which might lead to an underestimation of the oxygen escape rates, while the high solar wind values might on the contrary lead to an overestimation. Together with the omission of the intrinsic magnetic field, these results might hence not depict a realistic scenario of non-thermal escape at early Mars.

In another recent study, Sakai et al. (2018) studied the effect of a weak intrinsic magnetic field on the atmospheric escape at Mars. They used the same three-dimension multispecies single-fluid magnetohydrodynamic model as the previously mentioned study by Terada et al. (2009), but simulated Mars once with a weak magnetic field of 100 nT and once without any intrinsic magnetic field, both for present-day solar wind conditions. They found an increase in the tailward flux of atmospheric ions for the weak intrinsic magnetic field of $O^+$ and $CO_2^+$ by a factor of 2 and one order of magnitude, respectively, while the escape rate of $O^+$ staid the same. Altogether, the loss increased from $\sim 3.3 \times 10^{24} s^{-1}$ to $\sim 4.1 \times 10^{24} s^{-1}$, indicating that the intrinsic magnetic field encourages the escape of heavier ion species in the lower ionosphere through escape mainly via outflow along open field lines at the poles and magnetic reconnection between the planetary and solar wind magnetic fields at the flank of the magnetopause. However, this study did neither simulate the behaviour of atmospheric escape for a strong intrinsic magnetic field nor for an enhanced solar plasma and radiation environment and an extended Martian atmosphere, as would be expected early-on in the evolution of the solar system.

However, Sakata et al. (2020) very recently presented a follow-up study in which they applied the same model to simulate the effect of an intrinsic magnetic field around ancient Mars of different field strengths on atmospheric escape. As input, they used again the same model atmospheres by Kulikov et al. (2007), EUV flux, and solar wind characteristics as in the previously discussed study by Terada et al. (2009). They found that the ion escape increases if the ionospheric plasma pressure together with the pressure of the intrinsic magnetic field cannot counterbalance the incident solar wind dynamic pressure, i.e. as long as the system is in the so-called "overpressure" state similarly to present-day Venus and Mars (e.g., Luhmann et al., 1987). Due to this effect, the $O_2^+$ loss rate, for instance, is increased by up to a factor of 6 from the unmagnetized case to an equatorial dipole field strength of 1000 nT. For cases, where the intrinsic magnetic field is strong enough to sustain the solar wind dynamic pressure, the escape rates plummet by two orders of magnitude, indicating that a sufficiently strong intrinsic magnetic field provides a shield for ancient Mars against strong atmospheric escape (Sakata et al., 2020). Sakata et al. (2020) further note that their results likely underestimate the escape at Mars since their single-fluid MHD model cannot reproduce the observed ion escape due to the Martian polar plume, as simulated with the multispecies MHD model by Dong et al. (2015, 2017, 2018a), and which accounts for 20% - 30% of ion escape at present-day Mars (Dong et al., 2015).

Since the simulations by Sakata et al. (2020) are based on the background atmosphere of Kulikov et al. (2007), as mentioned before, it has again to be noted that their results likely underestimate atmospheric escape at ancient Mars. As for the simulations performed by Terada et al. (2009) and Boeswetter et al. (2010), this atmosphere neglects the dissociation of $CO_2$ by the strong EUV flux at that time (of 100 $EUV_{\odot}$ in their scenarios). According to Tian et al. (2009) $CO_2$ would be mainly dissociated in the upper atmosphere under EUV fluxes ≥ 10 $EUV_{\odot}$ which reduces the infrared cooling of $CO_2$ in the thermosphere. Consequently, this would lead to a significantly expanded upper atmosphere (see also Fig. 7) and potentially to higher escape rates of the lighter carbon atoms. With the exobase of the atmosphere at 20 $EUV_{\odot}$ already increasing to several Martian radii (Tian et al., 2009), the interplay between ionosphere, exobase, intrinsic magnetic field and the solar wind would change significantly since this scenario would strongly alter the pressure equilibrium between the ionospheric plasma and the intrinsic magnetic field on the one hand and the solar wind ram pressure on the other hand. No robust estimate on the specific escape rates under such a scenario can be derived by the already existing studies and further research is urgently warranted.

For sputtering, another non-thermal escape process as described above, there is indeed one study that investigated the effect of an intrinsic magnetic field at Mars on sputtering induced escape (Hutchins et al., 1997). Hutchins et al. (1997) examined the implications of a Martian paleomagnetic field at 1, 3, and 6 $EUV_{\odot}$ on the sputtering loss of Argon and Neon with scenarios ranging up to magnetopause standoff distances as far as 14 000 km from the Martian surface (~ 4 $R_{Mars}$), and found escape through sputtering being significantly reduced. Based on the sputtering simulations by Luhmann et al. (1992) and the upper atmosphere model by Zhang et al. (1993), they found a decrease in escape rates by two orders of magnitude or more even for standoff distances significantly below 2000 km. This reduction, according to Hutchins et al. (1997), is mainly attributed due to (i), the magnetosphere deflecting the solar wind around the atmosphere, thereby reducing solar wind-induced ionization processes, and (ii), the shielding of ions produced in the upper atmosphere from the magnetic field of the solar wind. However, since for EUV fluxes prevalent during the first few 100 Myr, the Martian atmosphere is expected to be significantly expanded even above the most distant standoff distance (~ 4 $R_{Mars}$), investigated by Hutchins et al. (1997), their results cannot be simply extrapolated to the pre-Noachian. It remains unclear how sputtering, but also ion pickup, reacts in such a case. However, for an atmosphere that reaches or even exceeds its magnetopause standoff-distance, neither argument (i), the deflection of the solar wind around the atmosphere, nor argument (ii), the shielding of ions in the upper atmosphere, fully account under such a scenario.

Another important parameter that enhances non-thermal escape at ancient Mars is the frequency of interplanetary CMEs hitting the young planet. These were likely significantly more frequent in the past (e.g., Airapetian et al., 2016) and can enhance atmospheric escape, not only at unmagnetized Mars (e.g., Ma et al., 2017), but potentially also at planets with intrinsic magnetic fields (e.g., Khodachenko et al., 2007). Recently, Kay et al. (2019) found that for young solar like stars including the Sun, CMEs are mainly focused on the ecliptic plane due to its coronal magnetic field deflecting the CMEs towards the astrospheric current sheet. They studied the 0.7 Gyr old Sun's twin $k^1$ Ceti and found the likelihood of a CME hitting the terrestrial planets in the Solar System to be 31 %, while it is only 6 – 16 % at present-day. Combined with the higher frequency of CMEs in the past, this should have a profound effect on non-thermal escape at early Mars as well, but to date, again, no studies exist that investigate the effect of frequent CMEs on the non-thermal escape of ancient magnetized Mars.

Another clue on non-thermal escape processes on early Mars might be gained through studies that investigate other terrestrial planets (Kislyakova et al., 2020) or even exoplanets (Airapetian et al., 2017a; Dong et al., 2017, 2018b, 2019). Recently, Kislyakova et al. (2020) studied the polar outflow at Earth back until the Great Oxidation Event (GOE) and beyond until 3.0 Ga with a Direct Simulation Monte Carlo (DSMC) model by also including theEUV flux induced increase in the exobase altitude. For a present-day atmospheric composition, they found a strong increase in polar outflow for oxygen and nitrogen from $2.1 \times 10^{26}\mathrm{s}^{-1}$ and $8.4 \times 10^{24}\mathrm{s}^{-1}$ at present-day to $1.6 \times 10^{27}\mathrm{s}^{-1}$ and $5.6 \times 10^{26}\mathrm{s}^{-1}$ at 2.5 Ga, respectively. If they decrease the amount of oxygen to 1 %, O escape at 2.5 Ga decreases to $1.0 \times 10^{26}\mathrm{s}^{-1}$ while N escape increases to even $2.9 \times 10^{27}\mathrm{s}^{-1}$. This rise is due to a higher atmospheric temperature and exobase level, since atomic oxygen that cools the upper atmosphere being less abundant in the second simulation. With such an escape rate, all of today's atmospheric nitrogen would have been lost within ~ 2.4 Gyr indicating that higher amounts of infrared-coolants such as $CO_2$ are additionally needed to reduce the exobase level. It further has to be noted that the inner boundary in their simulations (2 % below the exobase level) only increase from 1.075 $R_{Earth}$ at present-day to 1.127 $R_{Earth}$ for the present-day atmospheric composition and to 1.309 $R_{Earth}$ for the 1 % oxygen run, respectively, at 2.5 Ga.

Such an extensive polar outflow already at ancient Earth also indicates substantial escape through the polar caps of early Mars' paleo-magnetosphere. Even though Mars is farther away from the Sun than the Earth, it is only about 10 % by mass and the EUV flux during the first few 100 Myr at Mars were significantly higher than at 2.5 Ga at Earth. With an exobase altitude at ancient Mars of several $R_{Mars}$ (see Fig. 7), the polar outflow likely was a substantial non-thermal loss mechanism that cannot be neglected. It has further to be noted that studies on exoplanets also found significant non-thermal escape rates at Proxima Centauri B (Airapetian et al., 2017b; Dong et al., 2017) or the Trappist planets (Dong et al., 2018b). Due to the substantially higher irradiation and denser plasma environments at M-type stars, however, these results can hardly be compared with ancient Mars. But future research in this direction might also help to better understand non-thermal escape processes in the early Solar System.

Finally, photochemical escape at Mars was investigated by several different studies in the past (e.g., Fox, 2004; Gröller et al., 2014; Lee et al., 2015; Zhao and Tian, 2015; Amerstorfer et al., 2017; Lillis et al., 2017; Zhao et al., 2017), some of them also included – at least partially – the pre-Noachian into their study (Zhao and Tian, 2015; Amerstorfer et al., 2017; Zhao et al., 2017). While Zhao and Tian (2015) investigated the photochemical escape of oxygen with a focus on dissociative recombination of $O_2^+$ for 1, 3, 10, and 20 $EUV_{\odot}$, Amerstorfer et al. (2017) and Zhao et al. (2017) extended this work by including more source reactions for hot oxygen (such as dissociative recombination of $CO_2^+$) and by including hot carbon, since both elements are related to the loss of $CO_2$. In addition, Amerstorfer et al. (2017) also included the photodissociation of the molecule CO and the chemical reaction $O_2^+ + C \rightarrow CO^+ + 1$.

Fig. 10 illustrates the total escape rates of C and O based on the studies of Amerstorfer et al. (2017), Zhao and Tian (2015), and Zhao et al. (2017), for 1, 3, 6, 10 and 20 EUV$_{\odot}$; the different reactions included in these studies are depicted on the top of the figure. While in the work of Zhao and Tian (2015) and Zhao et al. (2017) the total escape rates of C and O decrease for high EUV fluxes, they show an increase for 10 and 20 EUV$_{\odot}$ in the study by Amerstorfer et al. (2017). This increase is due to the photodissociation of CO which is not included in the two other studies. For all other reactions, the escape rates are decreasing for 10 and 20 EUV$_{\odot}$, which is in agreement with the simulations of Zhao and Tian (2015) and Zhao et al (2017). Furthermore, if one compares the escape rates of O from Amerstorfer et al. (2017) for 10 EUV$_{\odot}$ with the photochemical escape rates simulated by Dong et al. (2018a) for the same EUV flux, they show comparable values, with the ones by Dong et al. (2018a) being slightly lower for hot O. Unfortunately, Dong et al. (2018a)did not simulate photochemical escape prior to 4.0 Ga.

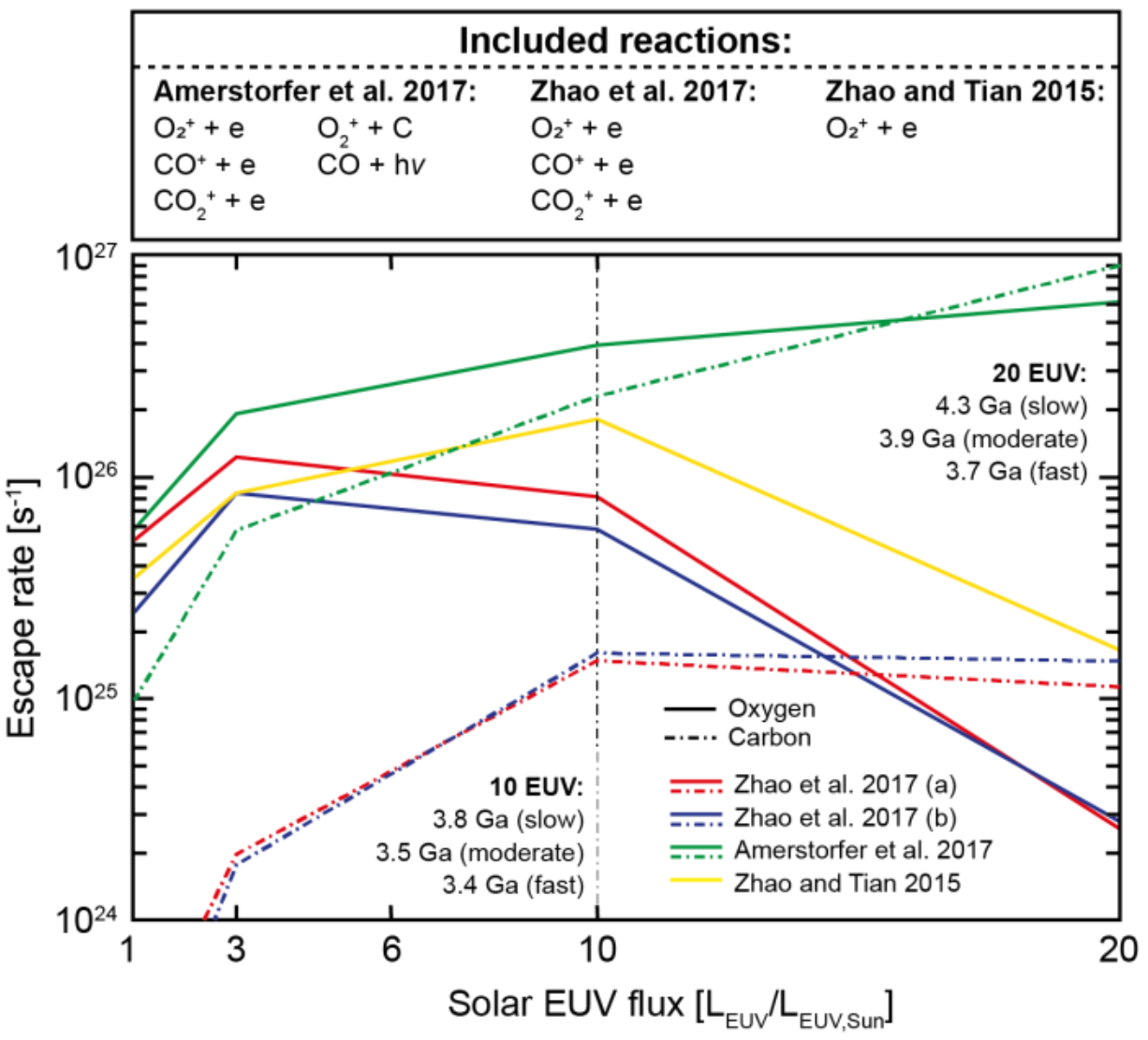

**Fig. 10. Escape rates of photochemical escape of C and O for 1, 3, 6, 10, and 20 times the present-day EUV flux, according to Amertstorfer et al. 2017, Zhao and Tian 2015, and Zhao et al 2017. The reacions included in the different studies are listed at the top. The respective times for 10 and 20 times the repesent-day EUV flux for slow, moderate and fast rotator are according to the solar EUV flux model by Tu et al. 2015. Zhao et al. (I) and Zhao et al. (II) are from the same study but with different implemented cross sections.**

While Zhao et al. (2017) found a total escape of only 20 mbar of $CO_2$ from 4.5 Ga until present, Amerstorfer et al. (2017) received significantly higher values with a total loss of $CO_2$ of ~140 mbar (slow rotator) to ~400 mbar (fast rotator) between 3.9 – 4.3 Ga. They do not provide escape rates older than 4.3 Ga, but in case that photodissociation proceeds to increase for EUV fluxes stronger than 20 times the present-day value, it can be expected that a similar or even higher amount of $CO_2$ could have potentially been lost from the atmosphere of Mars prior to 4.3 Ga. Photochemical escape can therefore not be neglected, if one wants to address the question of whether Mars had a dense atmosphere during the first ~400 million years or not.

Summarizing this subsection, a lack of comprehensive studies on non-thermal escape processes on magnetized planets under extreme solar and stellar environments in general, and on magnetized Mars in particular, can be regarded as its main conclusion. Besides photochemical escape on early Mars, studies on non-thermal escape during the pre-Noachian – the likely timeframe of its paleomagnetic filed – remain scarce. An important study (Sakata et al., 2020), however, has been published recently that deals with the effect of a Martian paleomagnetosphere. Even though Sakata et al. (2020) already retrieve substantial losses for weak intrinsic magnetic fields, it can be expected that these values are nevertheless underestimates since their model does not consider the dissociation of atmospheric $CO_2$ induced by the high EUV flux from the young Sun. Whether an intrinsic magnetic field provides a shield for an extended atmosphere remains unclear and studies on this topic would be from greatest importance. However, research on other magnetized (exo)planets and, generally, on the shielding of magnetospheres against atmospheric losses indicate that non-thermal escape processes on ancient Mars likely were substantial additional drivers of atmospheric erosion.

# 8 Other Processes affecting Atmospheric Stability

## *8.1 Photochemical Stability and Atmospheric Collapse*

It has, finally, to be emphasized that at least a thick $CO_2$-dominated atmosphere might even not have been photochemically stable under the conditions of early Mars as have been shown by Zahnle et al. (2008). While $CO_2$ will be readily dissociated into CO and O, the opposite reaction, i.e. the recombination of CO and O to $CO_2$ under relevant atmospheric temperature and pressure conditions (see Fig. 11a) takes place significantly slower, which will consequently convert a $CO_2$- into a CO-dominated atmosphere. This might in turn also affect atmospheric escape. Contrary to CO, $CO_2$ is acting as an infrared cooler in the upper atmosphere, which lowers the exobase level and therefore also thermal escape.

Besides photochemical stability, one also has to consider the stability of such an atmosphere on early Mars against the so-called "atmospheric collapse" related to changes in the obliquity and particularly to the lower bolometric luminosity of the young Sun. Atmospheric collapse can occur when the thermal energy that is deposited into the atmosphere of Mars is so low that the temperature falls below the sublimation temperature of $CO_2$, which potentially can lead to a complete freezing out of the atmosphere onto the surface (e.g., Leighton and Murray, 1966; Gierasch and Toon, 1973; Haberle et al., 1994; Forget et al., 2013; Soto et al., 2015). Through 3D global climate simulations, Forget et al. (2013) found that a $CO_2$-atmosphere would not have been able to raise the annual mean temperature on early Mars above 0°C. They further even predict the collaps of such an atmosphere and the permanent formation of $CO_2$-ice caps at the poles for pressures above 3 bar, and also for less than 1 bar in case that the obliquity of early Mars declined below 10° to 30° (see Fig. 11b). A higher obliquity increases the seasonal cycle and, consequently, also the temperature gradient between summer and winter. This in turn, leads to an increase in seasonal melting which lowers the chance of permanent $CO_2$-ice cap formation and, thus, of atmospheric collapse (Forget et al., 2013).

That early Mars had episodes of obliquities below 30° could indeed have been the case since its chaotic obliquity strongly varied throughout its lifetime (e.g., Laskar et al., 2004; Brasser and Walsh, 2011). Over the last 5 billion years, the Martian obliquity could have thus varied from less than 5° up to 70° with an average of 37.625°, as has been shown by Laskar et al. (2004). Brasser and Walsh (2011) further simulated the Martian obliquity during the Noachian eon with the assumption that the giant planets were on different orbits before the LHB. They found a most probable chaotic range of obliquity for Mars between 30° and 60° but with less probable values below 30° being more stable than above. However, more recent studies (e.g., Boehnke and Harrison, 2016; Morbidelli et al., 2018) are shading some doubt on the existence of the LHB that was induced by a late giant planet migration event. Some other recent works indeed indicate that such an instability likely already happened within the first ~ 100 Myr after the formation of the solar system (Nesvorný et al., n.d.; Mojzsis et al., 2019; Sousa et al., 2020).

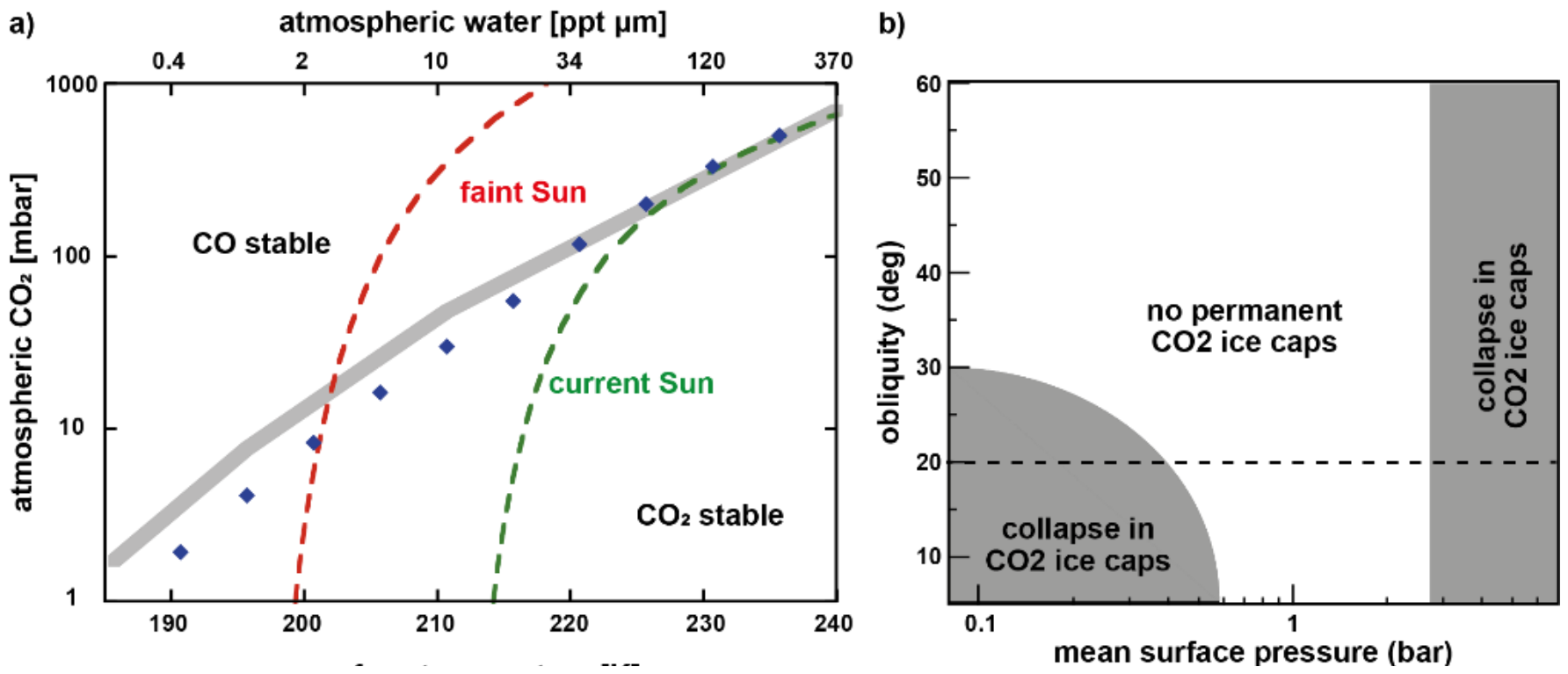

**Fig. 11. a) The photochemical stability of $CO_2$ according to Zahnle et al. (2008). Here, $CO_2$ is stable in the atmosphere to the bottom right side of the grey line (present-day UV and low oxygen escape) or blue diamonds (10 times the present-day UV and high oxygen escape). According to this model a thick $CO_2$-dominated atmosphere would not have been photochemically stable on early Mars. Figure from Zahnle et al. (2008). b) The stability of a Martian CO2-atmosphere against atmospheric collapse as simulated by Forget et al. (2013). The dotted line illustrates the results by Soto et al. 2015, who even found that for obliquities <20° any atmosphere below 3 bar would collapse. Data from Forget et al. (2013) and Soto et al. (2015).**

Therefore, the study by Brasser and Walsh (2011) might not be applicable onto most of the pre-Noachian eon.

Another work (Soto et al., 2015) that studied the parameter space for which a Martian $CO_2$ atmosphere might collapse with a general circulation model, found an even broader range of scenarios for which its atmosphere would not be stable. Even though they did not take into account the faint young Sun, their simulations show that for obliquities below 20° atmospheric collapse will occur for all pressures below 3 bar. Even up to an obliquity of < 45°, mid-range atmospheric masses continued to experience collapse (i.e. 0.3, 0.6, 1.2 bar in their simulations at 30°).

However, this study does not take into account additional heating sources such as dust activity and variations in ice albedo (Forget et al., 2013; Kahre et al., 2013). Additional greenhouse gases might further have aided to avoid the collapse of an early atmosphere. Ramirez et al. (2020) found in agreement with Ramirez et al. (2014) and Batalha et al. (2015) that it may be possible to keep the atmosphere from collapsing in the Noachian – both studies did not investigate the pre-Noachian eon – through $CO_2$-$H_2$ collision induced absorption for hydrogen concentrations as low as ~1%. Several other studies further investigated climate warming through $CO_2$-$H_2$ collisional heating (e.g., Hayworth et al., 2020) and how $H_2$ could have been enriched in the atmosphere of Noachian Mars (e.g., Chassefière et al., 2016; Tosca et al., 2018; Haberle et al., 2019). Moreover, another recent study (Ito et al., 2020) found that within an oxidized environment $H_2O_2$ gas of only 1 ppm in a 3 bar $CO_2$-atmosphere on early Mars could have raised its temperature above freezing. Whether these scenarios, however, would also be compatible during the pre-Noachian eon, might be questionable. As already discussed in Section 6, $CO_2$ likely gets dissociated (Tian et al., 2009) in the pre-Noachian eon due to the high EUV flux from the young Sun. The very lightweight $H_2$ molecule was, furthermomre, if not dissociated as well, highly susceptible to atmospheric escape – even more so than the heavier C atom.

Another molecule that was suggested to aid rising the atmospheric temperature on early-Mars above 273 K is $N_2$. Von Paris et al. (2013) found that 0.5 bar of $N_2$ in the late-Noachian atmosphere at 3.8 Ga could raise temperature by about 13 K due to pressure broadening of absorption lines and collision-induced $N_2$-$N_2$ absorption. However, an atmosphere with a high amount of nitrogen might not be stable under high EUV fluxes as well, as studies for the Earth's $N_2$-dominated (Tian et al., 2008b, 2008a), and early $CO_2$-$N_2$-dominated atmosphere (Johnstone et al., 2020), but also for exoplanets (C. P. Johnstone et al., 2019), indicate.

Another way of avoiding atmospheric collapse on early Mars was studied by Kahre et al. (2013). They investigated the role of dust and found that increasing the dustiness of the atmosphere could have only avoided the collapse of an otherwise unstable atmosphere for high $CO_2$ ice cap albedos (> 0.6), together with high obliquities (> 50°) and an atmospheric dust obtical depth of around unity or higher. They further point out that it yet not known at the time of their study whether such a combination could even occur in a self-consistent system.

To summarize, if a dense atmosphere would have survived thermal and non-thermal escape, it would be highly susceptible to be either photochemically transformed from $CO_2$ to CO and/or to collapse, thereby forming permanent CO/$CO_2$ polar ice caps.

## *8.2 Atmosphere-Surface Interactions*

Besides atmospheric escape, atmosphere-surface interactions and surface sinks are important processes that determine the long-term stability or depletion of an atmosphere. Signs of these interactions, such as carbon sequestration (Michalski and Niles, 2010; Ehlmann et al., 2013; Niles et al., 2013; e.g., Tomkinson et al., 2013; Edwards and Ehlmann, 2015; Wray et al., 2016), surface oxidation (e.g., Catalano, 2013; Dehouck et al., 2016; Chemtob et al., 2017), aqueous alterations (Chevrier et al., 2007; Hurowitz et al., 2010; Wade et al., 2017), or surface weathering (e.g., Dehouck et al., 2014; Baron et al., 2019), therefore, give important insights into the presence and characteristics of an atmosphere, or into the lack thereof. For Mars, however, most of these signatures date back not earlier than to the early Noachian eon at ~ 4.1 Ga, the age of the earliest surface features (e.g., Carr and Head III, 2010).

The sequestration of carbon provides an important sink, and it was suggested to be a main driver for the loss of a thick Noachian $CO_2$-atmosphere, with carbonates, consequently, predicted to be highly abundant in the Martian crust (e.g., Kahn, 1985; Pollack et al., 1987). However, Mars seems to be relatively avoid of carbonates (e.g., J. P. Bibring et al., 2006) even though localized shallow and deep burials of carbonates have indeed been discovered (e.g., Bandfield et al., 2003; Ehlmann et al., 2008; Michalski and Niles, 2010; Morris et al., 2010; Niles et al., 2013; Edwards and Ehlmann, 2015; Wray et al., 2016; Bultel et al., 2019). Shallow carbonates discovered at the Nili Fossae region were estimated to add up to an equivalent of 0.25 - 12 mbar $CO_2$ (Edwards and Ehlmann, 2015). Deep burials of carbonates exposed by impacts or tectonic activity were discovered by Michalski and Niles (2010) in the Leighton impact crater and by Wray et al. (2016) around the Huygens-basin. Depending on whether these outcrops are local phenomena or the exposure of a global buried layer, deep carbonates might comprise an equivalent of about 10 mbar to a maximum of 1 bar of $CO_2$ (Wray et al., 2016). Hu et al. (2015) further estimated an upper bound on stored $CO_2$ of even 1.4 bar by assuming 5 wt% of carbonates stored in the upper 500 m of the Martian crust. A more reasonable estimate, Hu et al. (2015) emphasize, would be 1 wt%, thereby resulting in an equivalent of 300 mbar. An abundance of 1 wt% was also found to be present in the Martian meteorite ALH84001 (Romanek et al., 1994). In the soil of the Phoenix landing site, carbonates were detected with an abundance of 3 – 5 % (Boynton et al., 2009).

Moreover, $CO_2$ ice deposits contain up to about 20 – 30 mbar (Phillips et al., 2011), $CO_2$ adsorbed in the regolith holds another 30 mbar or even less (Zent and Quinn, 1995), while carbonates in dust provide an additional ~ 20 mbar of $CO_2$ (Bandfield et al., 2003; Jakosky, 2019). All reservoirs taken together, therefore, comprise a minimum of less than 100 mbar to an unprobable maximum of about 1.5 bar of $CO_2$ that could have been sequestered from an ancient atmosphere, either in the Noachian eon or even earlier. However, the deeply buried carbon, which lies in a depth of 5 – 10 km, might be the only reservoir that dates back to the pre-Noachian eon (Wray et al., 2016). But whether it indeed originated from the sequestration of a pre-Noachian atmosphere remains unclear.

It further has to be noted that the lack of a substantial carbonate reservoir at Mars might also be attributed to the likely existence of acidic conditions during the Noachian eon (Bullock and Moore, 2007; e.g., Halevy et al., 2007; Fernández-Remolar et al., 2011; Peretyazhko et al., 2018). As pointed out by Halevy et al. (2007) and others, volcanically degassed $SO_2$ and $H_2S$ (see Section 5.2, and Fig. 6) might not only have aided to rise the temperature above freezing but through conversion of $SO_2$ to $H_2SO_4$, sulfuric acid, (e.g., Bullock and Moore, 2007) also prohibited carbonate formation, while, on the other hand, permitting phyllosilicates (Peretyazhko et al., 2018) which are relatively abundant on present-day Mars (e.g., J. P. Bibring et al., 2006; Thomas et al., 2017). Zolotov and Mironenko (2016) point out that the vast majority of these phyllosilicates could have been formed in the Noachian eon under a $CO_2$ atmospheric partial pressure below 0.15 bar. Moreover, weathering conditions could have prohibited the extensive formation of carbonates (Dehouck et al., 2014; Baron et al., 2019). Whether a dense atmosphere was present during the pre-Noachian eon can, therefore, hardly be deduced from today's existing carbonate reservoirs.

Another insight into the existence of an early atmosphere can be gained by the evolution of the oxidation of the Martian mantle. As pointed out in Sections 4 and 6, the hydrodynamic escape of the lighter carbon and hydrogen atoms could have potentially led to the accumulation of the heavier left-behind oxygen from the dissociation of $CO_2$, CO, and $H_2O$. An abiotic $O_2$-dominated atmosphere could have built up (Tian et al. 2009) through this process and, in such a scenario, it seems evident that the Martian crust should have been oxidized already during the pre-Noachian eon. The reduced soil provides an important sink for the highly reactive oxidant $O_2$, a process that, similarly, (biotically) oxidized the terrestrial mantle during the Archean eon (e.g., Kasting et al., 1993; Kadoya et al., 2020). On Mars, however, anoxic conditions likely prevailed until the late-Noachian eon (Chemtob et al., 2017).

As has been shown by Chemtob et al. (2017), late-Noachian sediments contain the mineral ferric smectite which originates from oxidation of its unoxidized progenitor ferrous smectite, a product of anoxic basalt weathering, by aqueous hydrogen peroxide. If the Martian surface would have already been transformed to oxidized conditions at that time, ferrous smectites could not have formed in significant amounts during the late-Noachian. That mostly anoxic conditions yet prevailed at least during the early-Noachian is also suggested by petrological analysis of the Martian meteorite ALH84001 which formed under an anoxic or only minimally oxic environment (King and McSween, 2005; Righter et al., 2008). Such a conclusion is also supported by Catalano (2013) who found that widespread oxidation mainly occurred after the formation of phyllosilicates in the Noachian eon; these were before assumed to have formed under oxidized conditions (Chevrier et al., 2007). In addition, the afore discussed lack of carbonates can be further interpreted as an indication for poorly oxidizing or reducing conditions during most of the Noachian eon (Dehouck et al., 2016).

Taken these indicators together, it seems likely that the Martian surface did not significantly oxidize before the late Noachian/Hesperian eon. An accumulated $O_2$-dominated atmosphere that existed over a recognizable timeframe at some point during the pre-Noachian eon can, therefore, be considered equally improbable.

In a more reasonable scenario, any residual oxygen from a catastrophically outgassed steam atmosphere either escaped itself hydrodynamically or non-thermally as well, or was simply dragged away by the escaping hydrogen and carbon atoms early-on. Any atmospheric nitrogen would have been lost similarly due to the high EUV fluxes of the pre-Noachian eon, as already pointed out earlier. Whether $CO_2$ accumulated at a later point in the pre-Noachian eon, at a time when the EUV flux from a potentially slowly rotating young Sun declined, which later got sequestered to form the deep carbonates cannot be excluded. Interestingly, Kurahashi-Nakamura and Tajika (2006) proposed a potential scenario in which an early $CO_2$-atmosphere collapsed and formed clathrates and carbonates that were sequestered into the subsurface of Mars. Such a setting would indeed be possible, if one considers the deep carbon reservoir and studies on atmospheric collapse (Forget et al., 2013; Soto et al., 2015) that were discussed in the previous subsection.

There is another species that deserves some attention. As already mentioned above, sulfur with a mass of ~ 32 u is expected to be volcanically outgassed, either in its reduced, $H_2S$, or oxidzed form, $SO_2$, on early Mars in significant amounts of up to several 100 mbar to even 1 bar of sulfur (Bullock and Moore, 2007; Craddock and Greeley, 2009; Gaillard and Scaillet, 2009; Johnson et al., 2009; Chassefière et al., 2013a; see also Fig. 6), at least during the late Noachian and Hesperian eons (Gaillard et al., 2013). Acidic conditions were most probably prevalent at this time and significant sulfate deposits were discovered on the Martian surface (Farquhar et al., 2000; Feldman et al., 2004; Squyres et al., 2004; Bibring et al., 2005). Whether sulfur could have even formed a thin sulfur-dominated atmosphere already earlier in the pre-Noachian eon? Due to its heavier mass compared to C (12 u), O (16 u), and N (14 u), it could have accumulated in an early atmosphere relative to these elements. Chassefière et al. (2013a) found that any degassed $SO_2$ could have been trapped in clathrates in the Noachian eon in case of $CO_2$ partial pressures being higher than about 0.5 bar. As the pressure declined, it would have been released back into the atmosphere and formed sulfate deposits later-on. If one assumes lower pressures in the pre-Noachian, sulfate would, therefore, potentially not have been trapped and, if degassed, could have either accumulated or escaped to space.

Any aqueous alterations due to liquid water on the surface of Mars that are yet existing date back to the Noachian eon and hardly can tell something about any pre-Noachian atmosphere. But it is nevertheless interesting to note that the Martian mantle could have sequestered significant amounts of water and it might contain by volume up to 9% hydrous mineral species produced as a consequence of surface reactions compared to only 4% at Earth (Wade et al., 2017). Chassefiére et al. (2013b) further estimated that about 500 m EGL of $H_2O$ could have been trapped in subsurface serpentine. This would provide a significant sink that could have been similarly or even more important than escape to space; but it would potentially also slowly oxidize the crust. As pointed out earlier already, however, McCubbin et al. (2016) suggest a relatively dry Martian mantle, while the crust, on the other hand, could contain up to 0.14 wt% of $H_2O$. It seems furthermore likely that water was predominantly available through the subsurface during the Noachian eon. Substantial clay formation by hydrothermal groundwater circulation, for instance, point towards subsurface water availability (e.g., Ehlmann et al., 2011).

To summarize, atmosphere-surface interactions can be, and likely were, an important sink on early Mars. Due to the oldest formations dating back to the early Noachian eon, however, any signs of a pre-Noachian atmosphere are difficult to infer. Some atmospheric $CO_2$ could have been indeed sequestered into deep carbonates, while potentially reducing conditions until the Noachian eon suggest no $O_2$-rich atmosphere early-on. Sulfur, on the other hand, could have built up to some extent; but whether this could have happened before ~ 4 Ga, and whether it would have been stable against escape remains by now unclear.

# 9 Discussion

To evaluate the question on whether Mars had a dense atmosphere during the first ~400 million years, the sources and sinks – that is, the various escape and sequestration processes, volatile delivery and its environmental and outgassing history – have to be thoroughly taken into account, as described in the previous chapters. Even though we do not yet have a good understanding on the efficiency of non-thermal escape processes at Mars during the first few 100 million years, which is also connected to the role of its early intrinsic magnetic field, several studies investigated other relevant factors such as outgassing of an early steam atmosphere, thermal escape of $H_2O$ and $CO_2$, volcanic degassing, the stability of an early atmosphere and its interaction with the surface, and erosion and delivery of volatiles by impactors. To get an insight into the evolution of the early Martian atmosphere one can at least combine those different sources and sinks that can be estimated reasonably based on simulations and on observational data.

Fig. 12 illustrates different scenarios for the evolution of an early Martian $CO_2$-atmosphere which includes thermal (based on Fig. 4 in Tian et al. 2009, and scaled back into the past linearly to their escape rate at 4.5 Ga) and photochemical escape (Fig. 5 in Amerstorfer et al. 2017, with escape rates set constant from 20 EUV, i.e. ~4.3 Ga for a slow rotator, back to 4.5 Ga; this likely underestimates early supra-thermal escape) as sinks, and impact delivery (data from Fig. 5, red line, in Pham and Karatekin (2016) but set to 0 Myr as a lower and from Fig. 5, dashed green line, in Lammer et al. (2013) as a maximum value) and volcanic degassing (data for lower values from Fig. 4b, solid line, in Grott et al. (2011) and for maximum values from Fig. 6, line for Manning et al. (2006), in Hirschmann et al. 2008; continuous $CO_2$ degassing fluxes, high and low, in Fig. 12b from Fig. 4 in Tian et al. (2009)) as sources of $CO_2$.

To calculate the atmospheric pressure at any time step (of 1 Myr), the available data from the relevant figures in the afore mentioned studies on atmospheric erosion on the one hand, and volcanic outgassing and impact delivery on the other hand, were converted to, either, $CO_2$ source, or $CO_2$ sink in bar per million years. For every time step the sources and sinks were summed up: If the result was negative, then atmosphere was lost within this time step and the total atmospheric pressure was reduced by this value. In case the total pressure was already at zero, it remained at zero since the total pressure cannot become negative. If the sum of sources and sinks were positive within a time step, the total atmospheric pressure was increased by this value. Note that the theoretical maximum cumulative loss through the sinks (> 200 bar) is higher than the cumulative delivery through the sources, but of course, there cannot be more atmosphere lost than provided by the sources – the theoretical maximum cumulative loss, therefore, only indicates the maximum amount that can indeed be lost. Due to this, an atmosphere can transiently or at the end nevertheless build up, in case that at some point, for one or several successive time steps, the cumulative sources exceed the cumulative sinks, which is indeed the case in Fig. 12b towards about 450 Myr.

The black lines illustrate atmospheric pressures as a sum over all sources and sinks, while the coloured lines are cumulative $CO_2$ pressures for the different sources and sinks. Fig. 12a shows the evolution of the $CO_2$-atmosphere in case that Mars experienced an early catastrophically outgassing of a steam atmosphere, which seems to be likely since there are robust signs of an early magma ocean on Mars (see also Chapter 4). Here, the dash-dotted line illustrates a minimum value of 5 bar for the catastrophically outgassed $CO_2$ and the solid line a maximum of 133 bar (based on Erkaev et al. (2014), Odert et al. (2018), and Elkins-Tanton (2008); see also Table 1). As can be seen, the $CO_2$ is lost within a few million years even for the maximum case mainly through the strong thermal escape of C. Later volatile delivery and volcanic outgassing cannot build up a secondary $CO_2$-atmosphere during the first 500 million years, not even if one assumes very high delivery of volatiles via impactors and high volcanic degassing. Note that Fig. 12a has a logarithmic x-axes, contrary to Fig. 12b; otherwise the minimum case (black dash-dotted line) would be hardly visible due to the rapid escape of the atmosphere within less than 5 Myr after the assumed outgassing of the steam atmosphere (at 10 Myr, as in Odert et al. 2018).

Fig. 12b assumes, similarly to the assumption in Tian et al. (2009), that most of the Martian volatile inventory was not outgassed catastrophically at the time of the magma ocean solidification but continuously via volcanic degassing, declining ex-

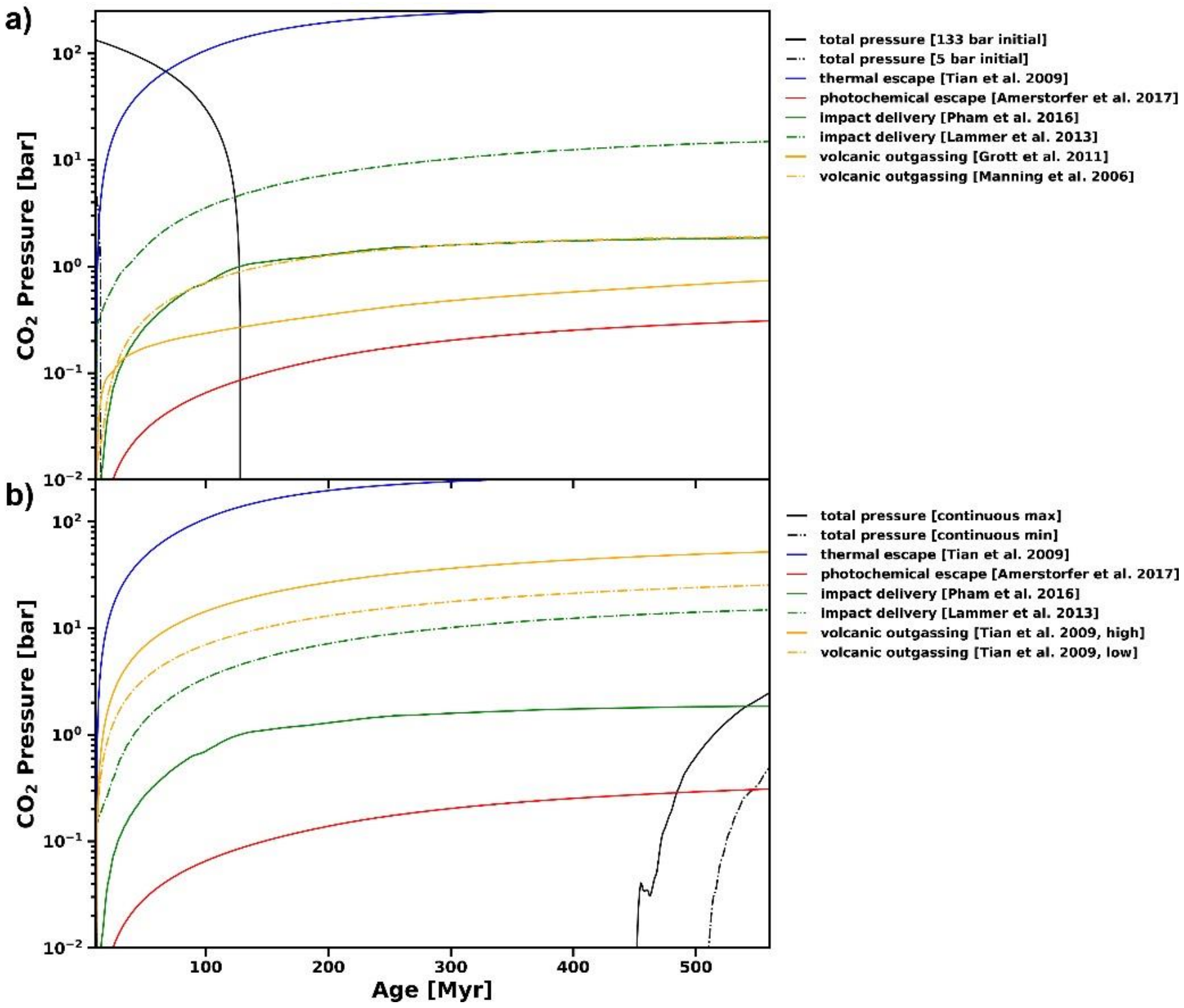

**Fig. 12. Evolution of the potential early Martian $CO_2$ atmosphere based on different studies of thermal and photochemical escape, volatile impact delivery, and volcanic degassing. The different lines show cumulative $CO_2$ pressures for different sources and sinks. Panel a) assumes a catastrophically outgassed steam atmosphere ranging from a minimum of 5 bar $CO_2$ (dash-dotted black line) to a maximum of 133 bar $CO_2$ (solid black line). Subsequent volcanic degassing and impacts deliver $CO_2$ to the atmosphere (Lammer et al. 2013 and Manning et al. 2006 were chosen to depict maximum values for impact delivery and volcanic degassing). The $CO_2$ is rapidly lost and no atmosphere can build up, not even for maximum values of delivery and outgassing. Panel b) shows cases of continuous outgassing of the whole Martian $CO_2$ inventory over its history based on Tian et al. (2009). The solid black line illustrates a maximum case (high $CO_2$ outgasssing and high impact delivery) while the dashed-dotted line illustrates a moderate case. No atmosphere can build up during the first 400 Myr, not even for favorable cases. This figure ignores other non-thermal escape processes assumes an EUV flux from a slow rotating Sun.**

ponentially over time. The rates for degassing are the same as in Tian et al. (2009),

i.e. a similar volatile inventory for Mars as for Earth and an inventory that is twice as big as the terrestrial one. For both inventories, however, no $CO_2$-atmosphere can build up before about 450 million years, not even if one assumes a significant delivery of volatiles via impactors. The Sun was assumed to be a slow rotator in both panels of Fig. 12.

There are, however, a few important points to make about these estimations.

First, for the evolution of the $CO_2$-atmosphere in Fig. 12 it was assumed that every thermally and photochemically escaping C atom is corresponding to one $CO_2$ molecule. The two respective oxygen atoms are then either escaping to space as well, accumulating in the atmosphere or are deposited into the soil (see discussion in the previous chapters). This assumption might, therefore, lead to the accumulation of O in the atmosphere; even though it seems likely that all or most of the oxygen will also escape to space but with a rate somewhat lower than that of C due to its higher mass (see also Tian et al. 2009). Also, oxygen from the escape of $H_2O$ could accumulate in addition. Fig. 12 further tells nothing about other elements that could be outgassed into the atmosphere, such as nitrogen, sulfur, or even argon. Particularly nitrogen, however, is highly susceptible to escape hydrodynamically as well; other elements can either escape themselves too, being dragged away into space by the hydrodynamically escaping hydrogen or carbon atoms, sequester into the soil, or potentially even accumulate to some extent in the atmosphere, as outlined in the previous sections.

Secondly, the thermal escape that shapes the evolution of the $CO_2$-atmosphere is based on the model of Tian et al. (2009) which leads to very high dissociation rates of $CO_2$ under such high EUV flux conditions. If $CO_2$ would not be dissociated then the thermal escape would be significantly lower due to i) $CO_2$ being clearly more massive than C and, more importantly, ii) $CO_2$ being an infrared cooler in the thermosphere which significantly decreases the exobase and therefore atmospheric escape. Even though the study by Tian et al. (2009) is the only investigation of thermal escape of $CO_2$ on early Mars it is not the only work studying the effect of high EUV fluxes on $CO_2$-dominated atmospheres. For instance, a recent study by Johnstone et al. (2020) on the atmosphere of early Earth also indicates that a $CO_2$-dominated atmosphere might be susceptible to strong thermal escape under very high EUV fluxes due to the dissociation of $CO_2$. An application of their model to early Mars would be favourable to confirm the results retrieved by Tian et al. (2009). For the feasibility of hydrodynamic escape in general see further also the discussion in Section 6.2.

Thirdly, other non-thermal escape processes than photochemical escape are not included in Fig. 12 since our knowledge about these processes over the early Martian history is yet scarce. Although several different studies (see Table 4 and Table 5) indicate that non-thermal escape became increasingly stronger over the past during the last ~4 billion years, its effect on early Mars cannot simply be extrapolated back in time due to its early intrinsic magnetic field. It is, however, very likely that non-thermal escape might have also been significant during the first 400 million years of the Martian history as the study by Sakata et al. (2020) and preliminary results by Dong et al. (2018c) may suggest. The early intrinsic magnetic field of Mars might therefore not have played a protective role to prevent strong non-thermal escape and the erosion of the atmosphere. In addition, Fig. 12 does not include the dragging of $CO_2$ by the hydrodynamically escaping hydrogen that originated from a catastrophically outgassed steam atmosphere. But even though (i) these two processes are not included, (ii) sources for $CO_2$ in Fig. 12 were considered to be very high, and (iii) the Sun was considered to be a slow rotator, no $CO_2$-atmosphere could have been maintained during the first 400 million years. Including these processes – dragging of $CO_2$ and non-thermal escape – will decrease the probability of a stable and dense $CO_2$ atmosphere around Mars during the pre-Noachian eon even further.

Whether any heavier element can accumulate on Mars within the first 400 million years, is comparatively more difficult to seek out than the fate of $CO_2$. The lighter hydrogen, if outgassed by volcanic degassing, certainly escaped hydrodynamically, the heavier nitrogen should have escaped significantly as well, since a nitrogen-dominated atmosphere already gets unstable for 10 $EUV_{\odot}$ or even less, as several other studies from the more massive Earth (e.g., Tian et al., 2008a, 2008b; Johnstone et al., 2020) and from exoplanets indicate (e.g., Airapetian et al., 2017a;

Johnstone et al., 2020). Sulfur, and potentially argon and/or other heavier noble gases remain, that could have probably accumulated in the atmosphere towards the end of the pre-Noachian eon to form a thin atmosphere before $CO_2$ finally started to prevail in the Noachian eon. But there are currently no studies existing that investigated the loss of these elements around early Mars, or, in detail, on any other terrestrial planet or exoplanet.

There is, however, one intriguing detail that deserves further attention. The measurement of $^{36}Ar/^{38}Ar$ in the Martian meteorite ALH84001 with an age of ~4.1 Ga (Lapen et al., 2010) revealed that the Martian argon isotopes were not yet fractionated at the time of the formation of this meteorite (Mathew and Marti, 2001). Although sputtering should be the main fractionation process at Mars (e.g., Slipski and Jakosky, 2016; Jakosky et al., 2017b; Lichtenegger et al., 2020), which potentially was not important until the time of the cessation of the Martian dynamo, it can be expected that Ar could have been fractionated through other escape processes early on, such as the dragging by hydrodynamically escaping hydrogen atoms (e.g., Odert et al., 2018) which indeed fractionated Ar on early Earth (Lammer et al., 2020). That no fractionation prior to ~4.1 Ga exists might point to the possibility that any previously fractionated argon isotopes escaped from the planet and that volcanic degassing replenished it later-on. Similarly, nitrogen isotopes in ALH84001 do not show strong sings of fractionation as well (Mathew and Marti, 2001); on the contrary to the present-day Martian nitrogen (Wong et al., 2013), which has to be fractionated later, for instance, through photochemical escape (e.g., Fox and Hać, 1997). This, in addition, suggests that nitrogen escaped hydrodynamically from early Mars, and was replenished later-on during the Noachian eon through volcanic degassing. Moreover, the very heavy xenon isotopes with masses ranging from 124 – 136 u were already completely fractionated at the time of the formation of ALH84001 (e.g., Cassata, 2017). This might illustrate that the escape of the easily ionizable xenon might be coupled with an intrinsic magnetic field since the lighter but hardly ionizable krypton shows almost no signatures of fractionation (e.g., Hébrard and Marty, 2014; Zahnle et al., 2019; Avice and Marty, 2020) But it also points to significant loss processes early in the Martian history, since xenon could otherwise hardly escape at all.

So, what can, finally, be done to shed better light on the early evolution of the Martian atmosphere?

From a modelling perspective, it would be helpful whether the intriguing results by Tian et al. (2009) could be confirmed by models from other scientists. In line with this, it must be better investigated whether there are any additional trace molecules that could potentially serve as infrared coolants in upper atmospheres and which would survive the high EUV flux of the young Sun. If so, such molecules could potentially cool the atmosphere and reduce its extension, thereby diminishing atmospheric escape. It is further important to study the thermal escape of other species from an early Martian atmosphere, such as sulfur, oxygen, nitrogen, argon, and molecules thereof. Combining such models with outgassing and surface sinks might give a good overview on which atmospheric combinations could have accumulated on Mars and at what time.

For this, however, the evolution of the young Sun must be better constrained as well. Even though it likely has been a slow rotator, further studies on this topic are warranted to get an increased knowledge on its early radiation and plasma environment, particularly also in view of itsevolution of solar mass loss and extreme events. To achieve this goal, comparative planetology will be needed, since the evolution of all the terrestrial planets must be reconstructed under the same single Sun. For this, it would be from highest priority to go back to Venus and measure in detail its isotopic noble gas composition. Without knowing its isotopic ratios, it will not be possible to reconstruct the evolution of Venus, which would in turn be a cornerstone for better understanding the young Sun (see, e.g., Lammer et al., 2020).

Another important next step will be a thorough investigation on the effect of the intrinsic magnetic field of Mars on its early non-thermal escape processes. This further includes not only a better understanding on the Martian paleomagnetosphere itself and the time of its cessation, but it also contains answering the question on whether, and under which conditions, a magnetosphere serves as a shield or a funnel. If non-thermal escape will be indeed enhanced by an intrinsic magnetic field with the characteristics prevalent at early Mars, then a significant additional sink would have acted over the first ~500 million years. With this, any dense atmosphere would be even more unlikely than without strong non-thermal escape. All those sources and sinks, from thermal to non-thermal escape, over volcanic outgassing, towards impact erosion and delivery, and, finally, to the relevant atmosphere-surface interactions, have to be taken together to retrieve a reliable model on the early evolution of its atmosphere.

From an observational perspective, there might be two aspects that can confirm or support any model based early Martian atmosphere. First, it will be important proceeding to investigate the Martian surface and subsurface environment. Finding any further residues of the pre-Noachian eon, will hopefully tell a lot about its early history, particularly about any interaction between the atmosphere and its surface. It would already be significant to better estimate the present carbon reservoirs in the Martian mantle and to find out, whether it stems from the sequestration of an early atmosphere or not. Further investigating Mars by planetary exploration will also tell us more about the outgassing history of the planet and will give insides onto its ancient magnetosphere.

The second aspect that will be exceedingly important in the future, is the empirical study of exoplanets and their environments. At present-day, we cannot directly observe the effects of the young Sun's environment on the atmospheres of the terrestrial planets. But there are numerous solar analogues such as $k^1$ Ceti that resemble the infant youth of the Sun, and with future instruments it will, at some point, hopefully be possible to study atmospheric escape at exoplanets around such young solar-like stars. Observing and researching these distant messengers will tell us a lot about the history of our own Sun and the planets surrounding it. It will also give us important insights into the evolution of early Mars and it might help us to better understand whether it indeed had no dense atmosphere or whether we will have to adopt our physics and assumptions about its distant history.

# 9 Conclusion

Early Mars might not have been able to maintain a stable and dense $CO_2$-dominated atmosphere during the first 400 million years. This is mainly due to an interplay of i) the environmental conditions of the early solar system such as the high EUV flux from the young Sun and ii) the physical characteristics of the planet itself since the small mass of Mars makes it highly susceptible to atmospheric escape. Thermal escape would have eroded any catastrophically outgassed steam atmosphere and hindered the build-up of a dense secondary atmosphere. Some oxygen might have been able to accumulate over time, but that the Martian mantle did most probably not oxidize already during the pre-Noachian makes such an accumulation unlikely; whether sulfur could have built-up to some extent before the Noachian eon remains unclear. After 400 to 500 million years – during the Noachian eon, a time for which there is some evidence for liquid water on Mars – a $CO_2$-dominated atmosphere might have been able to build up depending on volcanic degassing rates and the strength of non-thermal escape processes such as ion pickup and sputtering. To better understand the evolution of its early atmosphere, however, further investigations particularly into non-thermal escape processes must be made. This also includes the role of planetary magnetic fields as a shield for or a funnel against atmospheric escape, but also better constraining the early evolution of the Sun, and insights of exoplanets.

A warm and wet Mars as a potential early habitat for life likely requires a dense atmosphere. If it was not able to maintain such an atmosphere, it seems indeed unlikely that the planet was warm and wet early on and subsequently evolved to the present cold and dry body. Mars instead might just have always been a cold and dry planet that just episodically experienced liquid water 4 to 3.6 billion years ago due to volcanic degassing and/or impact heating.

**Acknowledgments** MS acknowledges the support of Europlanet 2020 RI. Europlanet 2020 RI has received funding from the European Union's Horizon 2020 research and innovation programme under grant agreement No 654208. We finally acknowledge two anonymous referees who, due to their comprehensive suggestions, helped to significantly enhance the value of this review.

## References

Acuna, M.H., Connerney, J.E.P., Ness, N.F., Lin, R.P., Mitchell, D., Carlson, C.W., McFadden, J., Anderson, K.A., Reme, H., Mazelle, C., Vignes, D., Wasilewski, P., Cloutier, P., 1999. Global Distribution of Crustal Magnetization Discovered by the Mars Global Surveyor MAG/ER Experiment. Science (80-. ). 284, 790–793. https://doi.org/10.1126/science.284.5415.790

Agee, C.B., Draper, D.S., 2004. Experimental constraints on the origin of Martian meteorites and the composition of the Martian mantle. Earth Planet. Sci. Lett. 224, 415–429. https://doi.org/10.1016/j.epsl.2004.05.022

Ahrens, T.J., J., T., 1993. Impact Erosion of Terrestrial Planetary Atmospheres. Annu. Rev. Earth Planet. Sci. 21, 525–555. https://doi.org/10.1146/annurev.ea.21.050193.002521

Airapetian, V.S., Glocer, A., Gronoff, G., Hébrard, E., Danchi, W., 2016. Prebiotic chemistry and atmospheric warming of early Earth by an active young Sun. Nat. Geosci. 9, 452–455. https://doi.org/10.1038/ngeo2719

Airapetian, V.S., Glocer, A., Khazanov, G. V, Loyd, R.O.P., France, K., Sojka, J., Danchi, W.C., Liemohn, M.W., 2017a. How Hospitable Are Space Weather Affected Habitable Zones ? The Role of Ion Escape. Astrophys. J. Lett. 836, 1–5. https://doi.org/10.3847/2041-8213/836/1/L3

Airapetian, V.S., Jackman, C.H., Mlynczak, M., Danchi, W., Hunt, L., 2017b. Atmospheric Beacons of Life from Exoplanets Around G and K Stars. Sci. Rep. 1–9. https://doi.org/10.1038/s41598-017-14192-4

Airapetian, V.S., Usmanov, A. V, 2016. Reconstructing the Solar Wind From Its Early History to Current Epoch. arXiv Prepr. https://doi.org/10.3847/2041-

8205/817/2/L24

Amerstorfer, U. V., Gröller, H., Lichtenegger, H., Lammer, H., Tian, F., Noack, L., Scherf, M., Johnstone, C., Tu, L., Güdel, M., 2017. Escape and evolution of Mars's CO 2 atmosphere: Influence of suprathermal atoms. J. Geophys. Res. Planets 122, 1321–1337. https://doi.org/10.1002/2016JE005175

Avice, G., Marty, B., 2020. Perspectives on Atmospheric Evolution from Noble Gas and Nitrogen Isotopes on Earth, Mars & Venus. Space Sci. Rev. https://doi.org/10.1007/s11214-020-00655-0

Bandfield, J.L., Glotch, T.D., Christensen, P.R., 2003. Spectroscopic identification of carbonate minerals in the martian dust. Science (80-. ). 301, 1084–1087. https://doi.org/10.1126/science.1088054

Barabash, S., Fedorov, A., Lundin, R., Sauvaud, J.-A., 2007. Martian Atmospheric Erosion Rates. Science (80-. ). 315, 501–503. https://doi.org/10.1126/science.1134358

Baratoux, D., Toplis, M.J., Monnereau, M., Sautter, V., 2013. The petrological expression of early Mars volcanism. J. Geophys. Res. Planets 118, 59–64. https://doi.org/10.1029/2012JE004234

Baron, F., Gaudin, A., Lorand, J.P., Mangold, N., 2019. New Constraints on Early Mars Weathering Conditions From an Experimental Approach on Crust Simulants. J. Geophys. Res. Planets 124, 1783–1801. https://doi.org/10.1029/2019JE005920

Batalha, N., Domagal-Goldman, S.D., Ramirez, R., Kasting, J.F., 2015. Testing the early Mars H<inf>2</inf>-CO<inf>2</inf> greenhouse hypothesis with a 1-D photochemical model. Icarus 258, 337–349. https://doi.org/10.1016/j.icarus.2015.06.016

Benedikt, M.R., Scherf, M., Lammer, H., Marcq, E., Odert, P., Leitzinger, M., Erkaev, N.V., 2020. Escape of rock-forming volatile elements and noble gases from planetary embryos. Icarus 347. https://doi.org/10.1016/j.icarus.2020.113772

Bibring, J., Langevin, Y., Mustard, J.F., Arvidson, R., 2006. Global Mineralogical and Aqueous Mars History Derived from OMEGA / Mars Express Data 312,

400–404.

Bibring, J.P., Langevin, Y., Gendrin, A., Gondet, B., Poulet, F., Berthé, M., Soufflot, A., Arvidson, R., Mangold, N., Mustard, J., Drossart, P., Erard, S., Fomi, O., Combes, M., Encrenaz, T., Fouchet, T., Merchiorri, R., Belluci, G.C., Altieri, F., Formisano, V., Bonello, G., Capaccioni, F., Cerroni, P., Coradini, A., Fonti, S., Kottsov, V., Ignatiev, N., Moroz, V., Titov, D., Zasova, L., Pinet, P., Douté, S., Schmitt, B., Sotin, C., Hauber, E., Hoffmann, H., Jaumann, R., Keller, U., Duxbury, T., Forget, F., 2005. Mars surface diversity as revealed by the OMEGA/Mars express observations. Science (80- . ). 307, 1576–1581. https://doi.org/10.1126/science.1108806

Bibring, J.P., Langevin, Y., Mustard, J.F., Poulet, F., Arvidson, Raymond, Gendrin, A., Gondet, B., Mangold, N., Pinet, P., Forget, F., Berthe, M., Gomez, C., Jouglet, D., Soufflot, A., Vincendon, M., Combes, M., Drossart, P., Encrenaz, T., Fouchet, T., Merchiorri, R., Belluci, G.C., Altieri, F., Formisano, V., Capaccioni, F., Cerroni, P., Coradini, A., Fonti, S., Korablev, O., Kottsov, V., Ignatiev, N., Moroz, V., Titov, D., Zasova, L., Loiseau, D., Pinet, Patrick, Douté, S., Schmitt, B., Sotin, C., Hauber, E., Hoffmann, H., Jaumann, R., Keller, U., Arvidson, Ray, Duxbury, T., Forget, François, Neukum, G., 2006. Global mineralogical and aqueous Mars history derived from OMEGA/Mars express data. Science (80-. ). 312, 400–404. https://doi.org/10.1126/science.1122659

Blackman, E.G., Tarduno, J.A., 2018. Mass, energy, and momentum capture from stellar winds by magnetized and unmagnetized planets: implications for atmospheric erosion and habitability. Mon. Not. R. Astron. Soc. Vol. 481, Issue 4, p.5146-5155 481, 5146–5155. https://doi.org/10.1093/mnras/sty2640

Boehnke, P., Harrison, T.M., 2016. Illusory Late Heavy Bombardments 2016. https://doi.org/10.1073/pnas.1611535113

Boesswetter, A., Lammer, H., Kulikov, Y., Motschmann, U., Simon, S., 2010. Non-thermal water loss of the early Mars: 3D multi-ion hybrid simulations. Planet. Space Sci. 58, 2031–2043. https://doi.org/10.1016/j.pss.2010.10.003

Bollard, J., Connelly, J.N., Whitehouse, M.J., Pringle, E.A., Bonal, L., Jørgensen,

J.K., Nordlund, Å., Moynier, F., Bizzarro, M., 2017. Early formation of planetary building blocks inferred from Pb isotopic ages of chondrules. Sci. Adv. vol. 3, issue 8, p. e1700407 3, e1700407. https://doi.org/10.1126/sciadv.1700407

Bouvier, L.C., Costa, M.M., Connelly, J.N., Jensen, N.K., Wielandt, D., Storey, M., Nemchin, A.A., Whitehouse, M.J., Snape, J.F., Bellucci, J.J., Moynier, F., Agranier, A., Gueguen, B., Schönbächler, M., Bizzarro, M., 2018. Evidence for extremely rapid magma ocean crystallization and crust formation on Mars. Nature 558, 586–589. https://doi.org/10.1038/s41586-018-0222-z

Boynton, W. V., Ming, D.W., Kounaves, S.P., Young, S.M.M., Arvidson, R.E., Hecht, M.H., Hoffman, J., Niles, P.B., Hamara, D.K., Quinn, R.C., Smith, P.H., Sutter, B., Catling, D.C., Morris, R. V., 2009. Evidence for calcium carbonate at the mars phoenix landing site. Science (80-. ). 325, 61–64. https://doi.org/10.1126/science.1172768

Brain, D.A., Baker, A.H., Briggs, J., Eastwood, J.P., Halekas, J.S., Phan, T.D., 2010. Episodic detachment of Martian crustal magnetic fields leading to bulk atmospheric plasma escape. Geophys. Res. Lett. 37, 3–7. https://doi.org/10.1029/2010GL043916

Brasser, R., Mojzsis, S.J., 2017. A colossal impact enriched Mars' mantle with noble metals. Geophys. Res. Lett. 44, 5978–5985. https://doi.org/10.1002/2017GL074002

Brasser, R., Mojzsis, S.J., Werner, S.C., Matsumura, S., Ida, S., 2016. Late veneer and late accretion to the terrestrial planets. Earth Planet. Sci. Lett. 455, 85–93. https://doi.org/10.1016/j.epsl.2016.09.013

Brasser, R., Ramon, 2013. The Formation of Mars: Building Blocks and Accretion Time Scale. Space Sci. Rev. 174, 11–25. https://doi.org/10.1007/s11214-012-9904-2

Brasser, R., Walsh, K.J., 2011. Stability analysis of the martian obliquity during the Noachian era. Icarus 213, 423–427. https://doi.org/10.1016/j.icarus.2011.02.024

Brasser, R., Werner, S.C., Mojzsis, S.J., 2020. Impact bombardment chronology of

the terrestrial planets from 4.5 Ga to 3.5 Ga. Icarus 338, 113514. https://doi.org/10.1016/j.icarus.2019.113514

Brien, D.P.O., Walsh, K.J., Morbidelli, A., Raymond, S.N., Mandell, A.M., 2014. Water delivery and giant impacts in the ‘ Grand Tack ’ scenario. Icarus 239, 74–84. https://doi.org/10.1016/j.icarus.2014.05.009

Bristow, T.F., Haberle, R.M., Blake, D.F., Des, D.J., Eigenbrode, J.L., Fairén, A.G., Grotzinger, J.P., Stack, K.M., Mischna, M.A., Rampe, E.B., Siebach, K.L., Sutter, B., Vaniman, D.T., Vasavada, A.R., 2016. Low Hesperian P CO2 constrained from in situ mineralogical analysis at Gale Crater , Mars. https://doi.org/10.1073/pnas.1616649114

Bullock, M.A., Moore, J.M., 2007. Atmospheric conditions on early Mars and the missing layered carbonates. Geophys. Res. Lett. 34, 2–7. https://doi.org/10.1029/2007GL030688

Bultel, B., Viennet, J.C., Poulet, F., Carter, J., Werner, S.C., 2019. Detection of Carbonates in Martian Weathering Profiles. J. Geophys. Res. Planets 124, 989–1007. https://doi.org/10.1029/2018JE005845

Cannon, K.M., Parman, S.W., Mustard, J.F., 2017. Primordial clays on Mars formed beneath a steam or supercritical atmosphere. Nature 552, 88–91. https://doi.org/10.1038/nature24657

Carlsson, E., Fedorov, A., Barabash, S., Budnik, E., Grigoriev, A., Gunell, H., Nilsson, H., 2006. Mass composition of the escaping plasma at Mars 182, 320–328. https://doi.org/10.1016/j.icarus.2005.09.020

Carr, M.H., 1986. Mars: A water-rich planet? Icarus 68, 187–216. https://doi.org/10.1016/0019-1035(86)90019-9

Carr, M.H., 1999. Retention of an atmosphere on early Mars. J. Geophys. Res. E Planets 104, 21897–21909. https://doi.org/10.1029/1999JE001048

Carr, M.H., Head III, J.W., 2010. Geologic history of Mars 294, 185–203. https://doi.org/10.1016/j.epsl.2009.06.042

Carr, M.H., Head, J.W., 2003. Oceans on Mars: An assessment of the observational evidence and possible fate. J. Geophys. Res. E Planets 108. https://doi.org/10.1029/2002je001963

Cassata, W.S., 2017. Meteorite constraints on Martian atmospheric loss and paleoclimate. Earth Planet. Sci. Lett. 479, 322–329. https://doi.org/10.1016/j.epsl.2017.09.034

Catalano, J.G., 2013. Thermodynamic and mass balance constraints on iron-bearing phyllosilicate formation and alteration pathways on early Mars. J. Geophys. Res. E Planets 118, 2124–2136. https://doi.org/10.1002/jgre.20161

Catling, D.C., Kasting, J.F., 2017. Atmospheric evolution on inhabited and lifeless worlds. Cambridge University Press. https://doi.org/10.1017/9781139020558

Chambers, J., 2001. Making More Terrestrial Planets. Icarus 152, 205–224. https://doi.org/10.1006/icar.2001.6639

Chambers, J.E., Wetherill, G.W., 1998. Making the Terrestrial Planets: N-Body Integrations of Planetary Embryos in Three Dimensions. Icarus 136, 304–327. https://doi.org/10.1006/icar.1998.6007

Chandrasekhar, S., 1961. Hydrodynamic and hydromagnetic stability. Int. Ser. Monogr. Physics, Oxford Clarendon, 1961. https://ui.adsabs.harvard.edu/abs/1961hhs..book.....C/abstract

Chassefière, E., 1996. Hydrodynamic Escape of Oxygen from Primitive Atmospheres : Icarus 124, 537–552.

Chassefière, E., Dartois, E., Herri, J.M., Tian, F., Schmidt, F., Mousis, O., Lakhlifi, A., 2013a. CO2–SO2 clathrate hydrate formation on early Mars. Icarus 223, 878–891. https://doi.org/10.1016/j.icarus.2013.01.001

Chassefière, E., E., 1996. Hydrodynamic escape of hydrogen from a hot water-rich atmosphere: The case of Venus. J. Geophys. Res. Planets 101, 26039–26056. https://doi.org/10.1029/96JE01951

Chassefière, E., Langlais, B., Quesnel, Y., Leblanc, F., 2013b. The fate of early Mars' lost water: The role of serpentinization. J. Geophys. Res. Planets 118, 1123–1134. https://doi.org/10.1002/jgre.20089

Chassefière, E., Lasue, J., Langlais, B., Quesnel, Y., 2016. Early Mars serpentinization-derived CH4 reservoirs, H2-induced warming and paleopressure evolution. Meteorit. Planet. Sci. 51, 2234–2245. https://doi.org/10.1111/maps.12784

Chassefière, E., Leblanc, F., 2004. Mars atmospheric escape and evolution; interaction with the solar wind. Planet. Space Sci. 52, 1039–1058. https://doi.org/10.1016/j.pss.2004.07.002

Chassefière, E., Leblanc, F., 2011a. Constraining methane release due to serpentinization by the observed D/H ratio on Mars. Earth Planet. Sci. Lett. 310, 262–271. https://doi.org/10.1016/j.epsl.2011.08.013

Chassefière, E., Leblanc, F., 2011b. Methane release and the carbon cycle on Mars. Planet. Space Sci. 59, 207–217. https://doi.org/10.1016/j.pss.2010.09.004

Chassefière, E., Leblanc, F., Langlais, B., 2007. The combined effects of escape and magnetic field histories at Mars. Planet. Space Sci. 55, 343–357. https://doi.org/10.1016/j.pss.2006.02.003

Chemtob, S.M., Nickerson, R.D., Morris, R. V., Agresti, D.G., Catalano, J.G., 2017. Oxidative Alteration of Ferrous Smectites and Implications for the Redox Evolution of Early Mars. J. Geophys. Res. Planets 122, 2469–2488. https://doi.org/10.1002/2017JE005331

Chevrier, V., Poulet, F., Bibring, J.P., 2007. Early geochemical environment of Mars as determined from thermodynamics of phyllosilicates. Nature 448, 60–63. https://doi.org/10.1038/nature05961

Claire, M.W., Sheets, J., Cohen, M., Ribas, I., Meadows, V.S., Catling, D.C., 2012. THE EVOLUTION OF SOLAR FLUX FROM 0.1 nm TO 160 μm: QUANTITATIVE ESTIMATES FOR PLANETARY STUDIES. Astrophys. J. 757, 95. https://doi.org/10.1088/0004-637X/757/1/95

Collinson, G., Mitchell, D., Glocer, A., Grebowsky, J., Peterson, W.K., Connerney, J., Andersson, L., Espley, J., Mazelle, C., Sauvaud, J.A., Fedorov, A., Ma, Y., Bougher, S., Lillis, R., Ergun, R., Jakosky, B., 2015. Electric Mars: The first direct measurement of an upper limit for the Martian polar wind electric potential. Geophys. Res. Lett. 42, 9128–9134. https://doi.org/10.1002/2015GL065084

Connerney, J.E.P., Acuña, M.H., Ness, N.F., Kletetschka, G., Mitchell, D.L., Lin, R.P., Reme, H., 2005. Tectonic implications of Mars crustal magnetism. Proc. Natl. Acad. Sci. U. S. A. 102, 14970–14975.

https://doi.org/10.1073/pnas.0507469102

Connerney, J.E.P., Acuna, M.H., Wasilewski, P.J., Ness, N.F., Reme, H., Mazelle, C., Vignes, D., Lin, R.P., Mitchell, D.L., Cloutier, P.A., 1999. Magnetic lineations in the ancient crust of Mars. Science (80-. ). 284, 794–798. https://doi.org/10.1126/science.284.5415.794

Craddock, R.A., Greeley, R., 1995. Estimates of the Amount and Timing of Gases Released into the Martian Atmosphere from Volcanic Eruptions. Abstr. Lunar Planet. Sci. Conf. Vol. 26, page 287, 26.

Craddock, R.A., Greeley, R., 2009. Minimum estimates of the amount and timing of gases released into the martian atmosphere from volcanic eruptions. Icarus 204, 512–526. https://doi.org/10.1016/j.icarus.2009.07.026

Cravens, T.E., Rahmati, A., Fox, J.L., Lillis, R., Bougher, S., Luhmann, J., Sakai, S., Deighan, J., Lee, Y., Combi, M., Jakosky, B., 2017. Hot oxygen escape from Mars: Simple scaling with solar EUV irradiance. J. Geophys. Res. Sp. Phys. 122, 1102–1116. https://doi.org/10.1002/2016JA023461

Curry, S.M., Liemohn, M., Fang, X., Ma, Y., Espley, J., 2013. The influence of production mechanisms on pick-up ion loss at Mars. J. Geophys. Res. Sp. Phys. 118, 554–569. https://doi.org/10.1029/2012JA017665

Curry, S.M., Luhmann, J., Ma, Y., Liemohn, M., Dong, C., Hara, T., 2015. Comparative pick-up ion distributions at Mars and Venus: Consequences for atmospheric deposition and escape. Planet. Space Sci. 115, 35–47. https://doi.org/10.1016/j.pss.2015.03.026

Dasgupta, R., Hirschmann, M.M., 2010. The deep carbon cycle and melting in Earth's interior. Earth Planet. Sci. Lett. 298, 1–13. https://doi.org/10.1016/j.epsl.2010.06.039

Dauphas, N., Pourmand, A., 2011. Hf-W-Th evidence for rapid growth of Mars and its status as a planetary embryo. Nature 473, 489–492. https://doi.org/10.1038/nature10077

De Niem, D., Kührt, E., Morbidelli, A., Motschmann, U., 2012. Atmospheric erosion and replenishment induced by impacts upon the earth and mars during a heavy bombardment. Icarus 221, 495–507.

https://doi.org/10.1016/j.icarus.2012.07.032

Debaille, V., Brandon, A.D., Yin, Q.Z., Jacobsen, B., 2007. Coupled 142Nd-143Nd evidence for a protracted magma ocean in Mars. Nature 450, 525–528. https://doi.org/10.1038/nature06317

Dehouck, E., Gaudin, A., Chevrier, V., Mangold, N., 2016. Mineralogical record of the redox conditions on early Mars. Icarus 271, 67–75. https://doi.org/10.1016/j.icarus.2016.01.030

Dehouck, E., Gaudin, A., Mangold, N., Lajaunie, L., Dauzères, A., Grauby, O., Le Menn, E., 2014. Weathering of olivine under CO2 atmosphere: A martian perspective. Geochim. Cosmochim. Acta 135, 170–189. https://doi.org/10.1016/j.gca.2014.03.032

Deng, J., Du, Z., Karki, B.B., Ghosh, D.B., Lee, K.K.M., 2020. A magma ocean origin to divergent redox evolutions of rocky planetary bodies and early atmospheres. Nat. Commun. 11, 1–7. https://doi.org/10.1038/s41467-020-15757-0

Donahue, T.M., 2004. Accretion, loss, and fractionation of martian water. Icarus 167, 225–227. https://doi.org/10.1016/j.icarus.2003.09.015

Dong, C., Huang, Z., Lingam, M., 2019. Role of Planetary Obliquity in Regulating Atmospheric Escape: G-dwarf versus M-dwarf Earth-like Exoplanets. Astrophys. J. 882, L16. https://doi.org/10.3847/2041-8213/ab372c

Dong, C., Lee, Y., Ma, Y., Lingam, M., Bougher, S., Luhmann, J., Curry, S., Toth, G., Nagy, A., Tenishev, V., Fang, X., Mitchell, D., Brain, D., Jakosky, B., 2018a. Modeling Martian Atmospheric Losses over Time: Implications for Exoplanetary Climate Evolution and Habitability. Astrophys. J. Lett. Vol. 859, Issue 1, Artic. id. L14, 5 pp. (2018). 859. https://doi.org/10.3847/2041-8213/aac489

Dong, C., Lingam, M., Airapetian, V.S., Ma, Y., van der Holst, B., 2018b. Atmospheric escape from the TRAPPIST-1 planets and implications for habitability. Proc. Natl. Acad. Sci. U. S. A. 115, 260–265. https://doi.org/10.1073/pnas.1708010115

Dong, C., Lingam, M., Ma, Y., Cohen, O., 2017. Is Proxima Centauri b Habitable ?

A Study of Atmospheric Loss. Astrophys. J. Lett. 837, L26. https://doi.org/10.3847/2041-8213/aa6438

Dong, C., Luhmann, J.G., Ma, Y., Lingam, M., Bhattacharjee, A., Lee, Y., Bougher, S.W., Wang, L., Glocer, A., Strangeway, R.J., Curry, S., Fang, X., Toth, G., Nagy, A.F., Lillis, R.J., Mitchell, D.L., Brain, D., Jakosky, B.M., 2018c. Global dipole magnetic field: Boon or bane for the Martian atmospheric retention? AGU Fall Meet. Abstr. 2018, P31C-3736.

Dong, C., Ma, Y., Bougher, S.W., Toth, G., Nagy, A.F., Halekas, J.S., Dong, Y., Curry, S.M., Luhmann, J.G., Brain, D., Connerney, J.E.P., Espley, J., Mahaffy, P., Benna, M., McFadden, J.P., Mitchell, D.L., DiBraccio, G.A., Lillis, R.J., Jakosky, B.M., Grebowsky, J.M., 2015. Multifluid MHD study of the solar wind interaction with Mars' upper atmosphere during the 2015 March 8th ICME event. Geophys. Res. Lett. 42, 9103–9112. https://doi.org/10.1002/2015GL065944

Edberg, N.J.T., Nilsson, H., Futaana, Y., Stenberg, G., Lester, M., Cowley, S.W.H., Luhmann, J.G., Mcenulty, T.R., Opgenoorth, H.J., Fedorov, A., Barabash, S., Zhang, T.L., 2011. Atmospheric erosion of Venus during stormy space weather. J. Geophys. Res. Sp. Phys. 116, 1–10. https://doi.org/10.1029/2011JA016749

Edwards, C.S., Ehlmann, B.L., 2015. Carbon sequestration on Mars. Geology 43, 863–866. https://doi.org/10.1130/G36983.1

Egan, H., Jarvinen, R., Ma, Y., Brain, D., 2019. Planetary Magnetic Field Control of Ion Escape from Weakly Magnetized Planets. Mon. Not. R. Astron. Soc. Vol. 488, Issue 2, p.2108-2120 488, 2108–2120. https://doi.org/10.1093/mnras/stz1819

Ehlmann, B.L., Anderson, F.S., Andrews-Hanna, J., Catling, D.C., Christensen, P.R., Cohen, B.A., Dressing, C.D., Edwards, C.S., Elkins-Tanton, L.T., Farley, K.A., Fassett, C.I., Fischer, W.W., Fraeman, A.A., Golombek, M.P., Hamilton, V.E., Hayes, A.G., Herd, C.D.K., Horgan, B., Hu, R., Jakosky, B.M., Johnson, J.R., Kasting, J.F., Kerber, L., Kinch, K.M., Kite, E.S., Knutson, H.A., Lunine, J.I., Mahaffy, P.R., Mangold, N., McCubbin, F.M.,

Mustard, J.F., Niles, P.B., Quantin-Nataf, C., Rice, M.S., Stack, K.M., Stevenson, D.J., Stewart, S.T., Toplis, M.J., Usui, T., Weiss, B.P., Werner, S.C., Wordsworth, R.D., Wray, J.J., Yingst, R.A., Yung, Y.L., Zahnle, K.J., 2016. The sustainability of habitability on terrestrial planets: Insights, questions, and needed measurements from Mars for understanding the evolution of Earth-like worlds. J. Geophys. Res. Planets 121, 1927–1961. https://doi.org/10.1002/2016JE005134

Ehlmann, B.L., Berger, G., Mangold, N., Michalski, J.R., Catling, D.C., Ruff, S.W., Chassefière, E., Niles, P.B., Chevrier, V., Poulet, F., 2013. Geochemical consequences of widespread clay mineral formation in Mars' ancient crust. Space Sci. Rev. 174, 329–364. https://doi.org/10.1007/s11214-012-9930-0

Ehlmann, B.L., Mustard, J.F., Murchie, S.L., Bibring, J.P., Meunier, A., Fraeman, A.A., Langevin, Y., 2011. Subsurface water and clay mineral formation during the early history of Mars. Nature 479, 53–60. https://doi.org/10.1038/nature10582

Ehlmann, B.L., Mustard, J.F., Murchie, S.L., Poulet, F., Bishop, J.L., Brown, A.J., Calvin, W.M., Clark, R.N., Des Marais, D.J., Milliken, R.E., Roach, L.H., Roush, T.L., Swayze, G.A., Wray, J.J., 2008. Orbital identification of carbonate-bearing rocks on Mars. Science (80-. ). 322, 1828–1832. https://doi.org/10.1126/science.1164759

Elkins-Tanton, L.T., 2008. Linked magma ocean solidification and atmospheric growth for Earth and Mars. Earth Planet. Sci. Lett. 271, 181–191. https://doi.org/10.1016/j.epsl.2008.03.062

Elkins-Tanton, L.T., T., L., 2008. Linked magma ocean solidification and atmospheric growth for Earth and Mars. Earth Planet. Sci. Lett. 271, 181–191. https://doi.org/10.1016/j.epsl.2008.03.062

Elkins-Tanton, L.T., T., L., 2012. Magma Oceans in the Inner Solar System. Annu. Rev. Earth Planet. Sci. 40, 113–139. https://doi.org/10.1146/annurev-earth-042711-105503

Erkaev, N. V., Lammer, H., Elkins-Tanton, L.T., Stökl, A., Odert, P., Marcq, E., Dorfi, E.A., Kislyakova, K.G., Kulikov, Y.N., Leitzinger, M., Güdel, M., 2014. Escape of the martian protoatmosphere and initial water inventory. Planet. Space Sci. 98, 106–119. https://doi.org/10.1016/j.pss.2013.09.008

Erkaev, N. V., Lammer, H., Odert, P., Kulikov, Y.N., Kislyakova, K.G., 2015. Extreme hydrodynamic atmospheric loss near the critical thermal escape regime. Mon. Not. R. Astron. Soc. 448, 1916–1921. https://doi.org/10.1093/mnras/stv130

Erkaev, N. V., Scherf, M., Thaller, S.E., Lammer, H., Mezentsev, A. V., Ivanov, V.A., Mandt, K.E., 2020. Escape and evolution of Titan's N2 atmosphere constrained by 14N/15N isotope ratios. Mon. Not. R. Astron. Soc. in print.

Erkaev, N. V, Lammer, H., Elkins-Tanton, L.T., Stökl, A., Odert, P., Marcq, E., Dorfi, E.A., Kislyakova, K.G., Kulikov, Y.N., Leitzinger, M., Güdel, M., 2014a. Escape of the martian protoatmosphere and initial water inventory. Planet. Sp. Sci. Vol. 98, p. 106-119. 98, 106–119. https://doi.org/10.1016/j.pss.2013.09.008

Erkaev, N. V, Lammer, H., Elkins-Tanton, L.T., Stökl, A., Odert, P., Marcq, E., Dorfi, E.A., Kislyakova, K.G., Kulikov, Y.N., Leitzinger, M., Güdel, M., 2014b. Escape of the martian protoatmosphere and initial water inventory. Planet. Sp. Sci. Vol. 98, p. 106-119. 98, 106–119. https://doi.org/10.1016/j.pss.2013.09.008

Erkaev, N. V, Lammer, H., Odert, P., Kulikov, Y.N., Kislyakova, K.G., 2015. Extreme hydrodynamic atmospheric loss near the critical thermal escape regime. Mon. Not. R. Astron. Soc. Vol. 448, Issue 2, p.1916-1921 448, 1916–1921. https://doi.org/10.1093/mnras/stv130

Farquhar, J., Savarino, J., Jackson, T.L., Thiemens, M.H., 2000. Evidence of atmospheric sulphur in the martian regolith from sulphur isotopes in meteorites. Nature 404, 50–52. https://doi.org/10.1038/35003517

Feldman, W.C., Mellon, M.T., Maurice, S., Prettyman, T.H., Carey, J.W., Vaniman, D.T., Bish, D.L., Fialips, C.I., Chipera, S.J., Kargel, J.S., Elphic, R.C., Funsten, H.O., Lawrence, D.J., Tokar, R.L., 2004. Hydrated states of MgSO4 at equatorial latiudes on Mars. Geophys. Res. Lett. 31. https://doi.org/10.1029/2004GL020181

Fernández-Remolar, D.C., Sánchez-Román, M., Hill, A.C., Gómez-Ortíz, D., Ballesteros, O.P., Romanek, C.S., Amils, R., 2011. The environment of early Mars and the missing carbonates. Meteorit. Planet. Sci. 46, 1447–1469. https://doi.org/10.1111/j.1945-5100.2011.01238.x

Forget, F., Wordsworth, R., Millour, E., Madeleine, J.B., Kerber, L., Leconte, J., Marcq, E., Haberle, R.M., 2013. 3D modelling of the early martian climate under a denser CO 2 atmosphere: Temperatures and CO 2 ice clouds. Icarus 222, 81–99. https://doi.org/10.1016/j.icarus.2012.10.019

Fossati, L., Erkaev, N. V, Lammer, H., Cubillos, P.E., Odert, P., Juvan, I., Kislyakova, K.G., Lendl, M., Kubyshkina, D., Bauer, S.J., 2017. Aeronomical constraints to the minimum mass and maximum radius of hot low-mass planets. Astron. & Astrophys. Vol. 598, id.A90, 9 pp. 598, A90. https://doi.org/10.1051/0004-6361/201629716

Fox, J.L., 2004. CO 2 + dissociative recombination: A source of thermal and nonthermal C on Mars. J. Geophys. Res. 109, A08306. https://doi.org/10.1029/2004JA010514

Fox, J.L., 2009. Morphology of the dayside ionosphere of Mars: Implications for ion outflows. J. Geophys. Res. 114, E12005. https://doi.org/10.1029/2009JE003432

Fox, J.L., Hać, A., 1997. The 15N/14N isotope fractionation in dissociative recombination of N2+. J. Geophys. Res. Planets 102, 9191–9204. https://doi.org/10.1029/97JE00086

Fox, J.L., Hać, A.B., 2014. The escape of O from mars: Sensitivity to the elastic cross sections. Icarus 228, 375–385. https://doi.org/10.1016/j.icarus.2013.10.014

Fränz, H.B., Trainer, M.G., Malespin, C.A., Mahaffy, P.R., Atreya, S.K., Becker, R.H., Benna, M., Conrad, P.G., Eigenbrode, J.L., Freissinet, C., Manning, H.L.K., Prats, B.D., Raaen, E., Wong, M.H., 2017. Initial SAM calibration gas experiments on Mars: Quadrupole mass spectrometer results and implications. Planet. Space Sci. 138, 44–54. https://doi.org/10.1016/j.pss.2017.01.014

Fulton, B.J., Petigura, E.A., 2018. The California- Kepler Survey. VII. Precise Planet Radii Leveraging Gaia DR2 Reveal the Stellar Mass Dependence of the Planet Radius Gap . Astron. J. 156, 264. https://doi.org/10.3847/1538-3881/aae828

Futaana, Y., Stenberg Wieser, G., Barabash, S., Luhmann, J.G., 2017. Solar Wind Interaction and Impact on the Venus Atmosphere. Space Sci. Rev. 212, 1453–1509. https://doi.org/10.1007/s11214-017-0362-8

Gaillard, F., Michalski, J., Berger, G., McLennan, S.M., Scaillet, B., 2013. Geochemical reservoirs and timing of sulfur cycling on Mars. Space Sci. Rev. 174, 251–300. https://doi.org/10.1007/s11214-012-9947-4

Gaillard, F., Scaillet, B., 2009. The sulfur content of volcanic gases on Mars. Earth Planet. Sci. Lett. 279, 34–43. https://doi.org/10.1016/j.epsl.2008.12.028

Garenne, A., Beck, P., Montes-Hernandez, G., Chiriac, R., Toche, F., Quirico, E., Bonal, L., Schmitt, B., 2014. The abundance and stability of "water" in type 1 and 2 carbonaceous chondrites (CI, CM and CR). Geochim. Cosmochim. Acta 137, 93–112. https://doi.org/10.1016/j.gca.2014.03.034

Genda, H., Abe, Y., 2003. Survival of a proto-atmosphere through the stage of giant impacts: The mechanical aspects. Icarus 164, 149–162. https://doi.org/10.1016/S0019-1035(03)00101-5

Gierasch, P.J., Toon, O.B., 1973. Atmospheric Pressure Variation and the Climate of Mars. J. Atmos. Sci. 30, 1502–1508. https://doi.org/10.1175/1520-0469(1973)030<1502:apvatc>2.0.co;2

Gillmann, C., Lognonné, P., Moreira, M., 2011. Volatiles in the atmosphere of Mars: The effects of volcanism and escape constrained by isotopic data. Earth Planet. Sci. Lett. 303, 299–309. https://doi.org/10.1016/j.epsl.2011.01.009

Girazian, Z., Mahaffy, P., Lillis, R.J., Benna, M., Elrod, M., Fowler, C.M., Mitchell, D.L., 2017. Ion Densities in the Nightside Ionosphere of Mars: Effects of Electron Impact Ionization. Geophys. Res. Lett. 44, 11,248-11,256. https://doi.org/10.1002/2017GL075431

Gladstone, G.R., Stern, S.A., Ennico, K., Olkin, C.B., Weaver, H.A., Young, L.A., Summers, M.E., Strobel, D.F., Hinson, D.P., Kammer, J.A., Parker, A.H.,

Steffl, A.J., Linscott, I.R., Parker, J.W., Cheng, A.F., Slater, D.C., Versteeg, M.H., Greathouse, T.K., Retherford, K.D., Throop, H., Cunningham, N.J., Woods, W.W., Singer, K.N., Tsang, C.C.C., Schindhelm, E., Lisse, C.M., Wong, M.L., Yung, Y.L., Zhu, X., Curdt, W., Lavvas, P., Young, E.F., Tyler, G.L., Bagenal, F., Grundy, W.M., McKinnon, W.B., Moore, J.M., Spencer, J.R., Andert, T., Andrews, J., Banks, M., Bauer, B., Bauman, J., Barnouin, O.S., Bedini, P., Beisser, K., Beyer, R.A., Bhaskaran, S., Binzel, R.P., Birath, E., Bird, M., Bogan, D.J., Bowman, A., Bray, V.J., Brozovic, M., Bryan, C., Buckley, M.R., Buie, M.W., Buratti, B.J., Bushman, S.S., Calloway, A., Carcich, B., Conard, S., Conrad, C.A., Cook, J.C., Cruikshank, D.P., Custodio, O.S., Ore, C.M.D., Deboy, C., Dischner, Z.J.B., Dumont, P., Earle, A.M., Elliott, H.A., Ercol, J., Ernst, C.M., Finley, T., Flanigan, S.H., Fountain, G., Freeze, M.J., Green, J.L., Guo, Y., Hahn, M., Hamilton, D.P., Hamilton, S.A., Hanley, J., Harch, A., Hart, H.M., Hersman, C.B., Hill, A., Hill, M.E., Holdridge, M.E., Horanyi, M., Howard, A.D., Howett, C.J.A., Jackman, C., Jacobson, R.A., Jennings, D.E., Kang, H.K., Kaufmann, D.E., Kollmann, P., Krimigis, S.M., Kusnierkiewicz, D., Lauer, T.R., Lee, J.E., Lindstrom, K.L., Lunsford, A.W., Mallder, V.A., Martin, N., McComas, D.J., McNutt, R.L., Mehoke, D., Mehoke, T., Melin, E.D., Mutchler, M., Nelson, D., Nimmo, F., Nunez, J.I., Ocampo, A., Owen, W.M., Paetzold, M., Page, B., Pelletier, F., Peterson, J., Pinkine, N., Piquette, M., Porter, S.B., Protopapa, S., Redfern, J., Reitsema, H.J., Reuter, D.C., Roberts, J.H., Robbins, S.J., Rogers, G., Rose, D., Runyon, K., Ryschkewitsch, M.G., Schenk, P., Sepan, B., Showalter, M.R., Soluri, M., Stanbridge, D., Stryk, T., Szalay, J.R., Tapley, M., Taylor, A., Taylor, H., Umurhan, O.M., Verbiscer, A.J., Versteeg, M.H., Vincent, M., Webbert, R., Weidner, S., Weigle, G.E., White, O.L., Whittenburg, K., Williams, B.G., Williams, K., Williams, S., Zangari, A.M., Zirnstein, E., 2016. The atmosphere of Pluto as observed by New Horizons. Science (80-. ). 351, aad8866–aad8866. https://doi.org/10.1126/science.aad8866

Grewal, D.S., Dasgupta, R., Farnell, A., 2020. The speciation of carbon, nitrogen,

and water in magma oceans and its effect on volatile partitioning between major reservoirs of the Solar System rocky bodies. Geochim. Cosmochim. Acta 280, 281–301. https://doi.org/10.1016/j.gca.2020.04.023

Gröller, H., Lichtenegger, H., Lammer, H., Shematovich, V.I., 2014. Hot oxygen and carbon escape from the martian atmosphere. Planet. Sp. Sci. Vol. 98, p. 93-105. 98, 93–105. https://doi.org/10.1016/j.pss.2014.01.007

Grott, M., Baratoux, D., Hauber, E., Sautter, V., Mustard, J., Gasnault, O., Ruff, S.W., Karato, S.-I., Debaille, V., Knapmeyer, M., Sohl, F., Van Hoolst, T., Breuer, D., Morschhauser, A., Toplis, M.J., 2013. Long-Term Evolution of the Martian Crust-Mantle System. Space Sci. Rev. 174, 49–111. https://doi.org/10.1007/s11214-012-9948-3

Grott, M., Morschhauser, A., Breuer, D., Hauber, E., 2011. Volcanic outgassing of CO 2 and H 2 O on Mars. Earth Planet. Sci. Lett. 308, 391–400. https://doi.org/10.1016/j.epsl.2011.06.014

Guedel, M., 2007. The Sun in Time: Activity and Environment. Living Rev. Sol. Physics, Vol. 4, Issue 1, Artic. id. 3, 137 pp. 4. https://doi.org/10.12942/lrsp-2007-3

Gunell, H., Maggiolo, R., Nilsson, H., Stenberg Wieser, G., Slapak, R., Lindkvist, J., Hamrin, M., De Keyser, J., 2018. Why an intrinsic magnetic field does not protect a planet against atmospheric escape. Astron. Astrophys. 614, L3. https://doi.org/10.1051/0004-6361/201832934

Guo, J.H., 2019. The Effect of Photoionization on the Loss of Water of the Planet. Astrophys. J. 872, 99. https://doi.org/10.3847/1538-4357/aaffd4

Gupta, A., Schlichting, H.E., 2019. Sculpting the valley in the radius distribution of small exoplanets as a by-product of planet formation: The core-powered mass-loss mechanism. Mon. Not. R. Astron. Soc. 487, 24–33. https://doi.org/10.1093/mnras/stz1230

Haberle, R.M., Tyler, D., McKay, C.P., Davis, W.L., 1994. A Model for the Evolution of CO2 on Mars. Icarus 109, 102–120. https://doi.org/10.1006/icar.1994.1079

Haberle, R.M., Zahnle, K., Barlow, N.G., Steakley, K.E., 2019. Impact Degassing

of H2 on Early Mars and its Effect on the Climate System. Geophys. Res. Lett. 46, 13355–13362. https://doi.org/10.1029/2019GL084733

Halevy, I., Zuber, M.T., Schrag, D.P., 2007. A sulfur dioxide climate feedback on early Mars. Science (80-. ). 318, 1903–1907. https://doi.org/10.1126/science.1147039

Hansen, B.M.S., 2009. Formation of the terrestrial planets from a narrow annulus. Astrophys. J. 703, 1131–1140. https://doi.org/10.1088/0004-637X/703/1/1131

Hayworth, B.P.C., Kopparapu, R.K., Haqq-Misra, J., Batalha, N.E., Payne, R.C., Foley, B.J., Ikwut-Ukwa, M., Kasting, J.F., 2020. Warming early Mars with climate cycling: The effect of CO2-H2 collision-induced absorption. Icarus 345. https://doi.org/10.1016/j.icarus.2020.113770

Hébrard, E., Marty, B., 2014. Coupled noble gas – hydrocarbon evolution of the early Earth atmosphere upon solar UV irradiation. Earth Planet. Sci. Lett. 385, 40–48. https://doi.org/10.1016/j.epsl.2013.10.022

Hirschmann, M.M., Withers, A.C., 2008. Ventilation of CO2 from a reduced mantle and consequences for the early Martian greenhouse. Earth Planet. Sci. Lett. 270, 147–155. https://doi.org/10.1016/j.epsl.2008.03.034

Hood, L.L., Harrison, K.P., Langlais, B., Lillis, R.J., Poulet, F., Williams, D.A., 2010. Magnetic anomalies near Apollinaris Patera and the Medusae Fossae Formation in Lucus Planum , Mars. Icarus 208, 118–131. https://doi.org/10.1016/j.icarus.2010.01.009

Hu, R., Kass, D.M., Ehlmann, B.L., Yung, Y.L., 2015. Tracing the fate of carbon and the atmospheric evolution of Mars. Nat. Commun. 6, 1–9. https://doi.org/10.1038/ncomms10003

Humayun, M., Nemchin, A., Zanda, B., Hewins, R.H., Grange, M., Kennedy, A., Lorand, J.P., Göpel, C., Fieni, C., Pont, S., Deldicque, D., 2013. Origin and age of the earliest Martian crust from meteorite NWA 7533. Nature 503, 513–516. https://doi.org/10.1038/nature12764

Hunten, D.M., Pepin, R.O., Walker, J.C.G., 1987. Mass fractionation in hydrodynamic escape. Icarus 69, 532–549. https://doi.org/10.1016/0019-

1035(87)90022-4

Hurowitz, J.A., Fischer, W.W., Tosca, N.J., Milliken, R.E., 2010. Origin of acidic surface waters and the evolutionof atmospheric chemistry on early Mars. Nat. Geosci. 3, 323–326. https://doi.org/10.1038/ngeo831

Hutchins, K.S., Jakosky, B.M., Luhmann, J.G., 1997. Impact of a paleomagnetic field on sputtering loss of Martian atmospheric argon and neon. J. Geophys. Res. E Planets 102, 9183–9189. https://doi.org/10.1029/96JE03838

Ito, Y., Hashimoto, G.L., Takahashi, Y.O., Ishiwatari, M., Kuramoto, K., 2020. H 2 O 2 -induced Greenhouse Warming on Oxidized Early Mars . Astrophys. J. 893, 168. https://doi.org/10.3847/1538-4357/ab7db4

Jakosky, B.M., 2019. The CO2 inventory on Mars. Planet. Space Sci. 175, 52–59. https://doi.org/10.1016/j.pss.2019.06.002

Jakosky, B.M., Brain, D., Chaffin, M., Curry, S., Deighan, J., Grebowsky, J., Halekas, J., Leblanc, F., Lillis, R., Luhmann, J.G., Andersson, L., Andre, N., Andrews, D., Baird, D., Baker, D., Bell, J., Benna, M., Bhattacharyya, D., Bougher, S., Bowers, C., Chamberlin, P., Chaufray, J.-Y., Clarke, J., Collinson, G., Combi, M., Connerney, J., Connour, K., Correira, J., Crabb, K., Crary, F., Cravens, T., Crismani, M., Delory, G., Dewey, R., DiBraccio, G., Dong, C., Dong, Y., Dunn, P., Egan, H., Elrod, M., England, S., Eparvier, F., Ergun, R., Eriksson, A., Esman, T., Espley, J., Evans, S., Fallows, K., Fang, X., Fillingim, M., Flynn, C., Fogle, A., Fowler, C., Fox, J., Fujimoto, M., Garnier, P., Girazian, Z., Groeller, H., Gruesbeck, J., Hamil, O., Hanley, K.G., Hara, T., Harada, Y., Hermann, J., Holmberg, M., Holsclaw, G., Houston, S., Inui, S., Jain, S., Jolitz, R., Kotova, A., Kuroda, T., Larson, D., Lee, Y., Lee, C., Lefevre, F., Lentz, C., Lo, D., Lugo, R., Ma, Y.-J., Mahaffy, P., Marquette, M.L., Matsumoto, Y., Mayyasi, M., Mazelle, C., McClintock, W., McFadden, J., Medvedev, A., Mendillo, M., Meziane, K., Milby, Z., Mitchell, D., Modolo, R., Montmessin, F., Nagy, A., Nakagawa, H., Narvaez, C., Olsen, K., Pawlowski, D., Peterson, W., Rahmati, A., Roeten, K., Romanelli, N., Ruhunusiri, S., Russell, C., Sakai, S., Schneider, N., Seki, K., Sharrar, R., Shaver, S., Siskind, D.E., Slipski, M., Soobiah, Y., Steckiewicz,

M., Stevens, M.H., Stewart, I., Stiepen, A., Stone, S., Tenishev, V., Terada, N., Terada, K., Thiemann, E., Tolson, R., Toth, G., Trovato, J., Vogt, M., Weber, T., Withers, P., Xu, S., Yelle, R., Yiğit, E., Zurek, R., 2018. Loss of the Martian atmosphere to space: Present-day loss rates determined from MAVEN observations and integrated loss through time. Icarus 315, 146–157. https://doi.org/10.1016/j.icarus.2018.05.030

Jakosky, Bruce M., Grebowsky, J.M., Luhmann, J.G., Brain, D.A., 2015. Initial results from the MAVEN mission to Mars. Geophys. Res. Lett. 42, 8791–8802. https://doi.org/10.1002/2015GL065271

Jakosky, B. M., Grebowsky, J.M., Luhmann, J.G., Connerney, J., Eparvier, F., Ergun, R., Halekas, J., Larson, D., Mahaffy, P., McFadden, J., Mitchell, D.F., Schneider, N., Zurek, R., Bougher, S., Brain, D., Ma, Y.J., Mazelle, C., Andersson, L., Andrews, D., Baird, D., Baker, D., Bell, J.M., Benna, M., Chaffin, M., Chamberlin, P., Chaufray, Y.Y., Clarke, J., Collinson, G., Combi, M., Crary, F., Cravens, T., Crismani, M., Curry, S., Curtis, D., Deighan, J., Delory, G., Dewey, R., DiBraccio, G., Dong, C., Dong, Y., Dunn, P., Elrod, M., England, S., Eriksson, A., Espley, J., Evans, S., Fang, X., Fillingim, M., Fortier, K., Fowler, C.M., Fox, J., Gröller, H., Guzewich, S., Hara, T., Harada, Y., Holsclaw, G., Jain, S.K., Jolitz, R., Leblanc, F., Lee, C.O., Lee, Y., Lefevre, F., Lillis, R., Livi, R., Lo, D., Mayyasi, M., McClintock, W., McEnulty, T., Modolo, R., Montmessin, F., Morooka, M., Nagy, A., Olsen, K., Peterson, W., Rahmati, A., Ruhunusiri, S., Russell, C.T., Sakai, S., Sauvaud, J.A., Seki, K., Steckiewicz, M., Stevens, M., Stewart, A.I.F., Stiepen, A., Stone, S., Tenishev, V., Thiemann, E., Tolson, R., Toublanc, D., Vogt, M., Weber, T., Withers, P., Woods, T., Yelle, R., 2015. MAVEN observations of the response of Mars to an interplanetary coronal mass ejection. Science (80-. ). 350, 0210. https://doi.org/10.1126/science.aad0210

Jakosky, B.M., Pepin, R.O., Johnson, R.E., Fox, J.L., 1994. Mars Atmospheric Loss and Isotopic Fractionation by Solar-Wind-Induced Sputtering and Photochemical Escape. Icarus 111, 271–288.

https://doi.org/10.1006/icar.1994.1145

Jakosky, B.M., Slipski, M., Benna, M., Mahaffy, P., Elrod, M., Yelle, R., Stone, S., Alsaeed, N., 2017a. Mars' atmospheric history derived from upper-atmosphere measurements of $^{38}$Ar/ $^{36}$Ar. Science (80-. ). 355, 1408–1410. https://doi.org/10.1126/science.aai7721

Jakosky, B.M., Slipski, M., Benna, M., Mahaffy, P., Elrod, M., Yelle, R., Stone, S., Alsaeed, N., 2017b. Mars' atmospheric history derived from upper-atmosphere measurements of 38Ar/36Ar. Science (80-. ). 355. https://doi.org/10.1126/science.aai7721

Jakosky, B.M.M., Brain, D., Chaffin, M., Curry, S., Deighan, J., Grebowsky, J., Halekas, J., Leblanc, F., Lillis, R., Luhmann, J.G.G., Andersson, L., Andre, N., Andrews, D., Baird, D., Baker, D., Bell, J., Benna, M., Bhattacharyya, D., Bougher, S., Bowers, C., Chamberlin, P., Chaufray, J.-Y.Y., Clarke, J., Collinson, G., Combi, M., Connerney, J., Connour, K., Correira, J., Crabb, K., Crary, F., Cravens, T., Crismani, M., Delory, G., Dewey, R., DiBraccio, G., Dong, C., Dong, Y., Dunn, P., Egan, H., Elrod, M., England, S., Eparvier, F., Ergun, R., Eriksson, A., Esman, T., Espley, J., Evans, S., Fallows, K., Fang, X., Fillingim, M., Flynn, C., Fogle, A., Fowler, C., Fox, J., Fujimoto, M., Garnier, P., Girazian, Z., Groeller, H., Gruesbeck, J., Hamil, O., Hanley, K.G.G., Hara, T., Harada, Y., Hermann, J., Holmberg, M., Holsclaw, G., Houston, S., Inui, S., Jain, S., Jolitz, R., Kotova, A., Kuroda, T., Larson, D., Lee, Y., Lee, C., Lefevre, F., Lentz, C., Lo, D., Lugo, R., Ma, Y.-J.J., Mahaffy, P., Marquette, M.L.L., Matsumoto, Y., Mayyasi, M., Mazelle, C., McClintock, W., McFadden, J., Medvedev, A., Mendillo, M., Meziane, K., Milby, Z., Mitchell, D., Modolo, R., Montmessin, F., Nagy, A., Nakagawa, H., Narvaez, C., Olsen, K., Pawlowski, D., Peterson, W., Rahmati, A., Roeten, K., Romanelli, N., Ruhunusiri, S., Russell, C., Sakai, S., Schneider, N., Seki, K., Sharrar, R., Shaver, S., Siskind, D.E.E., Slipski, M., Soobiah, Y., Steckiewicz, M., Stevens, M.H.H., Stewart, I., Stiepen, A., Stone, S., Tenishev, V., Terada, N., Terada, K., Thiemann, E., Tolson, R., Toth, G., Trovato, J., Vogt, M., Weber, T., Withers, P., Xu, S., Yelle, R., Yiğit, E.,

Zurek, R., 2018. Loss of the Martian atmosphere to space: Present-day loss rates determined from MAVEN observations and integrated loss through time. Icarus 315, 146–157. https://doi.org/10.1016/j.icarus.2018.05.030

Jambon, A., Zimmermann, J.L., 1990. Water in oceanic basalts: evidence for dehydration of recycled crust. Earth Planet. Sci. Lett. 101, 323–331. https://doi.org/10.1016/0012-821X(90)90163-R

Johnson, S.S., Pavlov, A.A., Mischna, M.A., 2009. Fate of SO2 in the ancient Martian atmosphere: Implications for transient greenhouse warming. J. Geophys. Res. E Planets 114, 1–16. https://doi.org/10.1029/2008JE003313

Johnstone, C.P., Güdel, M., Brott, I., Lüftinger, T., 2015a. Stellar Winds on the Main-Sequence II: the Evolution of Rotation and Winds. Astron. Astrophys. Vol. 577, id.A28, 25 pp. 577. https://doi.org/10.1051/0004-6361/201425301

Johnstone, C.P., Güdel, M., Lammer, H., Kislyakova, K.G., 2018. The Upper Atmospheres of Terrestrial Planets: Carbon Dioxide Cooling and the Earth's Thermospheric Evolution. Astron. Astrophys. Vol. 617, id.A107, 36 pp. 617. https://doi.org/10.1051/0004-6361/201832776

Johnstone, C.P., Güdel, M., Lüftinger, T., Toth, G., Brott, I., 2015b. Stellar Winds on the Main-Sequence I: Wind Model. Astron. Astrophys. Vol. 577, id.A27, 22 pp. 577. https://doi.org/10.1051/0004-6361/201425300

Johnstone, C P, Khodachenko, M.L., Lüftinger, T., Kislyakova, K.G., Lammer, H., Güdel, M., 2019. Extreme hydrodynamic losses of Earth-like atmospheres in the habitable zones of very active stars. Astron. & Astrophys. Vol. 624, id.L10, 5 pp. 624, L10. https://doi.org/10.1051/0004-6361/201935279

Johnstone, C. P., Khodachenko, M.L., Lüftinger, T., Kislyakova, K.G., Lammer, H., Güdel, M., 2019. Extreme hydrodynamic losses of Earth-like atmospheres in the habitable zones of very active stars. Astron. Astrophys. Vol. 624, id.L10, 5 pp. 624. https://doi.org/10.1051/0004-6361/201935279

Johnstone, C.P., Lammer, H., Kislyakova, K.G., Scherf, M., Güdel, M., 2020. High atmospheric carbon dioxide levels and low solar activity during the Earth's Archean. Nature, Submitt.

Kadoya, S., Catling, D.C., Nicklas, R.W., Puchtel, I.S., Anbar, A.D., 2020. Mantle

data imply a decline of oxidizable volcanic gases could have triggered the Great Oxidation. Nat. Commun. 11, 1–9. https://doi.org/10.1038/s41467-020-16493-1

Kahn, R., 1985. The evolution of CO2 on Mars. Icarus 62, 175–190. https://doi.org/10.1016/0019-1035(85)90116-2

Kahre, M.A., Vines, S.K., Haberle, R.M., Hollingsworth, J.L., 2013. The early Martian atmosphere: Investigating the role of the dust cycle in the possible maintenance of two stable climate states. J. Geophys. Res. E Planets 118, 1388–1396. https://doi.org/10.1002/jgre.20099

Kallio, E., Fedorov, A., Budnik, E., Säles, T., Janhunen, P., Schmidt, W., Koskinen, H., Riihelä, P., Barabash, S., Lundin, R., Holmström, M., Gunell, H., Brinkfeldt, K., Futaana, Y., Andersson, H., Yamauchi, M., Grigoriev, A., Sauvaud, J.-A., Thocaven, J.-J., Winningham, J.D., Frahm, R.A., Sharber, J.R., Scherrer, J.R., Coates, A.J., Linder, D.R., Kataria, D.O., Kozyra, J., Luhmann, J.G., Roelof, E., Williams, D., Livi, S., Curtis, C.C., Hsieh, K.C., Sandel, B.R., Grande, M., Carter, M., McKenna-Lawler, S., Orsini, S., Cerulli-Irelli, R., Maggi, M., Wurz, P., Bochsler, P., Krupp, N., Woch, J., Fränz, M., Asamura, K., Dierker, C., 2006. Ion escape at Mars: Comparison of a 3-D hybrid simulation with Mars Express IMA/ASPERA-3 measurements. Icarus 182, 350–359. https://doi.org/10.1016/j.icarus.2005.09.018

Kallio, E., Janhunen, P., 2002. Ion escape from Mars in a quasi-neutral hybrid model. J. Geophys. Res. 107, 1035. https://doi.org/10.1029/2001JA000090

Kass, D.M., Yung, Y.L., 1995. Loss of atmosphere from Mars due to solar wind-induced sputtering. Science (80-. ). 268, 697–699. https://doi.org/10.1126/science.7732377

Kass, D.M., Yung, Y.L., 1996. Effects of Induced Sputtering on delta (13) C and AR in the Martian Atmosphere. AAS/Division Planet. Sci. Meet. Abstr. #28 03.04.

Kasting, J.F., Eggler, D.H., Raeburn, S.P., 1993. Mantle redox evolution and the oxidation state of the Archean atmosphere. J. Geol. 101, 245–257.

https://doi.org/10.1086/648219

Kasting, J.F., Pollack, J.B., 1983. Loss of water from Venus. I. Hydrodynamic escape of hydrogen. Icarus 53, 479–508. https://doi.org/10.1016/0019-1035(83)90212-9

Kay, C., Airapetian, V.S., Lüftinger, T., Kochukhov, O., 2019. Frequency of Coronal Mass Ejection Impacts with Early Terrestrial Planets and Exoplanets around Active Solar-like Stars. Astrophys. J. 886, L37. https://doi.org/10.3847/2041-8213/ab551f

Khodachenko, M.L., Ribas, I., Lammer, H., Grietßmeier, J.M., Leitner, M., Selsis, F., Eiroa, C., Hanslmeier, A., Biernat, H.K., Farrugia, C.J., Rucker, H.O., 2007. Coronal Mass Ejection (CME) activity of low mass M stars as an important factor for the habitability of terrestrial exoplanets. I. CME impact on expected magnetospheres of Earth-like exoplanets in close-in habitable zones. Astrobiology 7, 167–184. https://doi.org/10.1089/ast.2006.0127

King, P.L., McSween, J.Y., 2005. Effects of H2O, pH, and oxidation state on the stability of Fe minerals on Mars. J. Geophys. Res. E Planets. https://doi.org/10.1029/2005JE002482

Kislyakova, K.G., Johnstone, C.P., Scherf, M., Lammer, H., Holmström, M., Alexeev, I., Khodachenko, M.L., Güdel, M., 2020. Evolution of the Earth ' s polar wind escape from mid-Archean to present. J. Geophys. Res. Sp. Phys. submitted, 1–17.

Kite, E.S., 2019. Geologic Constraints on Early Mars Climate, Space Science Reviews. Springer Nature B.V. https://doi.org/10.1007/s11214-018-0575-5

Kite, E.S., Williams, J., Lucas, A., Aharonson, O., 2014. Low palaeopressure of the martian atmosphere estimated from the size distribution of ancient craters 7, 335–339. https://doi.org/10.1038/NGEO2137

Kokubo, E., Ida, S., 1998. Oligarchic Growth of Protoplanets. Icarus 131, 171–178. https://doi.org/10.1006/icar.1997.5840

Koskinen, T.T., Yelle, R. V., Lavvas, P., Lewis, N.K., 2010. Characterizing the thermosphere of HD209458b with UV transit observations. Astrophys. Journal, Vol. 723, Issue 1, pp. 116-128 (2010). 723, 116–128.

https://doi.org/10.1088/0004-637X/723/1/116

Krasnopolsky, V.A., Feldman, P.D., 2001. Detection of Molecular Hydrogen in the Atmosphere of Mars. Science (80-. ). 294, 1914–1917. https://doi.org/10.1126/science.1065569

Kubyshkina, D., Vidotto, A.A., Fossati, L., Farrell, E., 2020. Coupling thermal evolution of planets and hydrodynamic atmospheric escape in MESA. Mon. Not. R. Astron. Soc. https://doi.org/10.1093/mnras/staa2815

Kulikov, Y.N., Lammer, H., Lichtenegger, H.I.M., Penz, T., Breuer, D., Spohn, T., Lundin, R., Biernat, H.K., 2007. A Comparative Study of the Influence of the Active Young Sun on the Early Atmospheres of Earth, Venus, and Mars. Space Sci. Rev. 129, 207–243. https://doi.org/10.1007/s11214-007-9192-4

Kulikov, Y.N., Lammer, H., Lichtenegger, H.I.M., Terada, N., Ribas, I., Kolb, C., Langmayr, D., Lundin, R., Guinan, E.F., Barabash, S., Biernat, H.K., 2006. Atmospheric and water loss from early Venus. Planet. Space Sci. 54, 1425–1444. https://doi.org/10.1016/j.pss.2006.04.021

Kurahashi-Nakamura, T., Tajika, E., 2006. Atmospheric collapse and transport of carbon dioxide into the subsurface on early Mars. Geophys. Res. Lett. 33, 1–5. https://doi.org/10.1029/2006GL027170

Kurokawa, H., Kurosawa, K., Usui, T., 2018. A lower limit of atmospheric pressure on early Mars inferred from nitrogen and argon isotopic compositions. Icarus 299, 443–459. https://doi.org/10.1016/j.icarus.2017.08.020

Kurokawa, H., Sato, M., Ushioda, M., Matsuyama, T., Moriwaki, R., Dohm, J.M., Usui, T., 2014. Evolution of water reservoirs on Mars: Constraints from hydrogen isotopes in martian meteorites. Earth Planet. Sci. Lett. 394, 179–185. https://doi.org/10.1016/j.epsl.2014.03.027

Lammer, H., Bauer, S.J., 1991. Nonthermal atmospheric escape from Mars and Titan. J. Geophys. Res. Sp. Phys. 96, 1819–1825. https://doi.org/10.1029/90JA01676

Lammer, H., Chassefière, E., Karatekin, Ö., Morschhauser, A., Niles, P.B., Mousis, O., Odert, P., Möstl, U. V., Breuer, D., Dehant, V.V., Grott, M., Gröller, H., Hauber, E., Pham, L.B.S.L.B.S., Chassefi??re, E., Karatekin, ??zg??r,

Morschhauser, A., Niles, P.B., Mousis, O., Odert, P., M??stl, U. V., Breuer, D., Dehant, V.V., Grott, M., Gr??ller, H., Hauber, E., Pham, L.B.S.L.B.S., Chassefière, E., Karatekin, Ö., Morschhauser, A., Niles, P.B., Mousis, O., Odert, P., Möstl, U. V., Breuer, D., Dehant, V.V., Grott, M., Gröller, H., Hauber, E., Pham, L.B.S.L.B.S., 2013. Outgassing history and escape of the martian atmosphere and water inventory. Space Sci. Rev. 174, 113–154. https://doi.org/10.1007/s11214-012-9943-8

Lammer, H., Güdel, M., Kulikov, Y., Ribas, I., Zaqarashvili, T. V., Khodachenko, M.L., Kislyakova, K.G., Gröller, H., Odert, P., Leitzinger, M., Fichtinger, B., Krauss, S., Hausleitner, W., Holmström, M., Sanz-Forcada, J., Lichtenegger, H.I.M., Hanslmeier, A., Shematovich, V.I., Bisikalo, D., Rauer, H., Fridlund, M., 2012. Variability of solar/stellar activity and magnetic field and its influence on planetary atmosphere evolution, in: Earth, Planets and Space. Springer Berlin, pp. 179–199. https://doi.org/10.5047/eps.2011.04.002

Lammer, H., Kasting, J.F., Chassefi??re, E., Johnson, R.E., Kulikov, Y.N., Tian, F., 2008. Atmospheric escape and evolution of terrestrial planets and satellites. Space Sci. Rev. 139, 399–436. https://doi.org/10.1007/s11214-008-9413-5

Lammer, H., Kolb, C., Penz, T., Amerstorfer, U.V., Biernat, H.K., Bodiselitsch, B., 2003a. Estimation of the past and present Martian water-ice reservoirs by isotopic constraints on exchange between the atmosphere and the surface. Int. J. Astrobiol. 2, 195–202. https://doi.org/10.1017/S1473550403001605

Lammer, H., Leitzinger, M., Scherf, M., Odert, P., Burger, C., Kubyshkina, D., Johnstone, C., Maindl, T., Schäfer, C.M., Güdel, M., Tosi, N., Nikolaou, A., Marcq, E., Erkaev, N.V., Noack, L., Kislyakova, K.G., Fossati, L., Pilat-Lohinger, E., Ragossnig, F., Dorfi, E.A., 2020. Constraining the early evolution of Venus and Earth through atmospheric Ar, Ne isotope and bulk K/U ratios. Icarus 339. https://doi.org/10.1016/j.icarus.2019.113551

Lammer, H., Lichtenegger, H., Kolb, C., Ribas, I., Guinan, E.F.F., Abart, R., Bauer, S.J.J., 2003b. Loss of water from Mars: Implications for the oxidation of the soil. Icarus 165, 9–25. https://doi.org/10.1016/S0019-1035(03)00170-2

Lammer, H., Lichtenegger, H.I.M., Biernat, H.K., Erkaev, N. V., Arshukova, I.L.,

Kolb, C., Gunell, H., Lukyanov, A., Holmstrom, M., Barabash, S., Zhang, T.L., Baumjohann, W., 2006. Loss of hydrogen and oxygen from the upper atmosphere of Venus. Planet. Space Sci. 54, 1445–1456. https://doi.org/10.1016/j.pss.2006.04.022

Lammer, H., Lichtenegger, H.I.M., Kolb, C., Ribas, I., Guinan, E.F., Abart, R., Bauer, S.J., 2003c. Loss of water from Mars: Icarus 165, 9–25. https://doi.org/10.1016/S0019-1035(03)00170-2

Lammer, H., Stumptner, W., Bauer, S.J., 1996. Loss of H and O from Mars: Implications for the planetary water inventory. Geophys. Res. Lett. 23, 3353–3356. https://doi.org/10.1029/96GL03153

Langlais, B., Purucker, M., 2007. A polar magnetic paleopole associated with Apollinaris Patera , Mars 55, 270–279. https://doi.org/10.1016/j.pss.2006.03.008

Lapen, T.J., Righter, M., Brandon, A.D., Debaille, V., Beard, B.L., Shafer, J.T., Peslier, A.H., 2010. A younger age for ALH84001 and Its geochemical link to shergottite sources in mars. Science (80-. ). 328, 347–351. https://doi.org/10.1126/science.1185395

Laskar, J., Correia, A.C.M., Gastineau, M., Joutel, F., Levrard, B., Robutel, P., 2004. Long term evolution and chaotic diffusion of the insolation quantities of Mars 170, 343–364. https://doi.org/10.1016/j.icarus.2004.04.005

Leblanc, F., Johnson, R.E., 2001. Sputtering of the Martian atmosphere by solar wind pick-up ions. Planet. Space Sci. 49, 645–656. https://doi.org/10.1016/S0032-0633(01)00003-4

Leblanc, F., Johnson, R.E., 2002. Role of molecular species in pickup ion sputtering of the Martian atmosphere. J. Geophys. Res. 107, 5010. https://doi.org/10.1029/2000JE001473

Leblanc, F., Martinez, A., Chaufray, J.Y., Modolo, R., Hara, T., Luhmann, J., Lillis, R., Curry, S., McFadden, J., Halekas, J., Jakosky, B., 2018. On Mars's Atmospheric Sputtering After MAVEN's First Martian Year of Measurements. Geophys. Res. Lett. 45, 4685–4691. https://doi.org/10.1002/2018GL077199

Lebrun, T., Massol, H., Chassefière, E., Davaille, A., Marcq, E., Sarda, P., Leblanc, F., Brandeis, G., 2013. Thermal evolution of an early magma ocean in interaction with the atmosphere. J. Geophys. Res. Planets 118, 1155–1176. https://doi.org/10.1002/jgre.20068

Lee, Y., Combi, M.R., Tenishev, V., Bougher, S.W., Deighan, J., Schneider, N.M., McClintock, W.E., Jakosky, B.M., 2015. A comparison of 3-D model predictions of Mars' oxygen corona with early MAVEN IUVS observations. Geophys. Res. Lett. 42, 9015–9022. https://doi.org/10.1002/2015GL065291

Leighton, R.B., Murray, B.C., 1966. Behavior of carbon dioxide and other volatiles on Mars. Science (80-. ). 153, 136–144. https://doi.org/10.1126/science.153.3732.136

Lichtenegger, H.I.M., Dyadechkin, S., Scherf, M., Lammer, H., Adam, R., 2020. Non-Thermal Escape of the Martian CO 2 atmosphere over time : Constrained by Ar isotopes. Icarus submitted.

Liemohn, M.W., Curry, S.M., Fang, X., Ma, Y., 2013. Comparison of high-altitude production and ionospheric outflow contributions to O $^{+}$ loss at Mars. J. Geophys. Res. Sp. Phys. 118, 4093–4107. https://doi.org/10.1002/jgra.50388

Lillis, R.J., Brain, D.A., 2013. Nightside electron precipitation at Mars: Geographic variability and dependence on solar wind conditions. J. Geophys. Res. Sp. Phys. 118, 3546–3556. https://doi.org/10.1002/jgra.50171

Lillis, R.J., Deighan, J., Fox, J.L., Bougher, S.W., Lee, Y., Combi, M.R., Cravens, T.E., Rahmati, A., Mahaffy, P.R., Benna, M., Elrod, M.K., McFadden, J.P., Ergun, R.E., Andersson, L., Fowler, C.M., Jakosky, B.M., Thiemann, E., Eparvier, F., Halekas, J.S., Leblanc, F., Chaufray, J., 2017. Photochemical escape of oxygen from Mars: First results from MAVEN in situ data. J. Geophys. Res. Sp. Phys. 122, 3815–3836. https://doi.org/10.1002/2016JA023525

Lillis, R.J., Robbins, S., Manga, M., Halekas, J.S., Frey, H. V., 2013. Time history of the Martian dynamo from crater magnetic field analysis. J. Geophys. Res. E Planets 118, 1488–1511. https://doi.org/10.1002/jgre.20105

Luhmann, J.G., 1997. Correction to "The ancient oxygen exosphere of Mars:

Implications for atmosphere evolution" by Zhang et al. J. Geophys. Res. Planets 102, 1637–1637. https://doi.org/10.1029/96JE03440

Luhmann, J.G., Johnson, R.E., Zhang, M.H.G., 1992. Evolutionary impact of sputtering of the Martian atmosphere by $O^+$ pickup ions. Geophys. Res. Lett. 19, 2151–2154. https://doi.org/10.1029/92GL02485

Luhmann, J.G., Kozyra, J.U., 1991. Dayside pickup oxygen ion precipitation at Venus and Mars: Spatial distributions, energy deposition and consequences. J. Geophys. Res. 96, 5457. https://doi.org/10.1029/90JA01753

Luhmann, J.G., Russell, C.T., Scarf, F.L., Brace, L.H., Knudsen, W.C., 1987. Characteristics of the Marslike limit of the Venus-solar wind interaction. J. Geophys. Res. 92, 8545. https://doi.org/10.1029/ja092ia08p08545

Lundin, R., Barabash, S., Dubinin, E., Winningham, D., Yamauchi, M., 2011a. Low-altitude acceleration of ionospheric ions at Mars. Geophys. Res. Lett. 38, n/a-n/a. https://doi.org/10.1029/2011GL047064

Lundin, R., Barabash, S., Fedorov, A., Holmström, M., Nilsson, H., Sauvaud, J.A., Yamuchi, M., 2008. Solar forcing and planetary ion escape from Mars. Geophys. Res. Lett. 35, 1–5. https://doi.org/10.1029/2007GL032884

Lundin, R., Barabash, S., Yamauchi, M., Nilsson, H., Brain, D., 2011b. On the relation between plasma escape and the Martian crustal magnetic field. Geophys. Res. Lett. 38, n/a-n/a. https://doi.org/10.1029/2010GL046019

Lundin, R., Lammer, H., Ribas, I., 2007. Planetary Magnetic Fields and Solar Forcing: Implications for Atmospheric Evolution. Space Sci. Rev. 129, 245–278. https://doi.org/10.1007/s11214-007-9176-4

Lundin, R., Zakharov, A., Pellinen, R., Borg, H., Hultqvist, B., Pissarenko, N., Dubinin, E.M., Barabash, S.W., Liede, I., Koskinen, H., 1989. First measurements of the ionospheric plasma escape from Mars. Nature 341, 609–612. https://doi.org/10.1038/341609a0

Lunine, J.I., Chambers, J., Morbidelli, A., Leshin, L.A., 2003. The origin of water on Mars. Icarus 165, 1–8. https://doi.org/10.1016/S0019-1035(03)00172-6

Ma, Y., Nagy, A.F., Sokolov, I. V., Hansen, K.C., 2004. Three-dimensional, multispecies, high spatial resolution MHD studies of the solar wind

interaction with Mars. J. Geophys. Res. 109, A07211. https://doi.org/10.1029/2003JA010367

Ma, Y.J., Russell, C.T., Fang, X., Dong, C.F., Nagy, A.F., Toth, G., Halekas, J.S., Connerney, J.E.P., Espley, J.R., Mahaffy, P.R., Benna, M., McFadden, J., Mitchell, D.L., Andersson, L., Jakosky, B.M., 2017. Variations of the Martian plasma environment during the ICME passage on 8 March 2015: A time-dependent MHD study. J. Geophys. Res. Sp. Phys. 122, 1714–1730. https://doi.org/10.1002/2016JA023402

Mahaffy, P.R., Webster, C.R., Atreya, S.K., Franz, H., Wong, M., Conrad, P.G., Harpold, D., Jones, J.J., Leshin, L.A., Manning, H., Owen, T., Pepin, R.O., Squyres, S., Trainer, M., 2013. Abundance and isotopic composition of gases in the martian atmosphere from the Curiosity rover. Science (80-. ). 341, 263–266. https://doi.org/10.1126/science.1237966

Maindl, T.I., Dvorak, R., Lammer, H., Güdel, M., Schäfer, C., Speith, R., Odert, P., Erkaev, N. V, Kislyakova, K.G., Pilat-Lohinger, E., 2015. Impact induced surface heating by planetesimals on early Mars. Astron. & Astrophys. Vol. 574, id.A22, 7 pp. 574, A22. https://doi.org/10.1051/0004-6361/201424256

Mangold, N., Baratoux, D., Witasse, O., Sotin, T.E.C., 2016. Mars : a small terrestrial planet. Astron. Astrophys. Rev. 24, 1–107. https://doi.org/10.1007/s00159-016-0099-5

Manning, C. V., McKay, C.P., Zahnle, K.J., 2006. Thick and thin models of the evolution of carbon dioxide on Mars. Icarus 180, 38–59. https://doi.org/10.1016/j.icarus.2005.08.014

Massol, H., Hamano, K., Tian, F., Ikoma, M., Abe, Y., Chassefière, E., Davaille, A., Genda, H., Güdel, M., Hori, Y., Leblanc, F., Marcq, E., Sarda, P., Shematovich, V.I., Stökl, A., Lammer, H., 2016. Formation and Evolution of Protoatmospheres. Sp. Sci. Rev. Vol. 205, Issue 1-4, pp. 153-211 205, 153–211. https://doi.org/10.1007/s11214-016-0280-1

Masunaga, K., Seki, K., Brain, D.A., Fang, X., Dong, Y., Jakosky, B.M., McFadden, J.P., Halekas, J.S., Connerney, J.E.P., Mitchell, D.L., Eparvier,

F.G., 2017. Statistical analysis of the reflection of incident O+ pickup ions at Mars: MAVEN observations. J. Geophys. Res. Sp. Phys. 122, 4089–4101. https://doi.org/10.1002/2016JA023516

Mathew, K.J., Marti, K., 2001. Early evolution of martian volatiles: Nitrogen and noble gas components in ALH84001 and Chassigny. J. Geophys. Res. E Planets 106, 1401–1422. https://doi.org/10.1029/2000JE001255

Mccubbin, F.M., Boyce, J.W., Srinivasan, P., Santos, A.R., Elardo, S.M., Filiberto, J., Steele, A., Shearer, C.K., 2016. Heterogeneous distribution of H 2 O in the Martian interior : Implications for the abundance of H 2 O in depleted and enriched mantle sources 2060, 2036–2060. https://doi.org/10.1111/maps.12639

Mccubbin, F.M., Hauri, E.H., Elardo, S.M., Kaaden, K.E. Vander, Wang, J., Shearer, C.K., Mccubbin, F.M., Hauri, E.H., Elardo, S.M., Kaaden, K.E. Vander, Wang, J., Shearer, C.K., 2012. Hydrous melting of the martian mantle produced both depleted and enriched shergottites. https://doi.org/10.1130/G33242.1

Melosh, H.J., Vickery, A.M., 1989. Impact erosion of the primordial atmosphere of Mars. Nature 338, 487–489. https://doi.org/10.1038/338487a0

Michalski, J.R., Niles, P.B., 2010. Deep crustal carbonate rocks exposed by meteor impact on Mars. Nat. Geosci. 3, 751–755. https://doi.org/10.1038/ngeo971

Milbury, C., Schubert, G., Raymond, C.A., Smrekar, S.E., Langlais, B., 2012. The history of Mars ' dynamo as revealed by modeling magnetic anomalies near Tyrrhenus Mons and Syrtis Major 117, 1–18. https://doi.org/10.1029/2012JE004099

Mittelholz, A., Johnson, C.L., Feinberg, J.M., Langlais, B., Phillips, R.J., 2020. Timing of the martian dynamo: New constraints for a core field 4.5 and 3.7 Ga ago. Sci. Adv. 6, 1–8. https://doi.org/10.1126/sciadv.aba0513

Mojzsis, S.J., Brasser, R., Kelly, N.M., Abramov, O., Werner, S.C., 2019. Onset of Giant Planet Migration before 4480 Million Years Ago. Astrophys. J. 881, 44. https://doi.org/10.3847/1538-4357/ab2c03

Moore, T.E., Horwitz, J.L., 2007. Stellar ablation of planetary atmospheres. Rev.

Geophys. 45, 1–34. https://doi.org/10.1029/2005RG000194

Morbidelli, A., Lunine, J.I., O'Brien, D.P., Raymond, S.N., Walsh, K.J., 2012. Building Terrestrial Planets. Annu. Rev. Earth Planet. Sci. 40, 251–275. https://doi.org/10.1146/annurev-earth-042711-105319

Morbidelli, A., Nesvorny, D., Laurenz, V., Marchi, S., Rubie, D.C., Elkins-Tanton, L., Wieczorek, M., Jacobson, S., 2018. The timeline of the lunar bombardment: Revisited. Icarus 305, 262–276. https://doi.org/10.1016/j.icarus.2017.12.046

Morris, R. V., Ruff, S.W., Gellert, R., Ming, D.W., Arvidson, R.E., Clark, B.C., Golden, D.C., Siebach, K., Klingelhöfer, G., Schröder, C., Fleischer, I., Yen, A.S., Squyres, S.W., 2010. Identification of carbonate-rich outcrops on Mars by the spirit rover. Science (80-. ). 329, 421–424. https://doi.org/10.1126/science.1189667

Morschhauser, A., Grott, M., Breuer, D., 2011. Crustal recycling, mantle dehydration, and the thermal evolution of Mars. Icarus 212, 541–558. https://doi.org/10.1016/j.icarus.2010.12.028

Möstl, U. V., Erkaev, N. V., Zellinger, M., Lammer, H., Gröller, H., Biernat, H.K., Korovinskiy, D., 2011. The Kelvin-Helmholtz instability at Venus: What is the unstable boundary? Icarus 216, 476–484. https://doi.org/10.1016/j.icarus.2011.09.012

Musselwhite, D.S., Dalton, H.A., Kiefer, W.S., Treiman, A.H., 2006. Experimental petrology of the basaltic shergottite Yamato-980459: Implications for the thermal structure of the Martian mantle. Meteorit. Planet. Sci. 41, 1271–1290. https://doi.org/10.1111/j.1945-5100.2006.tb00521.x

Nesvorný, D., Vokrouhlický, D., Bottke, W.F., Levison, H.F., n.d. Evidence for very early migration of the Solar System planets from the Patroclus – Menoetius binary Jupiter Trojan. https://doi.org/10.1038/s41550-018-0564-3

Neukum, G., Jaumann, R., Hoffmann, H., Hauber, E., Head, J.W., Basilevsky, A.T., Ivanov, B.A., Werner, S.C., van Gasselt, S., Murray, J.B., McCord, T., Team, H.C.-I., 2004. Recent and episodic volcanic and glacial activity on Mars revealed by the High Resolution Stereo Camera. Nature 432, 971–979.

https://doi.org/10.1038/nature03231

Niem, D. De, Kührt, E., Morbidelli, A., Motschmann, U., 2012. Atmospheric erosion and replenishment induced by impacts upon the Earth and Mars during a heavy bombardment. Icarus 221, 495–507. https://doi.org/10.1016/j.icarus.2012.07.032

Niles, P.B., Catling, D.C., Berger, G., Chassefière, E., Ehlmann, B.L., Michalski, J.R., Morris, R., Ruff, S.W., Sutter, B., 2013. Geochemistry of carbonates on Mars: Implications for climate history and nature of aqueous environments. Space Sci. Rev. 174, 301–328. https://doi.org/10.1007/s11214-012-9940-y

Nilsson, H., Carlsson, E., Brain, D.A., Yamauchi, M., Holmström, M., Barabash, S., Lundin, R., Futaana, Y., 2010. Ion escape from Mars as a function of solar wind conditions: A statistical study. Icarus 206, 40–49. https://doi.org/10.1016/j.icarus.2009.03.006

Nilsson, H., Edberg, N.J.T., Stenberg, G., Barabash, S., Holmström, M., Futaana, Y., Lundin, R., Fedorov, A., 2011. Heavy ion escape from Mars, influence from solar wind conditions and crustal magnetic fields. Icarus 215, 475–484. https://doi.org/10.1016/j.icarus.2011.08.003

Notsu, Y., Shibayama, T., Maehara, H., Notsu, S., Nagao, T., Honda, S., Ishii, T.T., Nogami, D., Shibata, K., 2013. SUPERFLARES ON SOLAR-TYPE STARS OBSERVED WITH KEPLER II. PHOTOMETRIC VARIABILITY OF SUPERFLARE-GENERATING STARS: A SIGNATURE OF STELLAR ROTATION AND STARSPOTS. Astrophys. J. 771, 127. https://doi.org/10.1088/0004-637X/771/2/127

OBRIEN, D., MORBIDELLI, A., LEVISON, H., 2006. Terrestrial planet formation with strong dynamical friction. Icarus 184, 39–58. https://doi.org/10.1016/j.icarus.2006.04.005

Odert, P., Lammer, H., Erkaev, N. V, Nikolaou, A., Lichtenegger, H.I.M., Johnstone, C.P., Kislyakova, K.G., Leitzinger, M., Tosi, N., 2018. Escape and fractionation of volatiles and noble gases from Mars-sized planetary embryos and growing protoplanets. Icarus, Vol. 307, p. 327-346. 307, 327–346. https://doi.org/10.1016/j.icarus.2017.10.031

Odert, P., Leitzinger, M., Hanslmeier, A., Lammer, H., 2017. Stellar coronal mass ejections - I. Estimating occurrence frequencies and mass-loss rates. Mon. Not. R. Astron. Soc. 472, 876–890. https://doi.org/10.1093/MNRAS/STX1969

Owen, J.E., Wu, Y., 2015. Atmospheres of low-mass planets: the "boil-off" Astrophys. Journal, Vol. 817, Issue 2, Artic. id. 107, 14 pp. (2016). 817. https://doi.org/10.3847/0004-637X/817/2/107

Owen, J.E., Wu, Y., 2017. The Evaporation Valley in the Kepler Planets. Astrophys. J. 847, 29. https://doi.org/10.3847/1538-4357/aa890a

Penz, T., Erkaev, N. V., Biernat, H.K., Lammer, H., Amerstorfer, U. V., Guneir, H., Kallio, E., Barabash, S., Orsini, S., Milillo, A., Baumjohann, W., 2004. Ion loss on Mars caused by the Kelvin-Helmholtz instability. Planet. Space Sci. 52, 1157–1167. https://doi.org/10.1016/j.pss.2004.06.001

Pepin, R.O., 1994. Evolution of the Martian Atmosphere. Icarus 111, 289–304. https://doi.org/10.1006/icar.1994.1146

Peretyazhko, T.S., Niles, P.B., Sutter, B., Morris, R. V., Agresti, D.G., Le, L., Ming, D.W., 2018. Smectite formation in the presence of sulfuric acid: Implications for acidic smectite formation on early Mars. Geochim. Cosmochim. Acta 220, 248–260. https://doi.org/10.1016/j.gca.2017.10.004

Perez-de-Tejada, H., 1992. Solar wind erosion of the Mars early atmosphere. J. Geophys. Res. Sp. Phys. 97, 3159–3167. https://doi.org/10.1029/91ja01985

Pérez-de-Tejada, H., 1987. Plasma flow in the Mars magnetosphere. J. Geophys. Res. 92, 4713. https://doi.org/10.1029/JA092iA05p04713

Pham, L.B.S., Karatekin, 2016. Scenarios of atmospheric mass evolution on Mars influenced by asteroid and comet impacts since the late Noachian. Planet. Space Sci. 125, 1–11. https://doi.org/10.1016/j.pss.2015.09.022

Pham, L.B.S., Karatekin, Ö., Dehant, V., 2009. Effects of meteorite impacts on the atmospheric evolution of mars. Astrobiology 9, 45–54. https://doi.org/10.1089/ast.2008.0242

Pham, L.B.S., Karatekin, Ö., Dehant, V., 2011. Effects of impacts on the atmospheric evolution: Comparison between Mars, Earth, and Venus. Planet.

Space Sci. 59, 1087–1092. https://doi.org/10.1016/j.pss.2010.11.010

Pham, L. B.S., Karatekin, Ö., Dehant, V., 2011. Effects of impacts on the atmospheric evolution: Comparison between Mars, Earth, and Venus. Planet. Space Sci. 59, 1087–1092. https://doi.org/10.1016/j.pss.2010.11.010

Phillips, R.J., Davis, B.J., Tanaka, K.L., Byrne, S., Mellon, M.T., Putzig, N.E., Haberle, R.M., Kahre, M.A., Campbell, B.A., Carter, L.M., Smith, I.B., Holt, J.W., Smrekar, S.E., Nunes, D.C., Plaut, J.J., Egan, A.F., Titus, T.N., Seu, R., 2011. Massive CO2 ice deposits sequestered in the south polar layered deposits of Mars. Science (80-. ). 332, 838–841. https://doi.org/10.1126/science.1203091

Phillips, R.J., Zuber, M.T., Solomon, S.C., Golombek, M.P., Jakosky, B.M., Banerdt, W.B., Smith, D.E., Williams, R.M.E., Hynek, B.M., Aharonson, O., Hauck, S.A., 2001. Ancient Geodynamics and Global-Scale Hydrology on Mars. Science (80-. ). 291, 2587–2591. https://doi.org/10.1126/science.1058701

Plescia, J.B., 1990. Recent flood lavas in the Elysium region of Mars. Icarus 88, 465–490. https://doi.org/10.1016/0019-1035(90)90095-Q

Pollack, J.B., Kasting, J.F., Richardson, S.M., Poliakoff, K., 1987. The case for a wet, warm climate on early Mars. Icarus 71, 203–224. https://doi.org/10.1016/0019-1035(87)90147-3

Quesnel, Y., Sotin, C., Langlais, B., Costin, S., Mandea, M., Gottschalk, M., Dyment, J., 2009. Serpentinization of the martian crust during Noachian. Earth Planet. Sci. Lett. 277, 184–193. https://doi.org/10.1016/j.epsl.2008.10.012

Rahmati, A., Larson, D.E., Cravens, T.E., Lillis, R.J., Dunn, P.A., Halekas, J.S., Connerney, J.E., Eparvier, F.G., Thiemann, E.M.B., Jakosky, B.M., 2015. MAVEN insights into oxygen pickup ions at Mars. Geophys. Res. Lett. 42, 8870–8876. https://doi.org/10.1002/2015GL065262

Rai, N., Van Westrenen, W., 2013. Core-mantle differentiation in Mars. J. Geophys. Res. E Planets 118, 1195–1203. https://doi.org/10.1002/jgre.20093

Ramirez, R.M., Craddock, R.A., Usui, T., 2020. Climate Simulations of Early Mars

With Estimated Precipitation, Runoff, and Erosion Rates, Journal of Geophysical Research: Planets. https://doi.org/10.1029/2019JE006160

Ramirez, Ramses M., Kopparapu, R., Zugger, M.E., Robinson, T.D., Freedman, R., Kasting, J.F., 2014. Warming early Mars with CO 2 and H 2. Nat. Geosci. 7, 59–63. https://doi.org/10.1038/ngeo2000

Ramirez, Ramses M, Kopparapu, R., Zugger, M.E., Robinson, T.D., Freedman, R., Kasting, J.F., 2014. Warming early Mars with CO 2 and H 2 7, 59–63. https://doi.org/10.1038/ngeo2000

Ramstad, R., Barabash, S., Futaana, Y., Nilsson, H., Holmström, M., 2018. Ion Escape From Mars Through Time: An Extrapolation of Atmospheric Loss Based on 10 Years of Mars Express Measurements. J. Geophys. Res. Planets 3051–3060. https://doi.org/10.1029/2018JE005727

Ramstad, R., Barabash, S., Futaana, Y., Nilsson, H., Wang, X.D., Holmström, M., 2015. The Martian atmospheric ion escape rate dependence on solar wind and solar EUV conditions: 1. Seven years of Mars Express observations. J. Geophys. Res. E Planets 120, 1298–1309. https://doi.org/10.1002/2015JE004816

Ramstad, R., Barabash, S., Futaana, Y., Yamauchi, M., Nilsson, H., Holmström, M., 2017. Mars Under Primordial Solar Wind Conditions: Mars Express Observations of the Strongest CME Detected at Mars Under Solar Cycle #24 and its Impact on Atmospheric Ion Escape. Geophys. Res. Lett. 44, 10,805-10,811. https://doi.org/10.1002/2017GL075446

Ramstad, R., Futaana, Y., Barabash, S., Nilsson, H., Martin, S., Lundin, R., Schwingenschuh, K., 2013. Phobos 2 / ASPERA data revisited : Planetary ion escape rate from Mars near the 1989 solar maximum 40, 477–481. https://doi.org/10.1002/grl.50149

Raymond, S.N., O'Brien, D.P., Morbidelli, A., Kaib, N.A., 2009. Building the Terrestrial Planets: Constrained Accretion in the Inner Solar System. Icarus, Vol. 203, Issue 2, p. 644-662. 203, 644–662. https://doi.org/10.1016/j.icarus.2009.05.016

Ribas, I., 2005. Evolution of the Solar Activity over Time and Effects on Planetary

Atmospheres. I. High-Energy Irradiances (1-1700 Ã…). Astrophys. J. 622, 680. https://doi.org/10.1086/427977

Ribas, I., Guinan, E.F., Guedel, M., Audard, M., 2005. Evolution of the Solar Activity over Time and Effects on Planetary Atmospheres: I. High-energy Irradiances (1-1700 A). Astrophys. Journal, Vol. 622, Issue 1, pp. 680-694. 622, 680–694. https://doi.org/10.1086/427977

RIGHTER, K., CHABOT, N.L., 2011. Moderately and slightly siderophile element constraints on the depth and extent of melting in early Mars. Meteorit. Planet. Sci. 46, 157–176. https://doi.org/10.1111/j.1945-5100.2010.01140.x

Righter, K., Hervig, R., Kring, D., 1998. Accretion and core formation on Mars: molybdenum contents of melt inclusion glasses in three SNC meteorites. Geochim. Cosmochim. Acta 62, 2167–2177. https://doi.org/10.1016/S0016-7037(98)00132-X

Righter, K., Yang, H., Costin, G., Downs, R.T., 2008. Oxygen fugacity in the Martian mantle controlled by carbon: New constraints from the nakhlite MIL 03346. Meteorit. Planet. Sci. 43, 1709–1723. https://doi.org/10.1111/j.1945-5100.2008.tb00638.x

Robbins, S.J., Hynek, B.M., Lillis, R.J., Bottke, W.F., 2013. Large impact crater histories of Mars : The effect of different model crater age techniques. Icarus 225, 173–184. https://doi.org/10.1016/j.icarus.2013.03.019

Romanek, C.S., Grady, M.M., Wright, I.P., Mittlefehldt, D.W., Socki, R.A., Pillinger, C.T., Gibson, E.K., 1994. Record of fluid-rock interactions on Mars from the meteorite ALH84001. Nature 372, 655–657. https://doi.org/10.1038/372655a0

Rubie, D.C., Jacobson, S.A., Morbidelli, A., Brien, D.P.O., Young, E.D., Vries, J. De, Nimmo, F., Palme, H., Frost, D.J., 2015. Accretion and differentiation of the terrestrial planets with implications for the compositions of early-formed Solar System bodies and accretion of water. Icarus 248, 89–108. https://doi.org/10.1016/j.icarus.2014.10.015

Ruhunusiri, S., Halekas, J.S., McFadden, J.P., Connerney, J.E.P., Espley, J.R., Harada, Y., Livi, R., Seki, K., Mazelle, C., Brain, D., Hara, T., DiBraccio,

G.A., Larson, D.E., Mitchell, D.L., Jakosky, B.M., Hasegawa, H., 2016. MAVEN observations of partially developed Kelvin-Helmholtz vortices at Mars. Geophys. Res. Lett. 43, 4763–4773. https://doi.org/10.1002/2016GL068926

Saal, A.E., Hauri, E.H., Langmuir, C.H., Perfit, M.R., 2002. Vapour undersaturation in primitive mid-ocean-ridge basalt and the volatile content of Earth's upper mantle. Nature 419, 451–455. https://doi.org/10.1038/nature01073

Sackmann, I.-J., Boothroyd, A.I., 2002. Our Sun. V. A Bright Young Sun Consistent with Helioseismology and Warm Temperatures on Ancient Earth and Mars. Astrophys. Journal, Vol. 583, Issue 2, pp. 1024-1039. 583, 1024–1039. https://doi.org/10.1086/345408

Saito, H., Kuramoto, K., 2018. Formation of a hybrid-type proto-atmosphere on mars accreting in the solar nebula. Mon. Not. R. Astron. Soc. 475, 1274–1287. https://doi.org/10.1093/mnras/stx3176

Sakai, S., Seki, K., Terada, N., Shinagawa, H., Tanaka, T., Ebihara, Y., 2018. Effects of a Weak Intrinsic Magnetic Field on Atmospheric Escape From Mars. Geophys. Res. Lett. 45, 9336–9343. https://doi.org/10.1029/2018GL079972

Sakata, R., Seki, K., Sakai, S., Terada, N., Shinagawa, H., Tanaka, T., 2020. Effects of an Intrinsic Magnetic Field on Ion Loss From Ancient Mars Based on Multispecies MHD Simulations. J. Geophys. Res. Sp. Phys. 125, e26945. https://doi.org/10.1029/2019ja026945

Saxena, P., Killen, R.M., Airapetian, V., Petro, N.E., Curran, N.M., Mandell, A.M., 2019. Was the Sun a Slow Rotator? Sodium and Potassium Constraints from the Lunar Regolith. Astrophys. J. 876, L16. https://doi.org/10.3847/2041-8213/ab18fb

Schaufelberger, A., Wurz, P., Lammer, H., Kulikov, Y.N., 2012. Is hydrodynamic escape from Titan possible?, in: Planetary and Space Science. pp. 79–84. https://doi.org/10.1016/j.pss.2011.03.011

Schlichting, H.E., Sari, R., Yalinewich, A., 2015. Atmospheric mass loss during planet formation: The importance of planetesimal impacts. Icarus 247, 81–94.

https://doi.org/10.1016/j.icarus.2014.09.053

Segura, T.L., Toon, O.B., Colaprete, A., Zahnle, K., 2002. Environmental Effects of Large Impacts on Mars.

Shibata, K., Isobe, H., Hillier, A., Choudhuri, A.R., Maehara, H., Ishii, T.T., Shibayama, T., Notsu, S., Notsu, Y., Nagao, T., Honda, S., Nogami, D., 2013. Can superflares occur on our sun? Publ. Astron. Soc. Japan 65. https://doi.org/10.1093/pasj/65.3.49

Shizgal, B.D., Arkos, G.G., 1996. Nonthermal escape of the atmospheres of Venus, Earth, and Mars. Rev. Geophys. 34, 483–505. https://doi.org/10.1029/96RG02213

Shuvalov, V., V., 2009. Atmospheric erosion induced by oblique impacts. Meteorit. Planet. Sci. 44, 1095–1105. https://doi.org/10.1111/j.1945-5100.2009.tb01209.x

Shuvalov, V. V., Artemieva, N.A., 2001. Atmosheric Erosion and Radiation Impulse Induced by Impacts. Int. Conf. Catastrophic Events Mass Extinctions Impacts Beyond, 9-12 July 2000, Vienna, Austria, Abstr. no.3060. http://adsabs.harvard.edu/abs/2001caev.conf.3060S

Sleep, N.H., Zahnle, K., 1998. Refugia from asteroid impacts on early Mars and the early Earth impacts are large enough to vaporize the oceans J ), thermal buffering serves only to buffering does not occur on Relatively small impacts J ) frequently heat the surface 3 months 103.

Slipski, M., Jakosky, B.M., 2016. Argon isotopes as tracers for martian atmospheric loss. Icarus 272, 212–227. https://doi.org/10.1016/j.icarus.2016.02.047

Smithtro, C.G., Sojka, J.J., 2005. Behavior of the ionosphere and thermosphere subject to extreme solar cycle conditions. J. Geophys. Res. Sp. Phys. 110, A08306. https://doi.org/10.1029/2004JA010782

Soto, A., Mischna, M., Schneider, T., Lee, C., Richardson, M., 2015. Martian atmospheric collapse: Idealized GCM studies. Icarus 250, 553–569. https://doi.org/10.1016/j.icarus.2014.11.028

Sousa, R. De, Morbidelli, A., Raymond, S.N., Izidoro, A., Gomes, R., Vieira, E., 2020. Dynamical evidence for an early giant planet instability 339.

Spreiter, J.R., Stahara, S.S., 1980. A new predictive model for determining solar wind-terrestrial planet interactions. J. Geophys. Res. 85, 6769. https://doi.org/10.1029/JA085iA12p06769

Squyres, S.W., Grotzinger, J.P., Arvidson, R.E., Bell, J.F., Calvin, W., Christensen, P.R., Clark, B.C., Crisp, J.A., Farrand, W.H., Herkenhoff, K.E., Johnson, J.R., Klingelhöfer, G., Knoll, A.H., McLennan, S.M., McSween, H.Y., Morris, R. V., Rice, J.W., Rieder, R., Soderblom, L.A., 2004. In situ evidence for an ancient aqueous environment at Meridiani Planum, Mars. Science (80-. ). 306, 1709–1714. https://doi.org/10.1126/science.1104559

Stanley, B.D., Hirschmann, M.M., Withers, A.C., 2011. CO2 solubility in Martian basalts and Martian atmospheric evolution. Geochim. Cosmochim. Acta 75, 5987–6003. https://doi.org/10.1016/j.gca.2011.07.027

Stevenson, D.J., 2001. Mars' core and magnetism. Nature 412, 214–219. https://doi.org/10.1038/35084155

Stökl, A., Dorfi, E., Lammer, H., 2015. Hydrodynamic simulations of captured protoatmospheres around Earth-like planets. Astron. & Astrophys. Vol. 576, id.A87, 11 pp. 576, A87. https://doi.org/10.1051/0004-6361/201423638

Stökl, A., Dorfi, E.A., Johnstone, C.P., Lammer, H., 2016. Dynamical Accretion of Primordial Atmospheres around Planets with Masses between 0.1 and 5 M $_{{\ensuremath{\oplus}}}$ in the Habitable Zone. \apj 825, 86. https://doi.org/10.3847/0004-637X/825/2/86

Strobel, D.F., 2009. Titan's hydrodynamically escaping atmosphere: Escape rates and the structure of the exobase region. Icarus 202, 632–641. https://doi.org/10.1016/j.icarus.2009.03.007

Strobel, D.F., F., D., 2008. N2 escape rates from Pluto's atmosphere. Icarus 193, 612–619. https://doi.org/10.1016/j.icarus.2007.08.021

Svetsov, V. V., 2007. Atmospheric erosion and replenishment induced by impacts of cosmic bodies upon the Earth and Mars. Sol. Syst. Res. 41, 28–41. https://doi.org/10.1134/S0038094607010030

Svetsov, V. V., V., V., 2007. Atmospheric erosion and replenishment induced by impacts of cosmic bodies upon the Earth and Mars. Sol. Syst. Res. 41, 28–41.

https://doi.org/10.1134/S0038094607010030

Tang, H., Dauphas, N., 2014. 60Fe-60Ni chronology of core formation in Mars. Earth Planet. Sci. Lett. 390, 264–274. https://doi.org/10.1016/j.epsl.2014.01.005

Tarduno, J.A., Blackman, E.G., Mamajek, E.E., 2014. Detecting the oldest geodynamo and attendant shielding from the solar wind: Implications for habitability. Phys. Earth Planet. Inter. 233, 68–87. https://doi.org/10.1016/j.pepi.2014.05.007

Terada, N., Kulikov, Y.N., Lammer, H., Lichtenegger, H.I.M., Tanaka, T., Shinagawa, H., Zhang, T., 2009. Atmosphere and Water Loss from Early Mars Under Extreme Solar Wind and Extreme Ultraviolet Conditions. Astrobiology 9, 55–70. https://doi.org/10.1089/ast.2008.0250

Thomas, R.J., Hynek, B.M., Osterloo, M.M., Kierein-Young, K.S., 2017. Widespread exposure of Noachian phyllosilicates in the Margaritifer region of Mars: Implications for paleohydrology and astrobiological detection. J. Geophys. Res. Planets 122, 483–500. https://doi.org/10.1002/2016JE005183

Tian, F., Kasting, J.F., Liu, H.L., Roble, R.G., 2008a. Hydrodynamic planetary thermosphere model: 1. Response of the Earth's thermosphere to extreme solar EUV conditions and the significance of adiabatic cooling. J. Geophys. Res. E Planets 113, 1–19. https://doi.org/10.1029/2007JE002946

Tian, F., Kasting, J.F., Solomon, S.C., 2009. Thermal escape of carbon from the early Martian atmosphere. Geophys. Res. Lett. 36, 1–5. https://doi.org/10.1029/2008GL036513

Tian, F., Solomon, S.C., Qian, L., Lei, J., Roble, R.G., 2008b. Hydrodynamic planetary thermosphere model: 2. Coupling of an electron transport/energy deposition model. J. Geophys. Res. E Planets 113, 1–13. https://doi.org/10.1029/2007JE003043

Tomkinson, T., Lee, M.R., Mark, D.F., Smith, C.L., 2013. Sequestration of Martian CO2 by mineral carbonation. Nat. Commun. 4, 1–6. https://doi.org/10.1038/ncomms3662

Tosca, N.J., Ahmed, I.A.M., Tutolo, B.M., Ashpitel, A., Hurowitz, J.A., 2018.

Magnetite authigenesis and the warming of early Mars. Nat. Geosci. 11, 635–639. https://doi.org/10.1038/s41561-018-0203-8

Tu, L., Johnstone, C.P., Güdel, M., Lammer, H., 2015. The extreme ultraviolet and X-ray Sun in Time: High-energy evolutionary tracks of a solar-like star. Astron. Astrophys. 577, L3. https://doi.org/10.1051/0004-6361/201526146

Tucker, O.J., Johnson, R.E., 2009. Thermally driven atmospheric escape: Monte Carlo simulations for Titan's atmosphere. Planet. Space Sci. 57, 1889–1894. https://doi.org/10.1016/j.pss.2009.06.003

Valeille, A., Bougher, S.W., Tenishev, V., Combi, M.R., Nagy, A.F., 2010. Water loss and evolution of the upper atmosphere and exosphere over martian history. Icarus 206, 28–39. https://doi.org/10.1016/j.icarus.2009.04.036

Vaucher, J., Baratoux, D., Mangold, N., Pinet, P., Kurita, K., Grégoire, M., 2009. The volcanic history of central Elysium Planitia: Implications for martian magmatism. Icarus 204, 418–442. https://doi.org/10.1016/j.icarus.2009.06.032

Vervelidou, F., Lesur, V., Grott, M., Morschhauser, A., Lillis, R.J., 2017. Constraining the Date of the Martian Dynamo Shutdown by Means of Crater Magnetization Signatures. J. Geophys. Res. Planets 122, 2294–2311. https://doi.org/10.1002/2017JE005410

Villanueva, G.L., Mumma, M.J., Novak, R.E., Käufl, H.U., Hartogh, P., Encrenaz, T., Tokunaga, A., Khayat, A., Smith, M.D., 2015. Strong water isotopic anomalies in the martian atmosphere: Probing current and ancient reservoirs. Science (80-. ). 348, 218–221. https://doi.org/10.1126/science.aaa3630

Volkov, A.N., Johnson, R.E., 2013. Thermal escape in the hydrodynamic regime: Reconsideration of parker's isentropic theory based on results of kinetic simulations. Astrophys. J. 765, 90. https://doi.org/10.1088/0004-637X/765/2/90

Volkov, A.N., Johnson, R.E., Tucker, O.J., Erwin, J.T., 2011. THERMALLY DRIVEN ATMOSPHERIC ESCAPE : TRANSITION FROM HYDRODYNAMIC TO JEANS ESCAPE 24, 0–4. https://doi.org/10.1088/2041-8205/729/2/L24

Von Paris, P., Grenfell, J.L., Rauer, H., Stock, J.W., 2013. N2-associated surface warming on early Mars. Planet. Space Sci. 82–83, 149–154. https://doi.org/10.1016/j.pss.2013.04.009

Wade, J., Dyck, B., Palin, R.M., Moore, J.D.P., Smye, A.J., 2017. The divergent fates of primitive hydrospheric water on Earth and Mars. Nature 552, 391–394. https://doi.org/10.1038/nature25031

Walsh, K.J., Morbidelli, A., Raymond, S.N., O'Brien, D.P., Mandell, A.M., 2011. A low mass for Mars from Jupiter's early gas-driven migration. Nature 475, 206–209. https://doi.org/10.1038/nature10201

Wang, H., Weiss, B.P., Bai, X.N., Downey, B.G., Wang, Jun, Wang, Jiajun, Suavet, C., Fu, R.R., Zucolotto, M.E., 2017. Lifetime of the solar nebula constrained by meteorite paleomagnetism. Science (80-. ). 355, 623–627. https://doi.org/10.1126/science.aaf5043

Wang, Y.-C., Luhmann, J.G., Leblanc, F., Fang, X., Johnson, R.E., Ma, Y., Ip, W.-H., Li, L., 2014. Modeling of the $O^+$ pickup ion sputtering efficiency dependence on solar wind conditions for the Martian atmosphere. J. Geophys. Res. Planets 119, 93–108. https://doi.org/10.1002/2013JE004413

Wang, Y.C., Luhmann, J.G., Fang, X., Leblanc, F., Johnson, R.E., Ma, Y., Ip, W.H., 2015. Statistical studies on Mars atmospheric sputtering by precipitating pickup O+: Preparation for the MAVEN mission. J. Geophys. Res. Planets 120, 34–50. https://doi.org/10.1002/2014JE004660

Warren, A.O., Kite, E.S., Williams, J.P., Horgan, B., 2019. Through the Thick and Thin: New Constraints on Mars Paleopressure History 3.8 – 4 Ga from Small Exhumed Craters. J. Geophys. Res. Planets 124, 2793–2818. https://doi.org/10.1029/2019JE006178

Watson, A.J., Donahue, T.M., Walker, J.C.G., 1981. The dynamics of a rapidly escaping atmosphere: Applications to the evolution of Earth and Venus. Icarus 48, 150–166. https://doi.org/10.1016/0019-1035(81)90101-9

Webster, C.R., Mahaffy, P.R., Flesch, G.J., Niles, P.B., Jones, J.H., Leshin, L.A., Atreya, S.K., Stern, J.C., Christensen, L.E., Owen, T., Franz, H., Pepin, R.O., Steele, A., 2013. Isotope ratios of H, C, and O in CO2 and H2O of the martian

atmosphere. Science (80-. ). 341, 260–263. https://doi.org/10.1126/science.1237961

Werner, S.C., Tanaka, K.L., 2011. Redefinition of the crater-density and absolute-age boundaries for the chronostratigraphic system of Mars. Icarus 215, 603–607. https://doi.org/10.1016/j.icarus.2011.07.024

Wetzel, D.T., Rutherford, M.J., Jacobsen, S.D., Hauri, E.H., Saal, A.E., 2013. Degassing of reduced carbon from planetary basalts. Proc. Natl. Acad. Sci. 110, 8010–8013. https://doi.org/10.1073/pnas.1219266110

Wong, M.H., Atreya, S.K., Mahaffy, P.N., Franz, H.B., Malespin, C., Trainer, M.G., Stern, J.C., Conrad, P.G., Manning, H.L.K., Pepin, R.O., Becker, R.H., McKay, C.P., Owen, T.C., Navarro-González, R., Jones, J.H., Jakosky, B.M., Steele, A., 2013. Isotopes of nitrogen on Mars: Atmospheric measurements by Curiosity's mass spectrometer. Geophys. Res. Lett. 40, 6033–6037. https://doi.org/10.1002/2013GL057840

Wood, B.E., Mueller, H.-R., Zank, G.P., Linsky, J.L., 2002. Measured Mass Loss Rates of Solar-like Stars as a Function of Age and Activity. Astrophys. Journal, Vol. 574, Issue 1, pp. 412-425. 574, 412–425. https://doi.org/10.1086/340797

Wood, B.E., Mueller, H.-R., Zank, G.P., Linsky, J.L., Redfield, S., 2005. New Mass Loss Measurements from Astrospheric Lyman-alpha Absorption. Astrophys. Journal, Vol. 628, Issue 2, pp. L143-L146. 628, L143–L146. https://doi.org/10.1086/432716

Wordsworth, R., Forget, F., Millour, E., Head, J.W., Madeleine, J., Charnay, B., 2013. Global modelling of the early martian climate under a denser CO 2 atmosphere : Water cycle and ice evolution. Icarus 222, 1–19. https://doi.org/10.1016/j.icarus.2012.09.036

Wray, J.J., Murchie, S.L., Bishop, J.L., Ehlmann, B.L., Milliken, R.E., Wilhelm, M.B., Seelos, K.D., Chojnacki, M., 2016. Orbital evidence for more widespread carbonate-bearing rocks on Mars. J. Geophys. Res. Planets 121, 652–677. https://doi.org/10.1002/2015JE004972

Xiao, L., Huang, J., Christensen, P.R., Greeley, R., Williams, D.A., Zhao, J., He,

Q., 2012. Ancient volcanism and its implication for thermal evolution of Mars. Earth Planet. Sci. Lett. 323–324, 9–18. https://doi.org/10.1016/j.epsl.2012.01.027

Xu, S., Mitchell, D., Liemohn, M., Dong, C., Bougher, S., Fillingim, M., Lillis, R., McFadden, J., Mazelle, C., Connerney, J., Jakosky, B., 2016. Deep nightside photoelectron observations by MAVEN SWEA: Implications for Martian northern hemispheric magnetic topology and nightside ionosphere source. Geophys. Res. Lett. 43, 8876–8884. https://doi.org/10.1002/2016GL070527

Yamauchi, M., Hara, T., Lundin, R., Dubinin, E., Fedorov, A., Sauvaud, J.-A., Frahm, R.A., Ramstad, R., Futaana, Y., Holmstrom, M., Barabash, S., 2015. Seasonal variation of Martian pick-up ions: Evidence of breathing exosphere. Planet. Space Sci. 119, 54–61. https://doi.org/10.1016/j.pss.2015.09.013

Yelle, R. V., 2004. Aeronomy of extra-solar giant planets at small orbital distances. Icarus 170, 167–179. https://doi.org/10.1016/j.icarus.2004.02.008

Yoshida, T., Kuramoto, K., 2020. Sluggish hydrodynamic escape of early Martian atmosphere with reduced chemical compositions. Icarus 345, 113740. https://doi.org/10.1016/j.icarus.2020.113740

Zahnle, K., Haberle, R.M., Catling, D.C., Kasting, J.F., 2008. Photochemical instability of the ancient Martian atmosphere. J. Geophys. Res. E Planets 113, 1–16. https://doi.org/10.1029/2008JE003160

Zahnle, K., Kasting, J.F., Pollack, J.B., 1990. Mass fractionation of noble gases in diffusion-limited hydrodynamic hydrogen escape. Icarus 84, 502–527. https://doi.org/10.1016/0019-1035(90)90050-J

Zahnle, K.J., Gacesa, M., Catling, D.C., 2019. Strange messenger: A new history of hydrogen on Earth, as told by Xenon. Geochim. Cosmochim. Acta 244, 56–85. https://doi.org/10.1016/J.GCA.2018.09.017

Zahnle, K.J., Kasting, J.F., 1986. Mass fractionation during transonic escape and implications for loss of water from Mars and Venus. Icarus 68, 462–480. https://doi.org/10.1016/0019-1035(86)90051-5

Zent, A.P., Quinn, R.C., 1995. Simultaneous adsorption of CO2 and H2O under Mars-like conditions and application to the evolution of the Martian climate.

J. Geophys. Res. 100, 5341–5349. https://doi.org/10.1029/94JE01899

Zhang, H.L., Hirschmann, M.M., Cottrell, E., Withers, A.C., 2017. Effect of pressure on Fe3+/ΣFe ratio in a mafic magma and consequences for magma ocean redox gradients. Geochim. Cosmochim. Acta. https://doi.org/10.1016/j.gca.2017.01.023

Zhang, M. H. G., Luhmann, J.G., Bougher, S.W., Nagy, A.F., 1993. The ancient oxygen exosphere of Mars: Implications for atmosphere evolution. J. Geophys. Res. 98, 10915. https://doi.org/10.1029/93JE00231

Zhang, M. H.G., Luhmann, J.G., Bougher, S.W., Nagy, A.F., 1993. The ancient oxygen exosphere of Mars: implications for atmosphere evolution. J. Geophys. Res. 98, 915–923. https://doi.org/10.1029/93je00231

Zhao, J., Tian, F., 2015. Photochemical escape of oxygen from early Mars. Icarus 250, 477–481. https://doi.org/10.1016/j.icarus.2014.12.032

Zhao, J., Tian, F., Ni, Y., Huang, X., 2017. DR-induced escape of O and C from early Mars. Icarus 284, 305–313. https://doi.org/10.1016/j.icarus.2016.11.021

Zolotov, M.Y., Mironenko, M. V., 2016. Chemical models for martian weathering profiles: Insights into formation of layered phyllosilicate and sulfate deposits. Icarus 275, 203–220. https://doi.org/10.1016/j.icarus.2016.04.011